\documentclass{jfm}

\usepackage{graphicx}
\usepackage{epstopdf,epsfig}
\usepackage{newtxtext}
\usepackage{newtxmath}
\usepackage{natbib}
\usepackage{subcaption}
\usepackage{hyperref}
\hypersetup{
    colorlinks = true,
    urlcolor   = blue,
    citecolor  = black,
}

\newcommand{\RomanNumeralCaps}[1]
\linenumbers

\newcommand{\R}{\mathbb{R}}
\newcommand{\Z}{\mathbb{Z}}

\newcommand{\C}{\mathbb{C}}
\newcommand{\T}{\mathbb{T}}

\newcommand{\bu}{\boldsymbol{u}}

\newcommand{\bx}{\boldsymbol{x}}

\newcommand{\bk}{\boldsymbol{k}}
\newcommand{\be}{\boldsymbol{e}}
\newcommand{\bp}{\boldsymbol{p}}

\renewcommand{\div}{{\rm{div}\,}}
\newcommand{\divp}{{\rm{div}}_p}

\newcommand{\abs}[1]{\left\lvert #1 \right\rvert}
\newcommand{\norm}[1]{\left\lVert #1 \right\rVert}

\newcommand{\mc}[1]{\mathcal{#1}}

\allowdisplaybreaks

\graphicspath{{figures/}}

\title{Weakly nonlinear analysis of pattern formation in active suspensions}

\author{Laurel Ohm\aff{1}
  \corresp{\email{laurel.ohm@princeton.edu}}
 \and Michael J. Shelley\aff{2,}\aff{3}}

\affiliation{\aff{1}Department of Mathematics, Princeton University, Princeton, NJ 08540
\aff{2}Courant Institute of Mathematical Sciences, New York University, New York, NY 10012
\aff{3}Center for Computational Biology, Flatiron Institute, Simons Foundation, New York, NY 10010}

\begin{document}
\maketitle

\begin{abstract}
We consider the Saintillan--Shelley kinetic model of active rodlike particles in Stokes flow \citep{saintillan2008instabilitiesPRL,saintillan2008instabilitiesPOF}, for which the uniform, isotropic suspension of pusher particles is known to be unstable in certain settings. Through weakly nonlinear analysis accompanied by numerical simulations, we determine exactly how the isotropic steady state loses stability in different parameter regimes. We study each of the various types of bifurcations admitted by the system, including both subcritical and supercritical Hopf and pitchfork bifurcations. Elucidating this system's behavior near these bifurcations provides a theoretical means of comparing this model with other physical systems which transition to turbulence, and makes predictions about the nature of bifurcations in active suspensions that can be explored experimentally.
\end{abstract}



\section{Introduction}

Inherently nonequilibrium suspensions of active particles abound in biological and experimental settings \citep{gompper20202020,marchetti2013hydrodynamics}.
 For example, motile bacteria such as \emph{E. coli} and \emph{Bacillus subtilis} propel themselves through their surrounding fluid environment, interacting through their induced flow fields \citep{lauga2009hydrodynamics,lushi2014fluid,mendelson1999organized,pedley1992hydrodynamic}, while likewise immersed microtubule bundles slide and extend, driven by ATP-driven molecular motors \citep{decamp2015orientational,henkin2014tunable,needleman2017active,opathalage2019self,sanchez2012spontaneous}. These active suspensions are remarkable because, despite the near lack of inertial effects relative to viscous ones, the activity of the particles can produce large-scale coherent flows and even so-called \emph{active} or \emph{bacterial turbulence}, characterized by chaotic fluctuations in particle concentration and fluid velocity \citep{dombrowski2004self,doostmohammadi2017onset,dunkel2013fluid,gachelin2014collective,koch2011collective,nishiguchi2017long,peng2021imaging,simha2002hydrodynamic,sokolov2012physical,sokolov2007concentration,stenhammar2017role,thampi2014vorticity,zhang2010collective}. \\

Here we consider the kinetic model for a dilute suspension of active elongated particles developed by Saintillan and Shelley \citep{saintillan2008instabilitiesPRL,saintillan2008instabilitiesPOF}. We note that a similar model was developed independently by \cite{subramanian2009critical}, and that both models share many similarities with the Doi--Edwards model for passive polymers \citep{doi1981molecular,doi1986theory}. 
In the dilute limit, particles only interact with each other hydrodynamically by exerting an `active stress' on the surrounding fluid. Even in this setting, changes in particle density and activity are known to cause the suspension to  transition from a uniform, isotropic state to more complex states -- many of which are observed experimentally -- involving large-scale patterns in particle alignment and concentration. 
The kinetic model \citep{saintillan2008instabilitiesPRL,saintillan2008instabilitiesPOF} thus presents an opportunity to elucidate some of the fundamental mechanisms behind the transition to collective dynamics and active turbulence.  \\

To understand this transition, we perform a multiple timescales expansion to determine exactly how the uniform, isotropic steady state in the 2D kinetic model loses stability in different parameter regimes. The variety of predicted behaviors near the onset of instability, which we verify through numerical simulations, is surprisingly complex. Linear stability analysis alone shows that, depending on the (fixed) ratio of particle diffusivity to concentration, the uniform isotropic state can lose stability through either a pitchfork or Hopf bifurcation. Here the bifurcation parameter is a ratio of the particle swimming speed to the particle concentration and magnitude of active stress the particles exert. Our weakly nonlinear analysis shows that both the pitchfork and Hopf parameter regions can be further subdivided into subcritical and supercritical regions, again depending on the ratio of particle diffusivity to concentration.   \\

Numerically, we find hysteresis in the subcritical Hopf region, where far-from-isotropic quasiperiodic patterns of particle alignment are bistable with the uniform isotropic state. The patterns in this region are perhaps precursors to active turbulence.
However, the dimensionality of the initial perturbation to the isotropic steady state makes a difference. If the initial perturbation is one dimensional, i.e. purely in the $x$ or $y$ direction, then only a supercritical Hopf bifurcation can occur, and we numerically locate the stable limit cycle that arises. An example of a 2D limit cycle is also located numerically within the region in which both 1D and 2D perturbations give rise to a supercritical Hopf bifurcation. In the supercritical pitchfork setting, which includes immotile (but active) particles, we identify the stable steady states emerging just beyond the bifurcation. These steady states resemble the steady vortex found in \cite{wioland2013confinement}, although we consider periodic boundary conditions rather than confinement.  \\

A key takeaway is that even this simple kinetic model is capable of capturing many different types of transitions to collective behavior in an active suspension. The different bifurcations analyzed here can be compared with systematic numerical studies of phase transitions in other active suspension models \citep{forest2004flow,forest2004weak,giomi2011excitable,giomi2012banding,xiao2014near,yang2014spiral} and help explain the 1D and 2D patterns -- including limit cycles and other attractors -- located numerically in \cite{forest2014rheological,forest2013kinetic,forest2015kinetic} for a similar version of the kinetic model. The weakly nonlinear analysis performed here also facilitates comparison with the normal forms arising in more classical pattern formation processes in fluid mechanics, especially thermal convection \citep{crawford1991symmetry,cross1993pattern,knobloch1986oscillatory,pomeau1986front,schopf1993convection,swift1977hydrodynamic,swinney1981hydrodynamic}, but also other phenomena arising in complex fluids such as electrohydrodynamic convection in nematic liquid crystals \citep{bodenschatz1988electrically} and the transition from sub- to supercritical instability in viscoelastic pipe flow \citep{wan2021subcritical}. \\

The paper begins by introducing the kinetic model (section \ref{subsec:model}) and recapping the well-studied linear stability analysis (sections \ref{subsec:nondim}, \ref{subsec:linearstab}) and role of rotational diffusion (section \ref{subsec:DR}). Readers familiar with the model may wish to skip directly to the outline of results in section \ref{subsec:outline}, where the types of bifurcations are mapped out in greater detail.

\section{Background}
\subsection{Kinetic model of an active suspension}\label{subsec:model}
In the kinetic model \citep{saintillan2008instabilitiesPRL, saintillan2008instabilitiesPOF}, a suspension of $N$ elongated particles in a 2-dimensional periodic box of length $L$ is modeled as a number density $\Psi(\bx,\bp,t)$ of particles with center-of-mass position $\bx$ and orientation $\bp\in S^{1}$ at time $t$. Due to conservation of the number of particles, the density $\Psi$ evolves according to a Smoluchowski equation: 
\begin{equation}\label{DSS_dim1}
\p_t \Psi = -\bnabla\bcdot(\dot\bx\Psi) - \bnabla_p\bcdot(\dot\bp\Psi); \quad \bnabla_p := ({\boldsymbol I} -\bp\bp^{\rm T})\p_{\bp}.
\end{equation}
Here $\bnabla_p\bcdot$ denotes the divergence on the unit sphere. 
The translational and rotational fluxes are given by
\begin{align}
\dot \bx &= V_0\bp + \bu - D_T\bnabla(\log \Psi)  \label{DSS_dim2} \\
\dot \bp &= ({\boldsymbol I}-\bp\bp^{\rm T})(\bnabla \bu \, \bp) - D_R\bnabla_p(\log \Psi).  \label{DSS_dim3}
\end{align}
The translational velocity consists of a particle swimming term with speed $V_0$ in direction $\bp$, particle advection by the surrounding fluid flow, and translational diffusion. For simplicity we take the translational diffusion to be isotropic. The rotational velocity depends on a Jeffery term for the rotation of an elongated particle in Stokes flow \citep{jeffery1922motion}, written here in the infinitely slender limit, along with rotational diffusion.  \\

Finally, the surrounding fluid medium satisfies the Stokes equations with active forcing:
\begin{align}
-\mu\Delta \bu + \bnabla q &= \bnabla\bcdot {\boldsymbol \Sigma}^a, \; \bnabla\bcdot\bu =0 \label{DSS_dim4} \\
 {\boldsymbol \Sigma}^a &= a_0 \int_{S^{1}} \Psi(\bx,\bp,t)(\bp\bp^{\rm T}- \textstyle\frac{1}{2}{\boldsymbol I}) \, d\bp.  \label{DSS_dim5}
\end{align} 
Here $\bu(\bx,t)$ and $q(\bx,t)$ are the fluid velocity and pressure, $\mu$ is the fluid viscosity, and ${\boldsymbol \Sigma}^a(\bx,t)$ is the trace-free active stress exerted by the particles on the fluid. The active stress is the orientational average of the force dipoles exerted by the particles on the fluid, and the sign of the coefficient $a_0$ corresponds to the sign of the dipoles: $a_0>0$ for \emph{puller} particles, while $a_0<0$ for \emph{pushers}. 
Note that ${\boldsymbol \Sigma}^a$ vanishes when the particles are in complete nematic alignment.

\subsection{Nondimensionalization and quantities of interest}\label{subsec:nondim}
We choose to nondimensionalize equations \eqref{DSS_dim1}--\eqref{DSS_dim5} over slightly different characteristic velocity, length, and time scales from those commonly used in the literature \citep{ezhilan2013instabilities,hohenegger2010stability,saintillan2008instabilitiesPOF,saintillan2008instabilitiesPRL}. 
In particular, letting $N$ denote the number of particles in the system and $L$ denote the length of the periodic box in which the particles are suspended, we nondimensionalize the model \eqref{DSS_dim1}--\eqref{DSS_dim5} according to
\[\Psi'= \frac{\Psi}{N/L^{2}}, \quad \bx' = \frac{\bx}{L/2\upi}, \quad t' = \frac{|a_0| N}{L^{2}\mu} t, \quad \bu' = \frac{2\upi\mu L}{|a_0| N} \bu, \] 
which results in the following system of equations: 
\begin{equation}\label{DSS_nondim} 
\begin{aligned}
\p_{t'} \Psi' &= -\div'(\p_{t'}\bx'\Psi') - \bnabla_p(\p_{t'}\bp\Psi') \\
\p_{t'} \bx' &= \beta\bp + \bu' - D_T'\bnabla'(\log \Psi')  \\
\p_{t'} \bp &= ({\bf I}-\bp\bp^{\rm T})(\bnabla' \bu' \, \bp) - D_R'\bnabla_p(\log \Psi') \\
-\Delta' \bu' + \bnabla' q' &= \pm \int_{S^{1}} (\bp\bp^{\rm T}- \textstyle\frac{1}{2}{\bf I})\bnabla'\Psi'(\bx,\bp,t) \, d\bp,  \quad  \div'\bu' =0.  
\end{aligned} 
\end{equation}

Here we note the presence of three dimensionless parameters: the diffusion coefficients 
\[ D_T' = \frac{4\upi^2 \mu D_T}{\abs{a_0}N} \quad \text{and} \quad D_R' = \frac{\mu L^{2} D_R}{\abs{a_0}N},\]
and a nondimensional `swimming speed'
\begin{equation}\label{beta_def}
\beta= \frac{2\upi\mu V_0 L}{\abs{a_0}N}.
\end{equation}
We choose this nondimensionalization in order to easily incorporate the immotile state ($V_0=0$) into the analysis. Without swimming, equations \eqref{DSS_dim1}--\eqref{DSS_dim5} describe a suspension of \emph{shakers} \citep{ezhilan2013instabilities,stenhammar2017role}, particles which do not swim but still exert an active stress on the surrounding fluid. We also fix the domain to be the 2-dimensional torus $\T^{2}:=\R^{2}/(2\upi\Z)^{2}$. \\

The parameter $\beta$ contains more information than just the particle swimming speed: it is really the ratio of swimming speed to the active stress magnitude and particle concentration. Note that $\beta$ may be related to the more familiar nondimensionalization used in \cite{hohenegger2010stability,saintillan2008instabilitiesPOF}
 via $\beta= (2\upi \ell)/(|\alpha|L\nu)$, where $\ell$ is the length of typical swimmer, $\alpha=a_0/(V_0\mu\ell)$ is a dimensionless signed active stress coefficient ($\alpha>0$ for pullers and $\alpha<0$ for pushers), and $\nu=N\ell^{2}/L^{2}$ is the relative volume concentration of swimmers. \\

In \eqref{DSS_nondim}, the active stress coefficient in the Stokes equations is scaled to unit magnitude but retains the sign of the force dipole exerted by the particles on the fluid: $+1$ for puller particles and $-1$ for pusher particles. \\ 

Dropping the prime notation, the system \eqref{DSS_nondim} may be written more succinctly as 
\begin{equation}\label{DSS}
\begin{aligned}
\p_t \Psi &= -\beta \bp\bcdot\bnabla\Psi -\bu\bcdot\bnabla\Psi -\divp\big(({\bf I}-\bp\bp^{\rm T})(\bnabla\bu\bp)\Psi \big) + D_T\Delta \Psi +D_R \Delta_p\Psi  \\
-\Delta \bu + \bnabla q &= \pm \int_{S^{1}} (\bp\bp^{\rm T}- \textstyle\frac{1}{2}{\bf I})\bnabla \Psi(\bx,\bp,t) \, d\bp, \quad \div\bu =0 .
\end{aligned}
\end{equation}

One way to measure deviations of the swimmer density $\Psi$ from the uniform, isotropic steady state $\Psi_0=1/(2\upi)$ is to consider the relative entropy 
\begin{equation}\label{entropy}
\mc{S}(t) = \int_{\T^{2}}\int_{S^{1}} \frac{\Psi}{\Psi_0}\log\bigg(\frac{\Psi}{\Psi_0} \bigg) \, d\bp d\bx.
\end{equation}
Using \eqref{DSS}, the entropy can be shown to evolve according to 
\begin{equation}\label{dt_entropy}
\dot {\mc{S}}(t) = \mp \frac{4}{\Psi_0}\int_{\T^{2}} \abs{\mathsfbi{E}}^2 d\bx - \int_{\T^{2}}\int_{S^{1}} \big( D_T\abs{\bnabla(\log\Psi)}^2 + D_R \abs{\bnabla_p(\log(\Psi))}^2 \big) \frac{\Psi}{\Psi_0}  d\bp d\bx
\end{equation}
where $\mathsfbi{E}= \frac{1}{2}(\bnabla\bu+(\bnabla\bu)^{\rm T})$ (see \citep{saintillan2008instabilitiesPOF}). The first term in \eqref{dt_entropy} arises from the viscous dissipation of the active stress exerted by the particles and is negative for pullers and positive for pushers. The two diffusive terms are negative; hence we expect puller suspensions to always relax to isotropy over time. For pushers, however, we may expect to see some more interesting behaviors: indeed, simulations show that patterns and fluctuations in particle alignment and concentration arise in certain parameter regimes \citep{hohenegger2010stability,saintillan2008instabilitiesPOF,saintillan2008instabilitiesPRL,saintillan2013active,saintillan2015theory}. 
We aim to study the onset of pattern formation in these active pusher suspensions. \\

It will be useful to first define some system quantities that can be measured numerically and used to verify analytical predictions.
One such quantity is the active power input, defined for pusher particles by  
\begin{equation}\label{active_power}
\mc{P}(t) = \int_{\T^{2}} \int_{S^{1}} \bp^{\rm T}\mathsfbi{E}(\bx,t)\bp \, \Psi(\bx,\bp,t)\, d\bp d\bx.
\end{equation}
The sign is opposite for puller particles. 
This quantity can be understood as the perturbative power input due to the interaction of the active particles with the flow (as opposed to the power input of each individual particle). 
Using the Stokes equations in \eqref{DSS}, the active power balances the rate of viscous dissipation in the fluid: 
\begin{equation}\label{power_to_VD}
\mc{P}(t) = \int_{\T^{2}} 2 \abs{\mathsfbi{E}(\bx,t)}^2 \, d\bx.
\end{equation}
Note that $\mc{P}(t)=0$ for the uniform, isotropic steady state, and tracking the growth of $\mc{P}(t)$ (or, equivalently, the rate of viscous dissipation) serves as a measure of the instability of the uniform state \citep{saintillan2015theory}.  \\

We also define the concentration field 
\begin{equation}\label{concentration}
c(\bx,t) = \int_{S^{1}}\Psi(\bx,\bp,t)\, d\bp
\end{equation}
and nematic order tensor 
\begin{equation}\label{nematic_tensor}
\mathsfbi{Q}(\bx,t) = \frac{1}{c(\bx,t)}\int_{S^{1}}(\bp\bp^{\rm T}-{\bf I}/2)\Psi(\bx,\bp,t)\, d\bp .
\end{equation}

We can then define the scalar-valued nematic order parameter as 
\begin{equation}\label{sop_def}
\mc{N}(\bx,t) = \text{maximum eigenvalue of }\mathsfbi{Q}(\bx,t).
\end{equation}
The nematic order parameter measures the local degree of nematic alignment: 
$\mc{N}(\bx,t)=0$ when the particle orientations are isotropic and $\mc{N}(\bx,t)=1$ when all surrounding particles are exactly aligned.

\subsection{Linear stability: the eigenvalue problem}\label{subsec:linearstab}
From here, we restrict our attention to the pusher case in two spatial dimensions. We begin by recalling the results of the eigenvalue problem resulting from a linear stability analysis about the uniform, isotropic steady state $\Psi_0=1/(2\upi)$ and $\bu=0$ \citep{hohenegger2010stability,saintillan2008instabilitiesPOF,subramanian2011stability}. Linearizing \eqref{DSS} about this state, we obtain 
\begin{equation}\label{DSS_linear}
\begin{aligned}
\p_t \Psi &= - \beta\bp\bcdot\bnabla\Psi + 2 \bp^{\rm T}\bnabla\bu \bp  + D_T\Delta \Psi +  D_R \Delta_p\Psi  \\
- \Delta \bu + \bnabla q &= -\frac{1}{2\upi} \int_{S^1} (\bp\bp^{\rm T}- \textstyle\frac{1}{2}{\bf I})\bnabla \Psi(\bx,\bp,t) \, d\bp, \quad \div\bu =0.
\end{aligned}
\end{equation}

We insert the plane wave ansatz $\Psi = \psi(\bk,\bp){\rm e}^{{\rm i}\bk\bcdot\bx+\sigma t}$, $\sigma\in\C$, into \eqref{DSS_linear}, and choose coordinates such that the wavevector $\bk = k\be_x$ and the particle orientation $\bp = \cos\theta\be_x+\sin\theta\be_y$, $0\le \theta<2\upi$. Defining $\lambda_k:= \sigma + D_Tk^2$, we obtain an eigenvalue problem for $\lambda_k$ in the form of an integrodifferential equation on $S^1$: 
\begin{equation}\label{eigenval_prob} 
\lambda_k\psi(k,\theta) = -{\rm i}k\beta\cos\theta\psi(k,\theta) + \frac{1}{\upi} \cos\theta\sin\theta \int_0^{2\upi}\psi(k,\theta) \, \sin\theta\cos\theta \, d\theta + D_R\p_{\theta\theta} \psi(k,\theta).
\end{equation}

We note that while it may be more suggestive to write $\psi$ in terms of $\sin(2\theta)$ rather than $\sin\theta\cos\theta$, we find that the details of the weakly nonlinear calculations in the following sections are slightly simpler in terms of $\cos\theta=\bp\cdot\be_x$ and $\sin\theta=\bp\cdot\be_y$ only. \\

In the absence of rotational diffusion ($D_R=0$), we may solve \eqref{eigenval_prob} for $\psi(k,\theta)$ as 
\begin{align*}
\psi(k,\theta) = \frac{\cos\theta\sin\theta}{\lambda_k+{\rm i}k\beta\cos\theta}, 
\end{align*}
where $\lambda_k$ is such that
\begin{equation}\label{Fdef}
F[\psi]:=\int_0^{2\upi}\psi(k,\theta)\,\cos\theta\sin\theta\,d\theta = \upi. 
\end{equation}
In particular, $\lambda_k$ satisfies the implicit dispersion relation 
\begin{equation}\label{disp_rel}
\frac{\lambda_k\beta^2k^2+2\lambda_k^3 - 2\lambda_k^2\sqrt{\beta^2k^2+\lambda_k^2}}{\beta^4k^4} = 1.
\end{equation}

\begin{figure}
\centering
\begin{subfigure}{.45\textwidth}
    \includegraphics[scale=0.044]{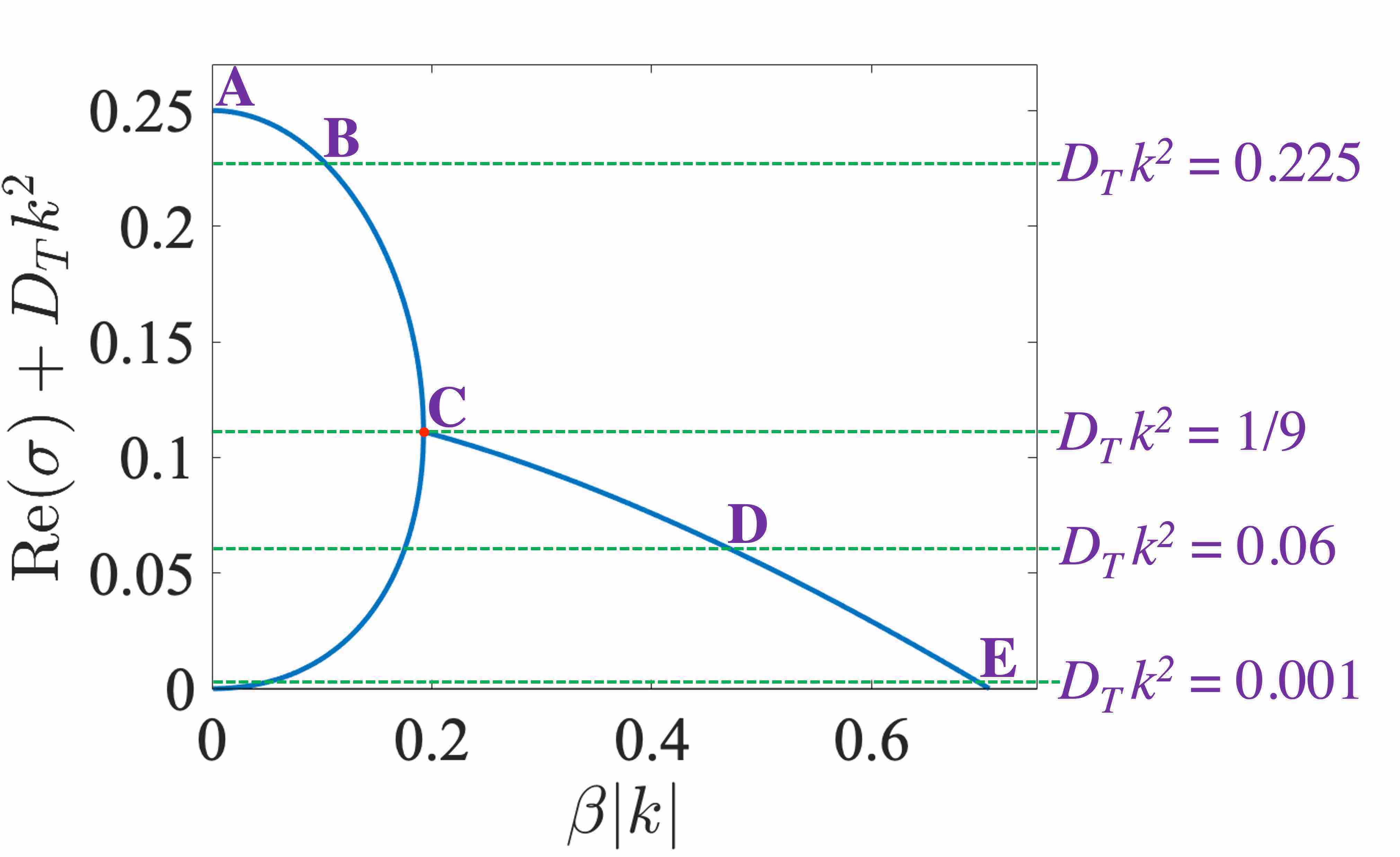}
    \caption{} 
     \label{subfig:disp_re}
   \end{subfigure}
   \begin{subfigure}{.41\textwidth}
      \includegraphics[scale=0.126]{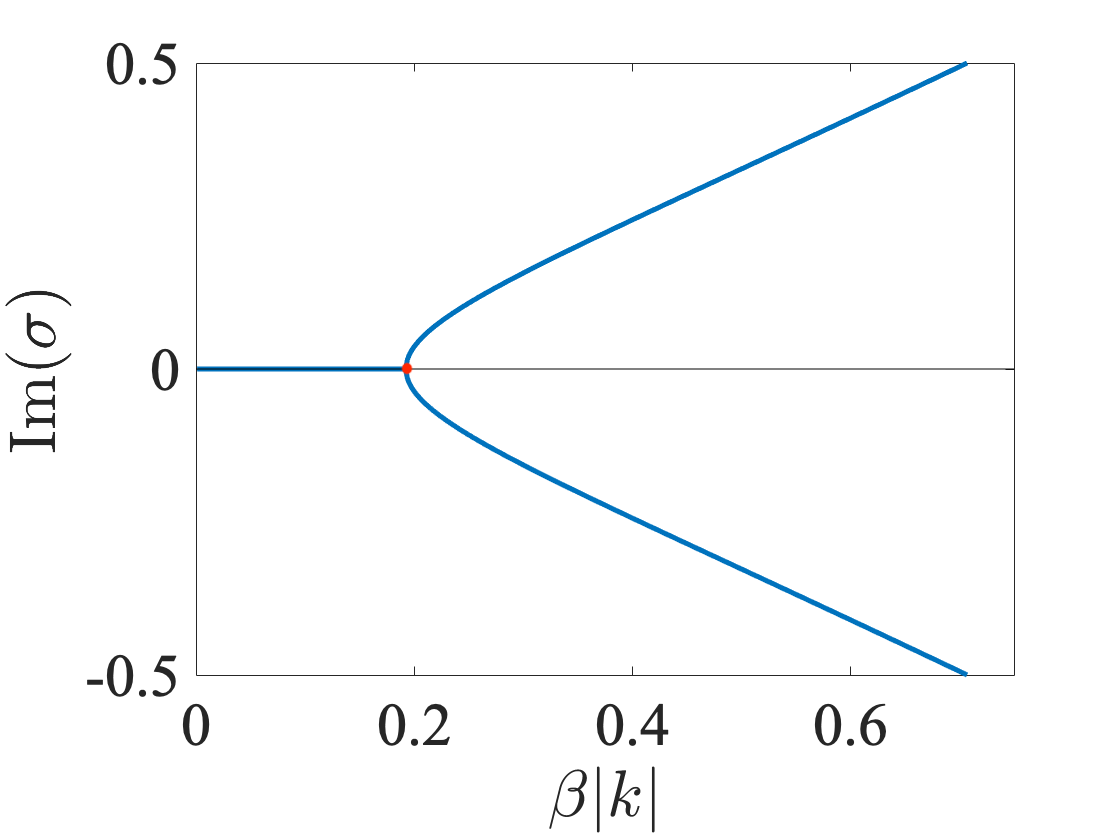}
      \caption{}
      \label{subfig:disp_im}
   \end{subfigure}
\caption{(a) Real part (shifted by $D_Tk^2$) and (b) imaginary part of the growth rate $\sigma(k)$ versus $\beta\abs{k}$. Figure (a) can be read as follows: fix a value of $0<D_Tk^2<0.25$. Look at the corresponding horizontal green line. As $\beta$ is lowered, the green line intersects the blue curve. This is where the eigenvalue $\sigma(k)$ becomes unstable.  \\
Different types of bifurcations are possible depending on the value of $D_Tk^2$: for example, point {\bf B} corresponds to a purely real eigenvalue crossing, while the presence of nonzero Im$(\sigma)$ signals a Hopf bifurcation at points {\bf D} and {\bf E}. At point {\bf C}, indicated by a red dot, two real eigenvalues meet and become complex. In the case of shakers ($\beta=0$), we may consider the real eigenvalue crossing at point {\bf A} along the $y$-axis.   }
\label{fig:dispersion}
\end{figure}

Recalling that $\lambda_k = \sigma+D_Tk^2$, we may then numerically solve for the relationship between $\sigma$ and $\beta$ (see figure \ref{fig:dispersion}).  \\

A similar calculation for $\bk=k\be_y$ shows that the eigenvalues $\sigma(k)$ plotted in figure \ref{fig:dispersion} also correspond to a $y$-direction eigenfunction whose eigenmodes are a 90-degree rotation (in $\theta$) of the $x$-direction eigenmodes. The solutions of \eqref{DSS_linear} arising from eigenfunctions of the linearized operator are thus given by 
\begin{equation}\label{xy_planewave}
\Psi(\bx,\theta,t) = c_x\psi_x(k,\theta){\rm e}^{{\rm i}kx+\sigma t} +c_y\psi_y(k,\theta){\rm e}^{{\rm i}ky+\sigma t}, \quad c_x^2+c_y^2=1, 
\end{equation}
and all scalar multiples of \eqref{xy_planewave}, where
\begin{align*}
\psi_x(k,\theta) = \frac{\cos\theta\sin\theta}{\sigma+D_Tk^2+{\rm i}k\beta\cos\theta},\quad \psi_y(k,\theta) = \frac{\cos\theta\sin\theta}{\sigma+D_Tk^2+{\rm i}k\beta\sin\theta}.
\end{align*}
In particular, besides point {\bf C}, each eigenvalue $\sigma(k)$ has a 4-dimensional eigenspace over $\C$, spanned by the $x$- and $y$-direction components with $\pm k$. When $\beta\abs{k}<\sqrt{3}/9$, there are two distinct real-valued eigenvalues $\sigma(k)$ which satisfy the required condition $\int_0^{2\upi}\psi_x \, \cos\theta\sin\theta d\theta=\int_0^{2\upi}\psi_y \, \cos\theta\sin\theta d\theta=\upi$, both of which then have a 4-dimensional eigenspace over $\R$. 
 \\

Note that the dispersion relation \eqref{disp_rel} is exact only in the absence of rotational diffusion, but for the purposes of this paper, we will consider $0<D_R\ll1$. This alters figure \ref{fig:dispersion}, but only slightly (see next section \ref{subsec:DR} for details). In return, we do not have to contend with the continuous spectrum present in the $D_R=0$ spectral analysis of \cite{subramanian2011stability}. 
Furthermore, when $D_R>0$, the $k=0$ mode is always linearly stable, as it satisfies the heat equation in $\bp$: if $\Psi=\Psi(\bp,t)$ then $\p_t\Psi=D_R\Delta_p\Psi$. \\

The eigenvalue relation \eqref{disp_rel} ceases to be valid for Re$(\lambda_k)\le0$; i.e. when Re$(\sigma(k))\le-D_Tk^2$ \citep{hohenegger2010stability,subramanian2011stability}.  
For fixed $0<D_T<1/4$, however, the eigenvalue analysis does capture a sign change in the real part of the growth rate $\sigma(k)$ as $\beta\abs{k}$ is varied. For each $k$, this sign change occurs where the blue curve in figure \ref{fig:dispersion} (corresponding to $\lambda_k$) intersects the green line corresponding to the fixed value of $D_Tk^2$.
Numerical simulations indicate that if the point $(D_Tk^2,\beta\abs{k})$ lies to the right of the blue contour in figure \ref{subfig:disp_re} for each $k$, then the uniform, isotropic steady state is stable to small perturbations (see sections \ref{sec:hopf} and \ref{sec:real_eval}). 
In this region, particle diffusion and swimming -- processes which tend to decrease order among the particles -- are large relative to both the active stress magnitude and the particle concentration, which favor local alignment.
As $\beta$ is decreased, Re($\sigma(k)$) crosses the green line corresponding to the (fixed) value of $D_Tk^2$ and the uniform, isotropic state becomes unstable. \\

 Since we are considering the system \eqref{DSS_nondim} on a periodic domain and have nondimensionalized $\bx$ according to the length of the domain, the $\abs{k}=1$ mode is the lowest nontrivial mode in the system and hence the first to lose stability as $\beta$ decreases. As noted in \cite{koch2011collective}, ``the finiteness of the domain may act to stabilize the system." Indeed, the stabilizing effects of mixing on $\T^d$ ($d=2,3$) due to swimming are studied in \cite{albritton2022stabilizing} and play a role in the patterns observed here.  \\

We aim to understand the onset of pattern formation in \eqref{DSS_nondim} by exploring the many different ways the $\abs{k}=1$ mode can lose stability as $\beta$ is decreased. From figure \ref{fig:dispersion}, we can see that depending on $D_T$, the type of bifurcation that we expect to see for the $\abs{k}=1$ mode will change. We aim to characterize all of the types of initial bifurcations admitted by the system through a weakly nonlinear analysis. 
According to the dispersion relation, if $1/9<D_T<1/4$, as $\beta$ decreases, the purely real eigenvalue $\sigma$ will change sign from positive to negative across the top branch of the blue curve, between points {\bf A} and {\bf C}. 
For $0<D_T<1/9$, however, we expect to see a Hopf bifurcation as $\beta$ crosses the blue curve roughly between points {\bf C} and {\bf E}, since for these values of $\beta$, $\sigma$ has nonzero imaginary part. \\

As noted, figure \ref{fig:dispersion} does not quite give the exact locations of the bifurcations that we will consider here, since we still need to consider the effects of (small) $D_R>0$. This will be the subject of the next section.

\subsection{Role of rotational diffusion}\label{subsec:DR}

The dispersion relation \eqref{disp_rel} was obtained in the absence of rotational diffusion; however, studying pattern formation near the isotropic steady state will require $D_R>0$. When $D_R=0$, the system \eqref{DSS} has infinitely many steady states; in particular, any spatially uniform swimmer distribution $\Psi=\Psi(\bp)$ is a steady state. The continuum of nearby steady states obscures the mechanism by which the isotropic state $\Psi=1/(2\upi)$ loses stability; indeed, any function $\Psi(\bp)$ belongs to the kernel of the linearized operator. When we add in $D_R>0$ (along with the assumption that the total number of swimmers is conserved), this kernel is eliminated. Thus we aim to determine when the effect of small $D_R>0$ can be considered as a perturbation of the dispersion relation \eqref{disp_rel} and figure \ref{fig:dispersion}. \\

In particular, given a wavenumber $k$, for small $D_R>0$, we wish to determine when an expansion (in $D_R$) of the form 
\begin{equation}\label{DR_expansion} 
\sigma = \sigma_0 + D_R\sigma_1 + O(D_R^2), \quad \psi = \psi_0 + D_R \psi_1 + O(D_R^2)
\end{equation}
is valid for some $(\sigma_1,\psi_1)$. \\ 

Plugging this expansion into the eigenvalue problem \eqref{eigenval_prob} and separating scales, at $O(1)$ we obtain the $D_R=0$ eigenfunctions and eigenvalue relation 
\begin{align*}
\psi_0(k,\theta) = \frac{\cos\theta\sin\theta}{\sigma_0+D_Tk^2+{\rm i}k\beta\cos\theta}
\end{align*}
where $\sigma_0$ is such that
\begin{equation}\label{Fpsi0}
F[\psi_0]=\int_0^{2\upi}\psi_0\cos\theta\sin\theta \, d\theta= \upi.
\end{equation}

At $O(D_R)$ we obtain the expression 
\begin{align*}
\psi_1 
 &= \frac{1}{\upi}\psi_0\int_0^{2\upi} \psi_1\cos\theta'\sin\theta' \, d\theta' - \frac{\sigma_1\psi_0}{\sigma_0+D_Tk^2+{\rm i}k\beta\cos\theta} + \frac{\p_\theta^2\psi_0}{\sigma_0+D_Tk^2+{\rm i}k\beta\cos\theta}.
\end{align*}

\begin{figure}
\centering      
    \begin{subfigure}{.45\textwidth}
     \includegraphics[scale=0.05]{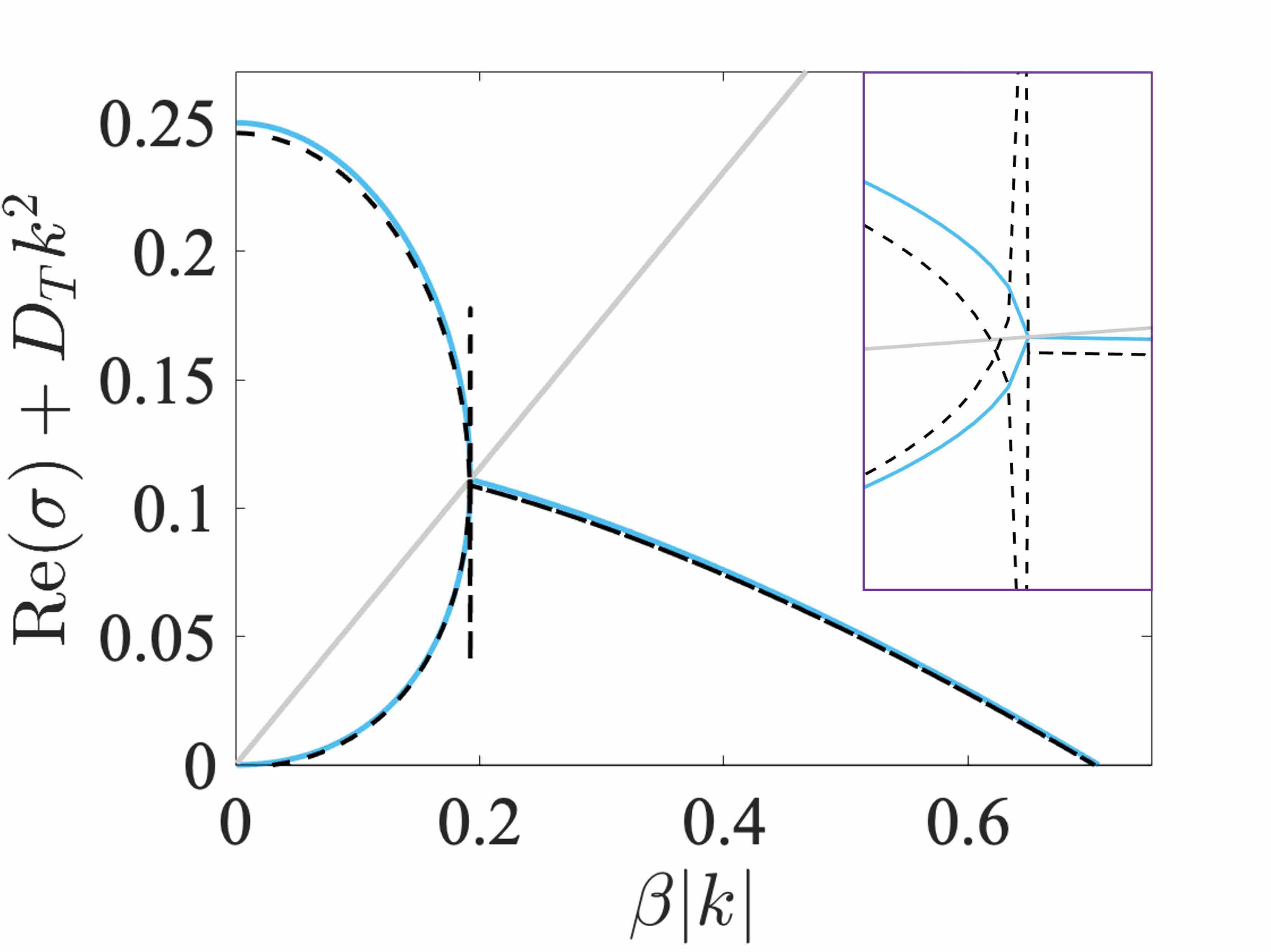}
     \caption{}
     \label{subfig:ReDisp_pert001}
   \end{subfigure}
    \begin{subfigure}{.45\textwidth}
     \includegraphics[scale=0.143]{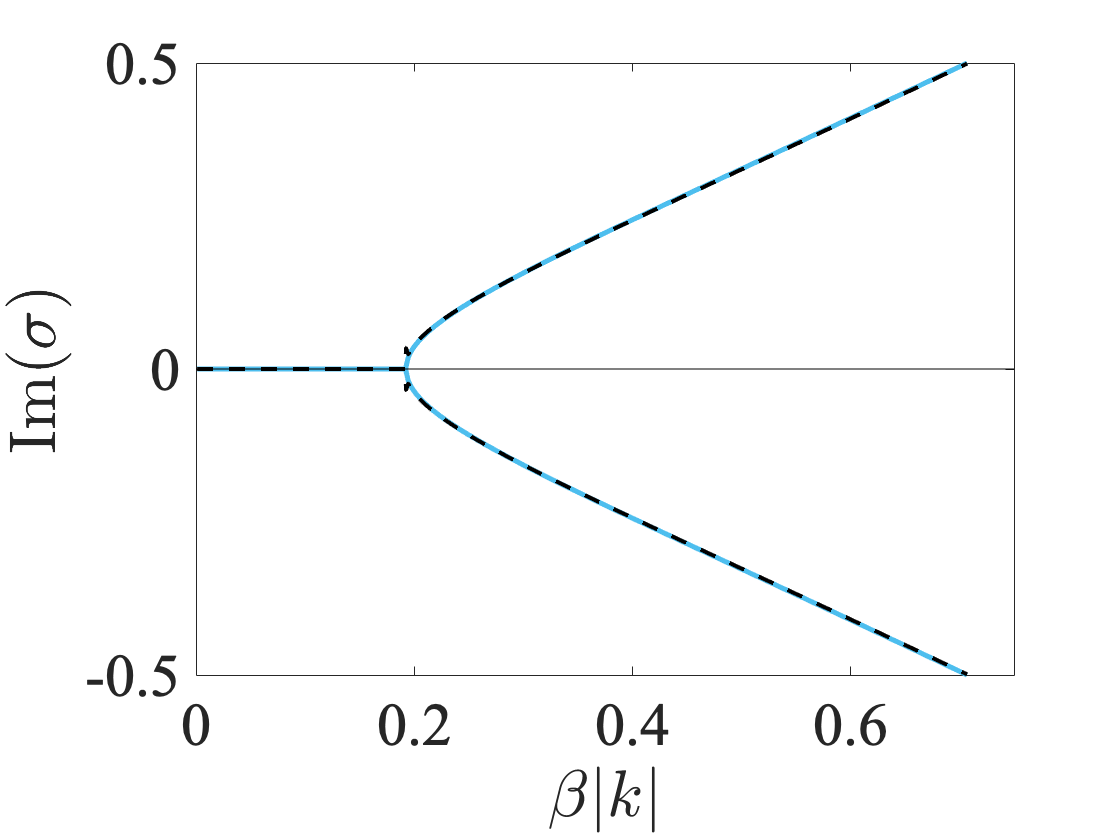}
     \caption{}
     \label{subfig:ImDisp_pert001}
   \end{subfigure}
\caption{The (a) real part and (b) imaginary part of the perturbed dispersion relation $\sigma_0+D_R\sigma_1$ (dotted black curve) is plotted for $D_R=0.001$ on top of the unperturbed relation with $D_R=0$ (solid light blue curve). The perturbative expression fails to be valid along the gray line $\sigma_0+D_Tk^2=\beta \abs{k}/\sqrt{3}$ plotted in (a). The inset in figure (a) shows in greater detail the behavior of the perturbative expression $\sigma_0+D_R\sigma_1$ near the point $(\sqrt{3}/9,1/9)$ where the gray line intersects the unperturbed expression. In particular, the perturbed expression blows up as $\beta\abs{k}$ approaches $\sqrt{3}/9$.  }
\label{fig:disp_pert}
\end{figure}

Taking $F[\bcdot]$ of both sides and using \eqref{Fpsi0} then yields an integral expression for $\sigma_1$:
\begin{align*}
 \sigma_1 \int_0^{2\upi}\frac{\cos^2\theta\sin^2\theta}{(\sigma_0+D_Tk^2+{\rm i}k\beta\cos\theta)^2}d\theta &= \int_0^{2\upi}\frac{\p_\theta^2\psi_0 \cos\theta\sin\theta}{\sigma_0+D_Tk^2+{\rm i}k\beta\cos\theta}d\theta.
\end{align*}
As long as Re$(\sigma_0+D_Tk^2)\neq 0$ (which holds for the $\abs{k}=1$ modes as long as $D_T\neq 0$), 
we obtain an expression for $\sigma_1$:
\begin{align*}
\sigma_1 &= \frac{\frac{5}{6}\beta^2k^2}{\beta^2k^2-3(D_Tk^2+\sigma_0)^2} + \frac{\frac{1}{2}\beta^2k^2}{\beta^2k^2+(D_Tk^2+\sigma_0)^2} \\
&\qquad + \frac{5(D_Tk^2+\sigma_0)^3 }{(\beta^2k^2-3(D_Tk^2+\sigma_0)^2)\sqrt{\beta^2k^2+(D_Tk^2+\sigma_0)^2}} -\frac{7}{3},
\end{align*}
which is finite as long as $\beta^2k^2\neq 3(D_Tk^2+\text{Re}(\sigma_0))^2$.
The line 
\begin{equation}\label{DR_condition}
\sigma_0+D_Tk^2=\frac{\beta \abs{k}}{\sqrt{3}}
\end{equation} 
is plotted along with the real part of the $D_R=0$ dispersion relation \eqref{disp_rel} in figure \ref{subfig:ReDisp_pert001}. Away from this line, for small $D_R>0$, we may consider the solutions of the eigenvalue problem \eqref{eigenval_prob} as perturbations of the $D_R=0$ eigenvalues and eigenfunctions.  
As we can see, this line \eqref{DR_condition} passes through the point {\bf C} from figure \ref{fig:dispersion} where the two real eigenvalues meet and become two complex conjugate eigenvalues. A very precise choice of $D_T$ and $D_R$ should correspond to a codimension 2 bifurcation at point {\bf C}; however, a different scaling than \eqref{DR_expansion} with respect to $D_R$ is likely needed to study this point in detail. We will not attempt to study point {\bf C} in detail here, and will instead focus on the more generic bifurcations between points {\bf A} and {\bf C} and between points {\bf C} and {\bf E}.  \\

In figures \ref{subfig:ReDisp_pert001} and \ref{subfig:ImDisp_pert001} we plot the dispersion relation for $\sigma_0+D_R\sigma_1$ on top of $\sigma_0$ using $D_R=0.001$. Away from the crossing with the line \eqref{DR_condition}, the $\sigma_0+D_R\sigma_1$ curve aligns very closely with the $\sigma_0$ curve. Hereafter, for most numerical purposes, we will take $D_R=0.001$ -- small enough to use figure \ref{fig:dispersion} as our roadmap for determining roughly where in the $(D_T,\beta)$ parameter space to look to see different system behaviors. 
We note that, while the expansion in $D_R$ is not rigorous, it is backed later on by the close agreement of the predicted behavior near bifurcation points (from amplitude equations) with the observed behavior from numerical simulations. In particular, it appears that, away from point ${\bf C}$, the small $D_R>0$ picture is largely captured by this expansion. 

\subsection{Outline of results}\label{subsec:outline}
\begin{figure}
     \includegraphics[scale=0.055]{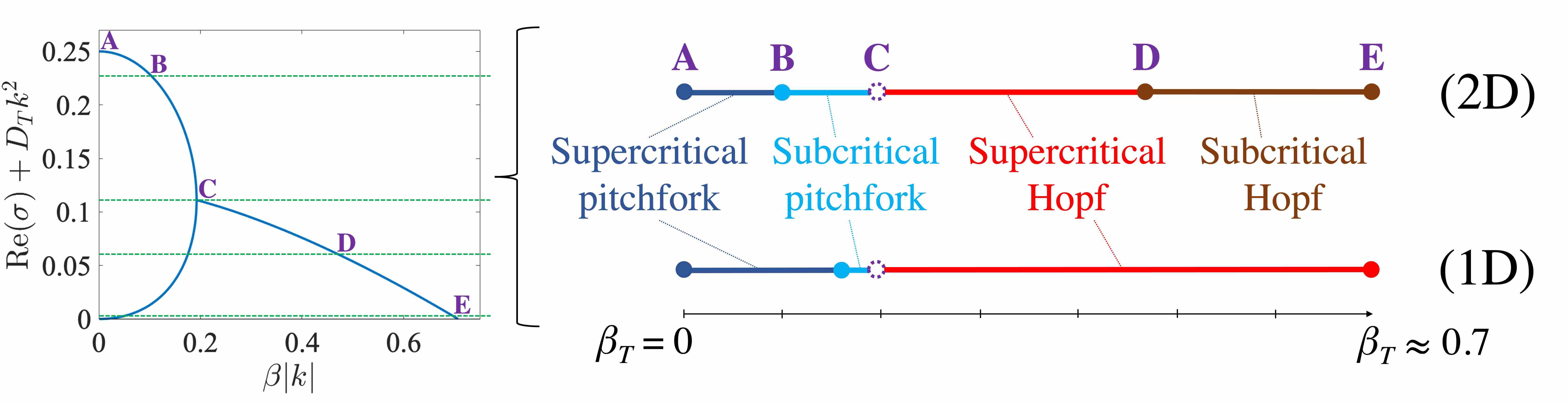}
\caption{Diagram of the various types of bifurcations through which the uniform isotropic steady state loses stability, depending on the location of the bifurcation value $\beta_T$. Here the subscript $T$ is used to reflect that the value of $\beta_T$ depends on the translational diffusion $D_T$ through the dispersion relation plotted in figure \ref{fig:dispersion}. Note that the letters {\bf A} -- {\bf E} correspond to the positions in figure \ref{subfig:disp_re}, which we also repeat here for clarity. The upper line labeled 2D corresponds to the evolution of initial perturbations to the uniform isotropic state with both components in both the $x$ and $y$ direction ($c_x$, $c_y\neq0$ in \eqref{xy_planewave}), while the lower line labeled 1D corresponds to perturbations with $x$ or $y$ component only ($c_y=0$ or $c_x=0$ in \eqref{xy_planewave}). }
\label{fig:bif_diag}
\end{figure}

The remainder of the paper is devoted to a weakly nonlinear analysis of the different possible bifurcations apparent in figure \ref{fig:dispersion}, which we map out in greater detail in figure \ref{fig:bif_diag}. We begin in section \ref{sec:noswim} by considering the immotile case $\beta=0$. We examine the real eigenvalue crossing at point {\bf A} for all values of $D_R$ for which a bifurcation occurs and show that the resulting pitchfork bifurcation is always supercritical. In section \ref{sec:hopf} we assume $D_R$ is very small and analyze the Hopf bifurcation occuring along the curve between points {\bf C} and ${\bf E}$. We show that for initial perturbations to the uniform isotropic state with both $x$ and $y$ components (i.e. both $c_x$, $c_y\neq 0$ in \eqref{xy_planewave}; see line labeled 2D in figure \ref{fig:bif_diag}), the bifurcation transitions from supercritical to subcritical at roughly point {\bf D}, but is always supercritical for initial perturbations with either $c_y=0$ or $c_x=0$ (labeled 1D in figure \ref{fig:bif_diag}).
 In section \ref{sec:real_eval}, we again consider very small $D_R$ and study real eigenvalue crossing occuring along the curve between points {\bf A} and {\bf C}. The pitchfork bifurcation also transitions from supercritical to subcritical in both the 2D and 1D cases, with the change occuring roughly at point {\bf B} in the case of an initial perturbation in both $x$ and $y$, and just before point {\bf C} in the case of an $x$-only or $y$-only initial perturbation. \\

Each section is accompanied by numerical simulations verifying the predicted behaviors near the different bifurcations and illustrating the various states that arise. The numerics are a pseudo-spectral implementation of equations \eqref{DSS_nondim} with time-stepping via a second order implicit-explicit backward differentiation scheme.

\section{Immotile particles: supercritical pitchfork bifurcation}\label{sec:noswim}
The simplest scenario for studying the onset of pattern formation in the model \eqref{DSS} is in the case $\beta=0$; i.e. the particles are immotile (or \emph{shakers} \citep{ezhilan2013instabilities,stenhammar2017role}) but still exert a dipolar force on the surrounding fluid. The uniform, isotropic steady state in a suspension of immotile particles undergoes a bifurcation to an alignment instability, indicated by point {\bf A} in figure \ref{fig:dispersion}, which we study in greater detail. We first show via weakly nonlinear analysis that the resulting pitchfork bifurcation is always supercritical, and we then numerically explore examples of the emerging nontrivial steady state. 

\subsection{Weakly nonlinear analysis}\label{subsec:noswim_wnl}

In the case of immotile particles, we can explicitly calculate the eigenvalues and eigenfunctions of the linearized operator when $D_R>0$.
The (purely real) eigenvalues $\sigma(k)$ and eigenmodes $\psi_x(k,\theta)$, $\psi_y(k,\theta)$ are given by 
\begin{equation}\label{noswim_eigen}
\sigma(k) = \frac{1}{4} - D_Tk^2 -4D_R, \quad \psi_x(k,\theta)=\psi_y(k,\theta)=\cos\theta\sin\theta.
\end{equation}

From the immotile dispersion relation \eqref{noswim_eigen}, when $D_T>\frac{1}{4} - 4D_R$, all eigenvalues $\sigma(k)$ are negative, but as $D_T$ is decreased, the $\abs{k}=1$ modes are the first to change sign as $D_T=D_T^* = \frac{1}{4} - 4D_R$ is crossed. We study the nature of this bifurcation for different $D_R$ via the method of multiple scales. For $0<\epsilon\ll 1$, we fix $D_T=\frac{1}{4}-4D_R-\epsilon^2$, so the $\abs{k}=1$ modes are just barely growing, and define the slow timescale $\tau=\epsilon^2 t$. We then assume the following expansions in $\epsilon$:
\begin{equation}\label{expansion}
\Psi = \frac{1}{2\upi}(1+\epsilon\Psi_1 + \epsilon^2\Psi_2 +\epsilon^3\Psi_3 +\cdots), \qquad \bu = \epsilon\bu_1 + \epsilon^2\bu_2+ \epsilon^3\bu_3+ \cdots.
\end{equation}

Plugging each of these expansions into equation \eqref{DSS} with $\beta=0$ and separating by orders of $\epsilon$, at $O(\epsilon)$ we obtain the $\beta=0$ version of the eigenvalue equation \eqref{eigenval_prob}, evaluated at the critical value $D_T^*=\frac{1}{4}-4D_R$ where $\sigma(1)=0$: 
\begin{equation}\label{Oeps_noswim}
\mc{L}[\Psi_1] :=  -2\bp^{\rm T}\bnabla\bu_1\bp - \bigg(\frac{1}{4}-4D_R\bigg)\Delta \Psi_1 - D_R\Delta_p\Psi_1=0.
\end{equation}

This equation is satisfied by the $\abs{k}=1$ modes of the eigenfunctions \eqref{xy_planewave}, where we recall that the eigenmodes in the immotile case are given by \eqref{noswim_eigen}. Thus $\Psi_1$ and $\bu_1$ have the form
\begin{equation}\label{noswim_psi1}
\begin{aligned}
\Psi_1 &= \cos\theta\sin\theta\big(c_xA_x(\tau){\rm e}^{{\rm i}x} +c_yA_y(\tau){\rm e}^{{\rm i}y} \big) + {\rm c.c.}, \\
\bu_1 &= - \frac{\rm i}{8} \big( c_x A_x(\tau) {\rm e}^{{\rm i}x}\be_y + c_y  A_y(\tau) {\rm e}^{{\rm i}y}\be_x\big) + {\rm c.c.}
\end{aligned}
\end{equation}
Here we have inserted the complex amplitudes $A_x(\tau)$, $A_y(\tau)$ which depend solely on the slow timescale $\tau$, and for which we aim to find an equation. Throughout, we use ${\rm c.c.}$ to denote the complex conjugate of each of the preceding terms. \\

At $O(\epsilon^2)$ we obtain the equation
\begin{equation}\label{Oeps2_noswim}
\mc{L}[\Psi_2] = -\bu_1\bcdot\bnabla\Psi_1 - \divp\big(({\bf I}-\bp\bp^{\rm T})(\bnabla\bu_1\bp)\Psi_1) \big) 
\end{equation}
where the operator $\mc{L}$ is as defined in \eqref{Oeps_noswim}. Using the expressions \eqref{noswim_psi1} for $\Psi_1$ and $\bu_1$, the right hand side of equation \eqref{Oeps2_noswim} can be calculated explicitly (see equation \eqref{Oeps2_noswim_RHS} in Appendix \ref{app:noswim}). Noting that the right hand side expression contains only exponential terms of the form ${\rm e}^{\pm2{\rm i}x}$, ${\rm e}^{\pm2{\rm i}y}$, and ${\rm e}^{\pm {\rm i}(x\pm y)}$, with no terms proportional to ${\rm e}^{\pm {\rm i}x}$ or ${\rm e}^{\pm {\rm i}y}$, equation \eqref{Oeps2_noswim} is solvable without any additional conditions on the coefficients $A_x$ and $A_y$. In particular, due to the form of the right hand side expression, we look for $\Psi_2$ and corresponding $\bu_2$ of the form
\begin{equation}\label{noswim_psi2}
\begin{aligned}
\Psi_2 &= \psi_{2,1}{\rm e}^{{\rm i}(x+y)}A_xA_y +\psi_{2,2}{\rm e}^{{\rm i}(x-y)}A_x\overline A_y + \psi_{2,3}A_x^2{\rm e}^{2{\rm i}x} +\psi_{2,4}A_y^2{\rm e}^{2{\rm i}y} \\
&\hspace{3cm}+ \psi_{2,5}\abs{A_x}^2+\psi_{2,6}\abs{A_y}^2 + {\rm c.c.}, \\
\bu_2&= - \frac{\rm i}{8\upi}\int_0^{2\upi}\bigg(\psi_{2,1}{\rm e}^{{\rm i}(x+y)}A_xA_y(\be_x-\be_y) + \psi_{2,2}{\rm e}^{{\rm i}(x-y)}A_x\overline A_y(\be_x+\be_y)\bigg) (\cos^2\theta - \sin^2\theta) \, d\theta  \\
&\qquad - \frac{\rm i}{4\upi}\int_0^{2\upi}\bigg( \psi_{2,3}{\rm e}^{2{\rm i}x} A_x^2\be_y + \psi_{2,4}{\rm e}^{2{\rm i}y} A_y^2\be_x \bigg)\sin\theta\cos\theta \, d\theta +{\rm c.c.},
\end{aligned}
\end{equation}
where each $\psi_{2,j}=\psi_{2,j}(\theta)$. Plugging these expressions \eqref{noswim_psi2} into the left hand side of \eqref{Oeps2_noswim} (see Appendix \ref{app:noswim}, equation \eqref{Oeps2_noswim_LHS} for the full expression), after matching exponents with the right hand side, we can explicitly solve for each $\psi_{2,j}$:
\begin{equation}\label{noswim_O2coeffs}
\begin{aligned}
\psi_{2,1} &= -\frac{c_xc_y}{1+16D_R}(\sin^4\theta+\cos^4\theta) -\frac{c_xc_y}{2(1-8D_R)}\sin\theta\cos\theta + \frac{3c_xc_y}{4(1+16D_R)}  \\
\psi_{2,2} &= -\frac{c_xc_y}{1+16D_R}(\sin^4\theta+\cos^4\theta) +\frac{c_xc_y}{2(1-8D_R)}\sin\theta\cos\theta + \frac{3c_xc_y}{4(1+16D_R)}  \\
\psi_{2,3} &= \frac{c_x^2}{4}\bigg(-2\cos^4\theta + \frac{3(1-16D_R)}{2(1-12D_R)}\cos^2\theta + \frac{3D_R}{1-12D_R}\bigg) \\
\psi_{2,4} &= \frac{c_y^2}{4}\bigg(-2\sin^4\theta + \frac{3(1-16D_R)}{2(1-12D_R)}\sin^2\theta + \frac{3D_R}{1-12D_R}\bigg) \\
\psi_{2,5} &= -\frac{c_x^2}{16D_R} \cos^4\theta + \frac{3c_x^2}{128D_R}\\
\psi_{2,6} &= -\frac{c_y^2}{16D_R}\sin^4\theta + \frac{3c_y^2}{128D_R}.
\end{aligned}
\end{equation}
Inserting each of the coefficients \eqref{noswim_O2coeffs} in the expression \eqref{noswim_psi2} for $\bu_2$, we have that each $\theta$-integral vanishes and therefore $\bu_2=0$. At $O(\epsilon^3)$ we thus obtain the following equation for $\Psi_3$:
\begin{equation}\label{Oeps3_noswim}
\begin{aligned}
\mc{L}[\Psi_3] &= -\p_\tau \Psi_1 -\bu_1\bcdot\bnabla\Psi_2 - \divp\big(({\bf I}-\bp\bp^{\rm T})(\bnabla\bu_1\bp)\Psi_2 \big) -\Delta\Psi_1 .
\end{aligned}
\end{equation}
 Letting $\mc{R}(\bx,\theta,\tau)$ denote the right hand side of \eqref{Oeps3_noswim}, we have that $\mc{R}$ may be calculated explicitly using \eqref{noswim_psi1} and \eqref{noswim_psi2}; in particular, $\mc{R}$ is of the form 
\begin{equation}\label{noswim_RHSform}
\begin{aligned}
\mc{R}(\bx,\theta,\tau) &= R_{x}(\theta,\tau) {\rm e}^{{\rm i}x} + R_{y}(\theta,\tau){\rm e}^{{\rm i}y} + R_{2x^+}(\theta,\tau){\rm e}^{{\rm i}(2x+y)} + R_{2y^+}(\theta,\tau){\rm e}^{{\rm i}(x+2y)} \\
&\quad  + R_{2x^-}(\theta,\tau){\rm e}^{{\rm i}(2x-y)}+ R_{2y^-}(\theta,\tau) {\rm e}^{{\rm i}(-x+2y)} + R_{3x}(\theta,\tau){\rm e}^{{\rm i}3x} + R_{3y}(\theta,\tau){\rm e}^{{\rm i}3y} + {\rm c.c.}
\end{aligned}
\end{equation}

By the Fredholm alternative, equation \eqref{Oeps3_noswim} admits a solution $\Psi_3$ as long as
\begin{equation}\label{noswim_fredholm}
\int_0^{2\upi}\int_{\T^2} \mc{R}(\bx,\theta,\tau) \Phi(\bx,\theta) d\bx d\theta = 0 \qquad \text{for all }  \Phi(\bx,\theta) \text{ such that } \mc{L}^*[\Phi] =0.
\end{equation}
Since the operator $\mc{L}$ defined in \eqref{Oeps_noswim} is self-adjoint in the immotile case, we have that any such $\Phi$ has the form $\Phi= \cos\theta\sin\theta (\alpha_x {\rm e}^{{\rm i}x}+\alpha_y{\rm e}^{{\rm i}y})+{\rm c.c.}$ for any $\alpha_x^2+\alpha_y^2=1$. \\

Thus \eqref{noswim_fredholm} is automatically satisfied for each term of $\mc{R}$ except for $R_{x}(\theta,\tau) {\rm e}^{{\rm i}x} + R_{y}(\theta,\tau){\rm e}^{{\rm i}y}+ {\rm c.c}$. The exact form of $R_x$ and $R_y$ is given in Appendix \ref{app:noswim}, equation \eqref{noswim_RxRy}. \\

Since the ratio $\alpha_x/\alpha_y$ is arbitrary, we need that both 
\begin{align*}
\int_0^{2\upi}R_{x}(\theta,\tau) d\theta =0 \quad \text{and}\quad
\int_0^{2\upi}R_{y}(\theta,\tau) d\theta =0.
\end{align*}

These two conditions together lead to a coupled system of ODEs for the amplitudes $A_x$, $A_y$: 
\begin{equation}\label{noswim_GLeqns}
\begin{aligned}
c_x\big( \quad \p_\tau A_x &= A_x + (c_x^2M_1\abs{A_x}^2 +c_y^2M_2\abs{A_y}^2)A_x \quad \big) \\
c_y\big( \quad \p_\tau A_y &= A_y + (c_y^2M_1\abs{A_y}^2 + c_x^2M_2\abs{A_x}^2)A_y \quad)
\end{aligned}
\end{equation}
where 
\begin{equation}\label{noswimM1andM2}
M_1= -\frac{3 (3 - 28D_R - 32D_R^2) }{1024 D_R (1 - 12 D_R)}, \quad M_2 = \frac{7 -136D_R -2432D_R^2}{1024D_R(1-8D_R)(1+16D_R)} .
\end{equation}

\begin{figure}
\centering
  \includegraphics[scale=0.15]{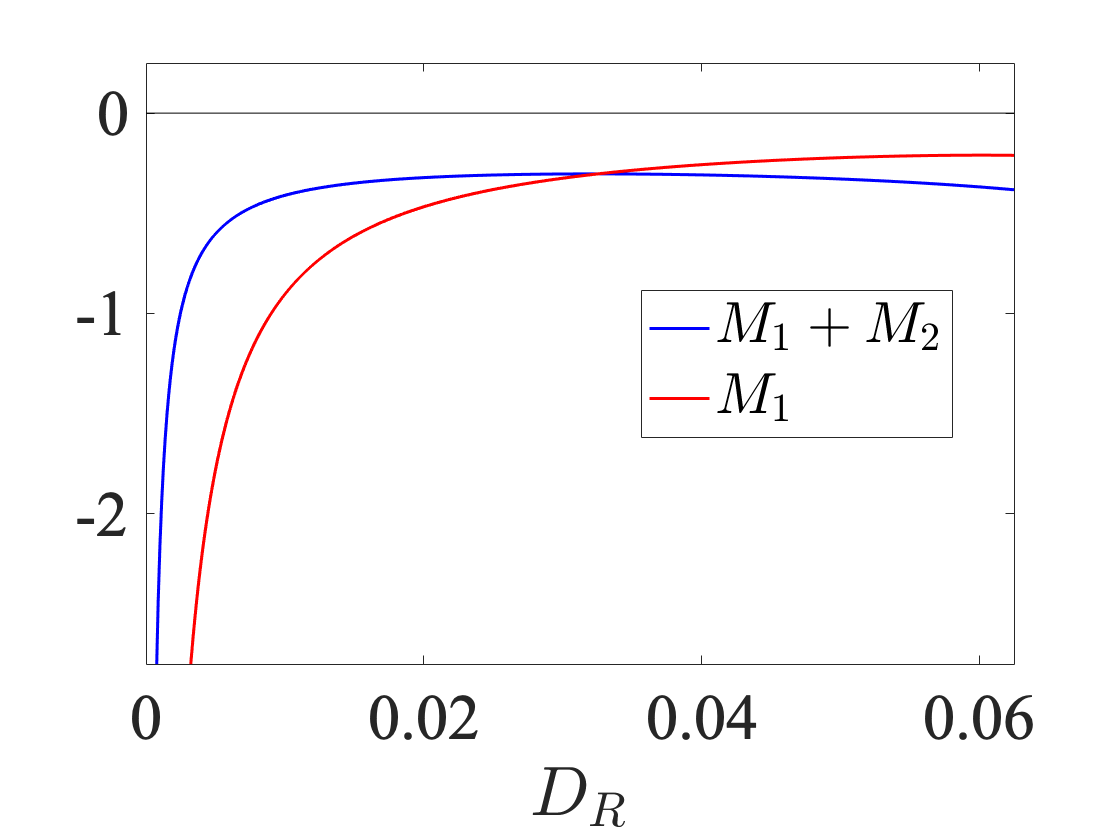} 
\caption{In the immotile setting, the coefficients $M_1+M_2$ \eqref{noswim_M1M2} and $M_1$ \eqref{noswimM1andM2} are both negative for all values of $D_R$ for which a bifurcation occurs ($0<D_R<1/16$), indicating that a supercritical pitchfork bifurcation occurs for both 2D ($x$ and $y$) and 1D ($x$-only) initial perturbations to the uniform isotropic state. } 
\label{fig:noswim_M1M2}
\end{figure}

If $M_1+M_2<0$, the system \eqref{noswim_GLeqns} has real, nonzero steady states of the form 
\begin{align*}
A_x &= \pm \frac{1}{c_x \sqrt{-(M_1+M_2)} }, \quad A_y = \pm \frac{1}{c_y \sqrt{-(M_1+M_2)} } .
\end{align*}
We have that 
\begin{equation}\label{noswim_M1M2}
M_1+M_2 = \frac{-1 -104D_R + 560D_R^2 +9600D_R^3 -6144D_R^4}{512D_R(1 -8D_R)(1 -12D_R)(1+16D_R)},
\end{equation}
which, as we can see from figure \ref{fig:noswim_M1M2}, is indeed negative for all values of $D_R$ for which a bifurcation occurs. Thus for any relevant level of rotational diffusion, the uniform, isotropic steady state loses stability through a supercritical pitchfork bifurcation and nontrivial stable steady states emerge. \\

To leading order in $\epsilon = \sqrt{D_T^*-D_T}=\sqrt{\frac{1}{4}-4D_R-D_T}$, the stable steady states which bifurcate from the uniform isotropic state are of the form 
\begin{equation}\label{noswim_steadystates}
\Psi = \frac{1}{2\upi} \pm \frac{\epsilon}{2\upi} \sqrt{\frac{512D_R(1 -8D_R)(1 -12D_R)(1+16D_R)}{1 +104D_R - 560D_R^2 -9600D_R^3 +6144D_R^4} }\cos\theta\sin\theta\big({\rm e}^{{\rm i}x} \pm {\rm e}^{{\rm i}y} \big) +{\rm c.c.} 
\end{equation}

Note that as long as the initial perturbation coefficients $c_x$ and $c_y$ are both nonzero, the form of \eqref{noswim_steadystates} does not depend on $c_x$ or $c_y$. If either $c_x=0$ or $c_y=0$ in \eqref{xy_planewave}, the bifurcating stable steady states take on a different form. Without loss of generality, we consider $c_y=0$. In this case, the coupled system \eqref{noswim_GLeqns} reduces to the single amplitude equation
\begin{equation}\label{noswim_GL_xonly}
\p_\tau A_x= A_x +M_1 \abs{A_x}^2A_x 
\end{equation}
where $M_1$ is as in \eqref{noswimM1andM2}. As shown in figure \ref{fig:noswim_M1M2}, $M_1$ is negative for all $0<D_R<1/16$, and therefore the uniform steady state still loses stability through a supercritical pitchfork bifurcation for all meaningful choices of $D_R$. To leading order in $\epsilon$, the stable steady states which emerge are of the form
\begin{equation}\label{noswim_steady_xonly}
\Psi(\bx,\theta) = \frac{1}{2\upi}\bigg(1 \pm \epsilon\sqrt{\frac{1024D_R(1-12D_R)}{3(3-28D_R-32D_R^2)}}\cos\theta\sin\theta {\rm e}^{{\rm i}x} \bigg) +{\rm c.c.}
\end{equation}

Numerical evidence of supercriticality along with simulated examples of the emerging steady states \eqref{noswim_steadystates} and \eqref{noswim_steady_xonly} are presented in the following section. \\

\subsection{Numerics}\label{subsec:noswim_numerics}
To study the immotile bifurcation numerically, we begin by checking for supercriticality. We first fix $D_R=0.0125$, so that by \eqref{noswim_eigen}, $D_T^*=0.2$ is the bifurcation value. Taking our initial condition to be a random, small-magnitude perturbation to the uniform isotropic state in both $x$ and $y$, we begin by running the simulation with $D_T=0.1$ until $t=500$. Then the value of $D_T$ is increased by $0.02$ every $100t$ until $D_T=0.3$. The bifurcation value $D_T^*=0.2$ is reached at $t=1000$. \\

\begin{figure}
\centering
  \begin{subfigure}{.45\textwidth}
   \includegraphics[scale=0.14]{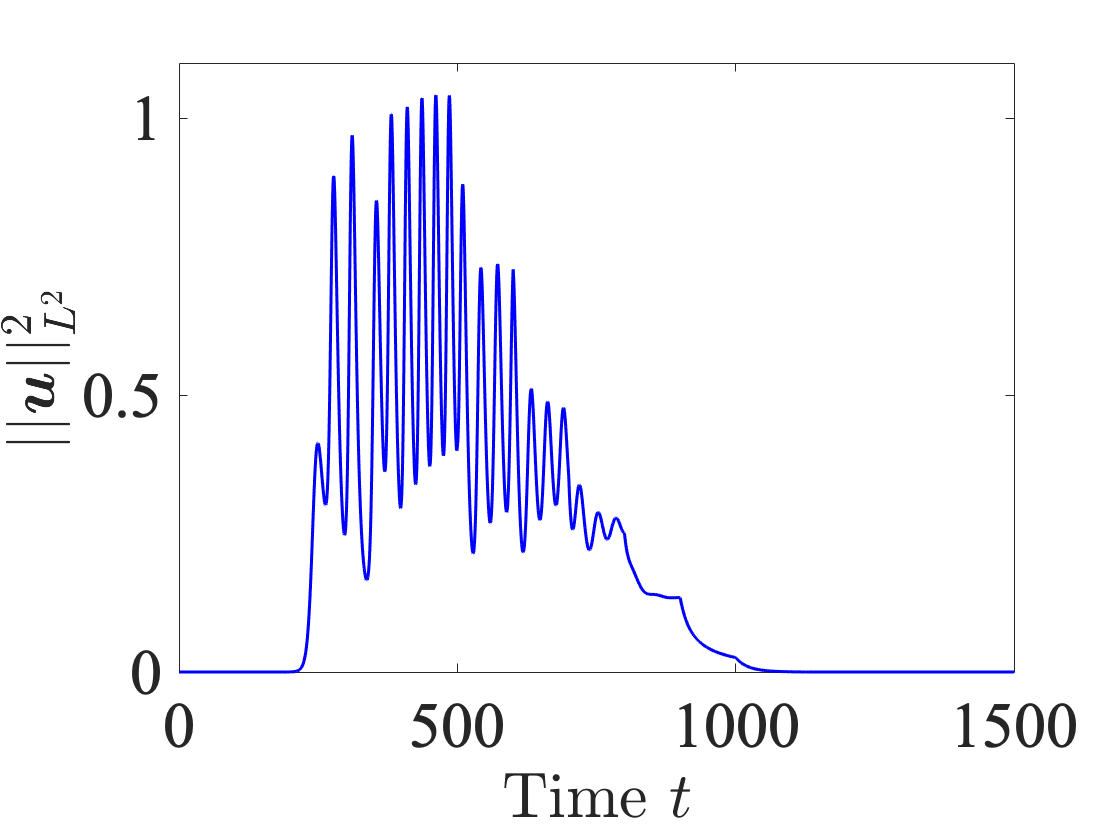}
    \caption{}
   \label{subfig:noswimL2}
   \end{subfigure}
    \begin{subfigure}{.45\textwidth}
   \includegraphics[scale=0.042]{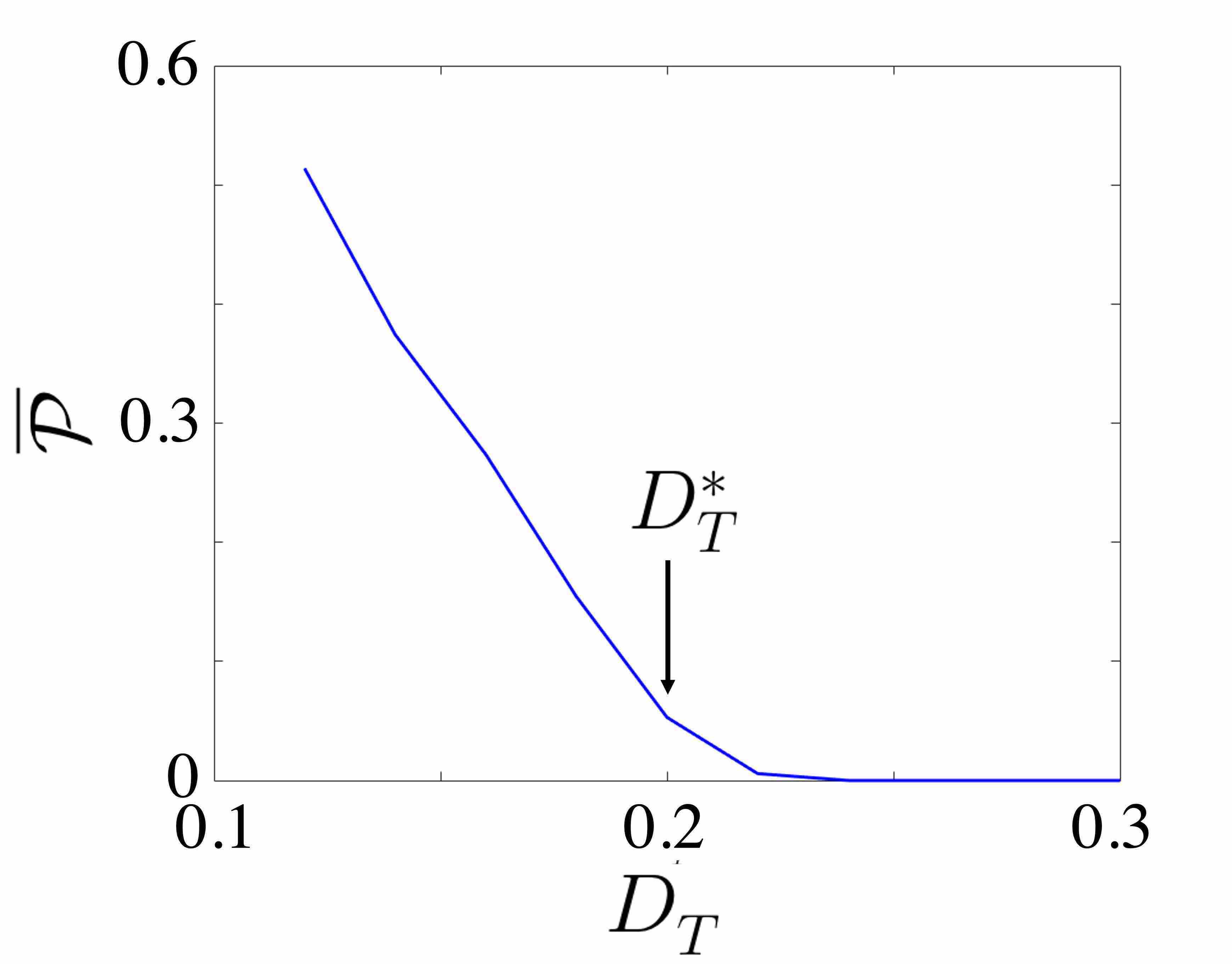}
   \caption{}
   \label{subfig:noswimVisDis}
   \end{subfigure}
\caption{Numerical evidence of supercriticality in the pitchfork bifurcation for immotile particles ($\beta=0$). Here $D_R=0.0125$ is fixed and the bifurcation occurs at $D_T^*=0.2$. The simulation is initiated with $D_T=0.1$ until $t=500$; then every $100t$ the value of $D_T$ is increased by $0.02$, so the bifurcation value is reached at $t=1000$. Figure (a) shows the $L^2$ norm of the fluid velocity field over time. 
Figure (b) shows the time-averaged viscous dissipation $\overline{\mc{P}}$ \eqref{timeavg_VD} for the different values of $D_T$. The apparently smooth decrease to zero as $D_T\to D_T^*$ indicates that the uniform isotropic steady state loses stability through a supercritical pitchfork bifurcation.  
 } 
\label{fig:noswim_supercrit}
\end{figure}

Over the course of the simulation we keep track of the $L^2$ norm of the velocity field 
\begin{align*}
\norm{\bu(\bcdot,t)}_{L^2(\T^2)}^2 = \int_{\T^2}\abs{\bu(\bx,t)}^2\,d\bx,
\end{align*} 
which is plotted continuously over time in figure \ref{subfig:noswimL2}. We also keep track of the time-averaged rate of viscous dissipation in the fluid for each value of $D_T$, which we recall from \eqref{power_to_VD} balances the active power input $\mc{P}$. Given a constant value of $D_T$ over the time interval $(t_1,t_2)$, we measure the value of 
\begin{equation}\label{timeavg_VD}
\overline{\mc{P}} = \frac{1}{t_2-t_1}\int_{t_1}^{t_2} \mc{P}(t)\, dt = \frac{1}{t_2-t_1}\int_{t_1}^{t_2}\int_{\T^2} 2\abs{\mathsfbi{E}(\bx,t)}^2 \, d\bx dt,
\end{equation}
which we consider as a function of $D_T$. In our case, $t_2-t_1=100$ for each different value of $D_T$. 
We plot $\overline{\mc{P}}$ in figure \ref{subfig:noswimVisDis} over the course of the simulation for the various values of $D_T$. As expected for a supercritical bifurcation, we see that $\overline{\mc{P}}$ smoothly decays to zero as $D_T$ increases toward the bifurcation value $D_T^*=0.02$, and remains zero after the bifurcation is reached. This smooth transition from nontrivial dynamics to the uniform, isotropic steady state as $D_T$ is slowly varied from a very unstable value through the bifurcation value and beyond may be contrasted with the hysteresis seen later in the subcritical Hopf region for motile particles (Section \ref{subsec:hopf_sub}).  \\

\begin{figure}
\centering
  \begin{subfigure}{.32\textwidth}
   \includegraphics[scale=0.29]{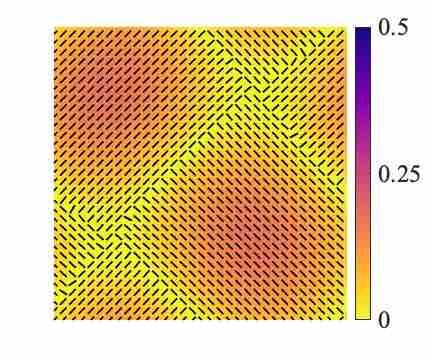}
    \caption{}
   \label{subfig:DT22xy}
   \end{subfigure}
    \begin{subfigure}{.32\textwidth}
   \includegraphics[scale=0.29]{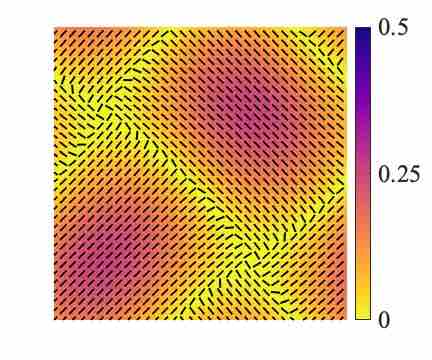}
   \caption{}
   \label{subfig:DT105xy}
   \end{subfigure}
     \begin{subfigure}{.32\textwidth}
   \includegraphics[scale=0.29]{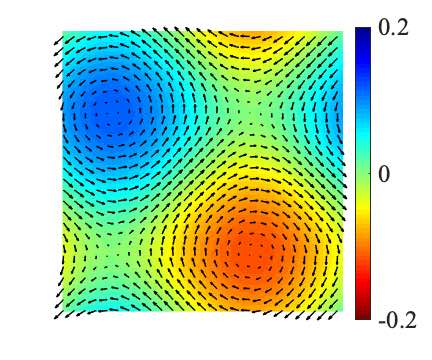}
   \caption{}
   \label{subfig:DT105xyvort}
   \end{subfigure}
\caption{
Plots of the nematic order parameter $\mc{N}(\bx,t)$ and the direction of local nematic alignment for (a) $(D_R,D_T)=(0.0025,0.22)$ ($D_T^*=0.24$) and (b) $(D_R,D_T)=(0.03125,0.105)$ ($D_T^*=0.125$) demonstrate the dependence of the emerging immotile steady state on $D_R$, as predicted by the form of \eqref{noswim_steadystates}. In both cases, $\epsilon^2=D_T^*-D_T=0.02$. Figure (c) shows the vorticity field $\omega(\bx,t)$ (colors) and velocity field $\bu(\bx,t)$ (arrows) for the same steady state pictured in figure (b). Note the clear extensile flow produced by the aligned dipoles in the top right and bottom left of the domain.
} 
\label{fig:xy_pert_differentDR}
\end{figure}

We next consider what the emerging stable steady states actually look like. Given $D_R$, we choose $D_T$ such that $\epsilon^2=D_T^*-D_T=0.02$. We initialize the simulation by perturbing the uniform, isotropic state with a small random-magnitude perturbation to a random assortment of the five lowest spatial modes in both $x$ and $y$ with random orientation $\theta$, and run until the dynamics settle into a steady state. The resulting nematic order parameter and direction of preferred local nematic alignment are plotted in figures \ref{subfig:DT22xy} and \ref{subfig:DT105xy} for $(D_R,D_T)=(0.0025,0.22)$ ($D_T^*=0.24$) and $(D_R, D_T)=(0.03125,0.105)$ ($D_T^*=0.125$), respectively. We see that the higher spatial modes decay over time, and the resulting steady state consists only of the unstable $\abs{k}=1$ mode. For $(D_R, D_T)=(0.03125,0.105)$ we also plot the fluid vorticity field 
\begin{equation}\label{vorticity}
\omega(\bx,t) = \bnabla\times \bu(\bx,t)
\end{equation}
and a sampling of the velocity field $\bu(\bx,t)$ throughout the domain. \\

 The supplementary video \texttt{Movie1} shows the system approaching an example of the steady state shown in figure \ref{fig:xy_pert_differentDR} as the bifurcation is approached from below. In this case $D_R=0.0125$ so $D_T^*=0.2$. The steady state is reached after approaching $D_T=0.18$ from below. \\

\begin{figure}
\centering
   \begin{subfigure}{.32\textwidth}
   \includegraphics[scale=0.29]{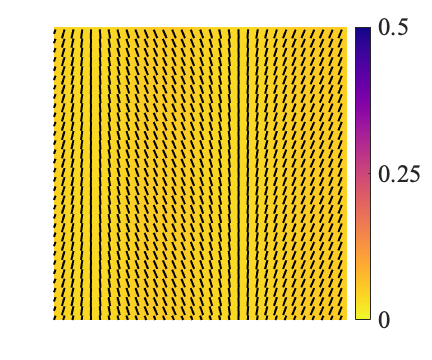}
    \caption{}
   \label{subfig:DT22xonly}
   \end{subfigure}
    \begin{subfigure}{.32\textwidth}
   \includegraphics[scale=0.29]{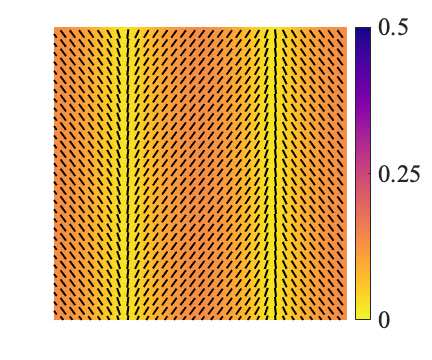}
   \caption{}
   \label{subfig:DT105xonly}
   \end{subfigure}
\begin{subfigure}{.32\textwidth}
   \includegraphics[scale=0.29]{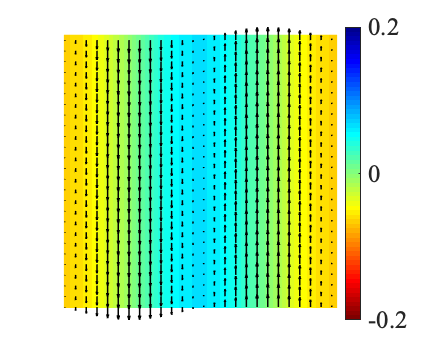}
   \caption{}
   \label{subfig:DT105xonly_vort}
   \end{subfigure}
\caption{In the case of an $x$-only initial perturbation, a similar dependence of the emerging steady state on $D_R$ can be seen in the two plots of the nematic order parameter $\mc{N}(\bx,t)$ and the direction of local nematic alignment for (a) $(D_R,D_T)=(0.0025,0.22)$ ($D_T^*=0.24$) and (b) $(D_R,D_T)=(0.03125,0.105)$ ($D_T^*=0.125$). Again, in both cases, $\epsilon^2=D_T^*-D_T=0.02$. Figure (c) shows the vorticity field $\omega(\bx,t)$ (colors) and velocity field $\bu(\bx,t)$ (arrows) for the $x$-only steady state with $(D_R,D_T)=(0.03125,0.105)$. 
} 
\label{fig:noswim_xonly_DR}
\end{figure}

We also consider the same parameter combinations as in figure \ref{fig:xy_pert_differentDR} for an $x$-only initial perturbation, and plot the results in figure \ref{fig:noswim_xonly_DR}. Given the form of the calculated steady state \eqref{noswim_steadystates} for a 2D ($x$ and $y$) perturbation and \eqref{noswim_steady_xonly} for a 1D ($x$-only) perturbation, we expect the particles to display a slight preference for the (nematic) orientations 
$\theta=\frac{\upi}{4}(\equiv\frac{5\upi}{4})$ and $\theta=\frac{3\upi}{4}(\equiv\frac{7\upi}{4})$ where $\pm\cos\theta\sin\theta$ is maximized. This preference is clearly visible in figures \ref{subfig:DT22xy} and \ref{subfig:DT105xy}
as well as in figures \ref{subfig:DT105xonly} and \ref{subfig:DT22xonly}. \\

 In addition, in the 2D case, figure \ref{fig:noswim_M1M2} suggests that given $\epsilon$ there should be a $D_R\in(0,\frac{1}{16})$ which maximizes the magnitude of the deviation from isotropy in the emerging stable steady state \eqref{noswim_steadystates}, and that as $D_R\to 0$ the correlation in particle alignment should disappear. We can see numerical evidence of this $D_R$ dependence in the difference in the magnitude of $\mc{N}(\bx,t)$ between figures \ref{subfig:DT105xy} and \ref{subfig:DT22xy}. \\

Similarly, in the case $c_y=0$, the $D_R$ dependence in \eqref{noswim_steady_xonly} suggests that as $D_R$ increases, the deviation from isotropy at a distance $\epsilon^2$ from the bifurcation continues to increase. This increase is shown in figures \ref{subfig:DT105xonly} and \ref{subfig:DT22xonly}.\\

\begin{figure}
\centering
   \begin{subfigure}{.45\textwidth}
   \includegraphics[scale=0.15]{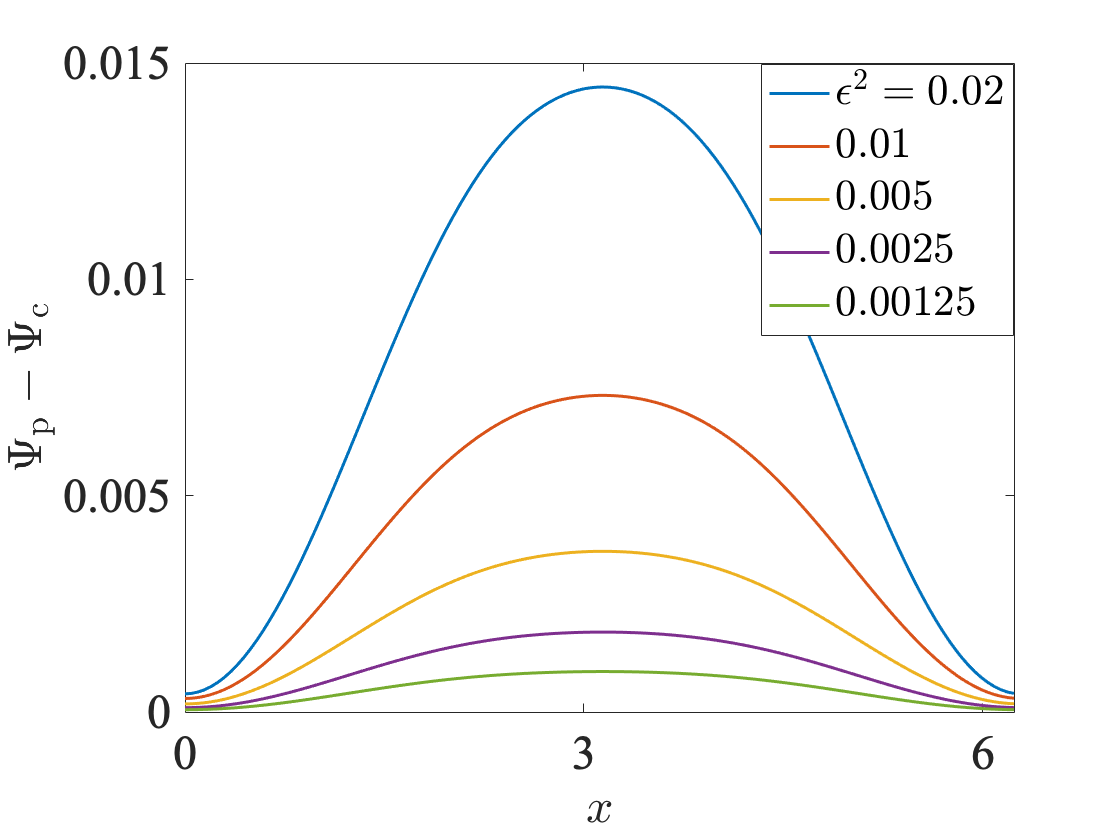}
    \caption{}
   \label{subfig:diffp17}
   \end{subfigure}
    \begin{subfigure}{.45\textwidth}
   \includegraphics[scale=0.15]{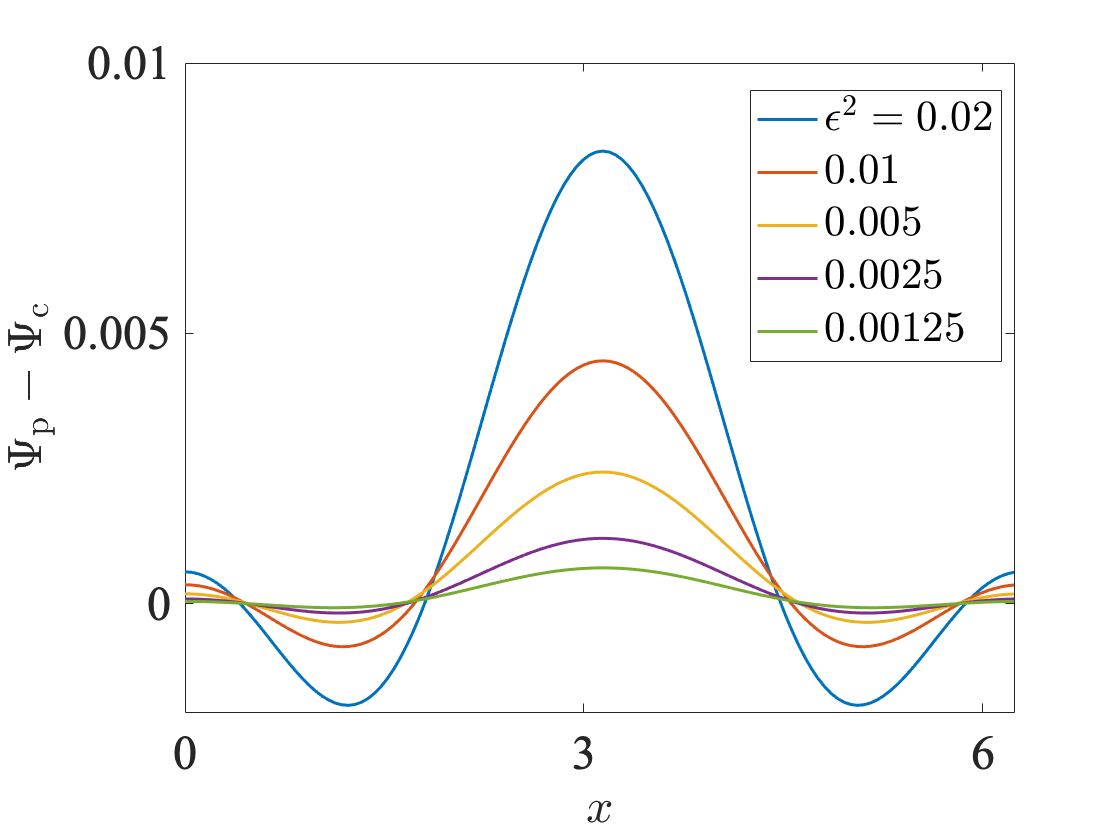}
   \caption{}
   \label{subfig:diffp24}
   \end{subfigure}
\caption{Difference between the predicted steady state $\Psi_{\rm p}$ \eqref{noswim_steadystates} and the computed steady state for \eqref{DSS} with $\beta=0$ for a fixed value of $y=\upi$ and two different fixed values of $\theta$: (a) $\theta=\upi$ and (b) $\theta=\frac{23\upi}{16}$. As expected, the difference between $\Psi_{\rm p}-\Psi_{\rm c}$ is $O(\epsilon^2)$. 
} 
\label{fig:noswim_comparison}
\end{figure}

Finally, we make a quantitative comparison between the expression \eqref{noswim_steadystates} and the numerical solution to the full system \eqref{DSS} with $\beta=0$ as $\epsilon^2=D_T^*-D_T$ is varied. We fix $D_R=0.03125$, so $D_T^*=0.125$, and consider five different values of $\epsilon^2$. For each $\epsilon$, we allow the system to reach a steady state, and in figure \ref{fig:noswim_comparison}, we plot the difference between the predicted steady state $\Psi_{\rm p}(x,y,\theta)$ \eqref{noswim_steadystates} and the computed steady state  $\Psi_{\rm c}(x,y,\theta)$ over a 1D slice of $x$-values for fixed $y$ and $\theta$. As $\epsilon$ is decreased, the pointwise difference between the predicted and computed steady states decreases like $\epsilon^2$, as expected. This is further corroborated by table \ref{tab:comparison}, which displays the maximum difference 
 $\max_{0\le x,y,\theta\le 2\upi}\abs{\Psi_{\rm p}(x,y,\theta)-\Psi_{\rm c}(x,y,\theta)}$.

\begin{table}
  \begin{center}
\def~{\hphantom{0}}
  \begin{tabular}{l|ccccc}
$\epsilon^2$ &  0.02 & 0.01 & 0.005 & 0.0025 & 0.00125 \\[3pt]
$\norm{\Psi_{\rm p}-\Psi_{\rm c}}_{L^\infty(\T^2\times S^1)}$  & 0.01667 & 0.00833 & 0.00421 & 0.00207 & 0.00107
  \end{tabular}
  \caption{Maximum difference between the predicted steady state $\Psi_{\rm p}$ \eqref{noswim_steadystates} and the computed steady state for \eqref{DSS} with $\beta=0$ for five different values of $\epsilon$. The difference scales with $\epsilon^2$, as expected. }
  \label{tab:comparison}
  \end{center}
\end{table}

\section{Motile particles: Hopf bifurcation}\label{sec:hopf}
When $\beta>0$, the system \eqref{DSS} can experience a much richer catalogue of bifurcations, as evidenced by figure \ref{fig:dispersion}. We focus first on the Hopf region (roughly between {\bf C} and {\bf E} on figure \ref{fig:dispersion}), where the eigenvalue corresponding to the $\abs{k}=1$ mode is complex-valued.  Fixing a very small level of rotational diffusion (e.g. $D_R=0.001$), we choose a value of translational diffusion $D_R\ll D_T<\frac{1}{9}$ such that for some value of $\beta=\beta_T$, we have $\sigma(\pm 1)=\pm {\rm i}b_T$; i.e. the real part of $\sigma(\pm1)$ vanishes and a Hopf bifurcation occurs. Here the subscript $T$ is used to denote that the values of $\beta_T$ and $b_T$ depend on the choice of $D_T$ through the implicit expression \eqref{disp_rel}. \\

We perform a weakly nonlinear analysis of the Hopf bifurcation for the $\abs{k}=1$ mode in Section \ref{subsec:hopf_wnl}. We show that for relatively large $D_T$, the bifurcation at $\beta=\beta_T$ is supercritical and a stable limit cycle arises just beyond the bifurcation. However, for small $D_T$, the bifurcation is subcritical for general initial perturbations in both $x$ and $y$, but supercritical for $x$-only perturbations. We find numerical evidence of hysteresis in this subcritical region, which we explore in Section \ref{subsec:hopf_sub}. In the supercritical region, we numerically locate an example of the emerging stable limit cycle (see Section \ref{subsec:hopf_super}).  
  
\subsection{Weakly nonlinear analysis}\label{subsec:hopf_wnl}
As in the immotile case, we consider $\beta=\beta_T-\epsilon^2$ for $\epsilon\ll1$, so the $\abs{k}=1$ modes are very slightly unstable. We again consider a slow timescale $\tau=\epsilon^2t$ and expand $\Psi$ and $\bu$ in $\epsilon$ as \eqref{expansion}. Plugging these expressions into \eqref{DSS}, at $O(\epsilon)$ we obtain the equation 
\begin{equation}\label{hopf_Oeps}
\mc{L}[\Psi_1] := \p_t\Psi_1 + \beta_T\bp\bcdot\bnabla\Psi_1 - 2\bp^{\rm T}\bnabla\bu_1\bp -D_T\Delta \Psi_1 - D_R\Delta_p \Psi_1=0.
\end{equation}

Note that when $\beta>0$ we no longer have an explicit expression for the eigenvalues and eigenfunctions of the linearized operator \eqref{DSS_linear} when $D_R>0$, and therefore the following analysis must be performed in the limit of very small rotational diffusion, which we treat perturbatively using Section \ref{subsec:DR}.  \\

Using the form \eqref{xy_planewave} of the eigenvalues and eigenfunctions of the linearized operator when $D_R=0$, and using the perturbative expression for $D_R>0$ from Section \ref{subsec:DR}, we have that $\Psi_1$ and $\bu_1$ are given up to $O(D_R)$ by
\begin{equation}\label{hopf_psi1}
\begin{aligned}
\Psi_1 &= c_x\psi_{x,1}(\theta)A_x(\tau){\rm e}^{{\rm i}x+{\rm i}b_Tt} +c_y\psi_{y,1}(\theta)A_y(\tau){\rm e}^{{\rm i}y+{\rm i}b_Tt} + {\rm c.c.} \\
&\quad \psi_{x,1}(\theta) = \frac{\cos\theta\sin\theta}{D_T+ {\rm i}b_T+ {\rm i}\beta_T\cos\theta} + O(D_R), \quad 
\psi_{y,1}(\theta) = \frac{\cos\theta\sin\theta}{D_T+ {\rm i}b_T + {\rm i}\beta_T\sin\theta} + O(D_R) \\
\bu_1 &= -\frac{\rm i}{2} \big( c_x A_x {\rm e}^{{\rm i}x+{\rm i}b_Tt}\be_y + c_y A_y{\rm e}^{{\rm i}y+{\rm i}b_Tt}\be_x\big) + {\rm c.c.} + O(D_R).
\end{aligned}
\end{equation}
Again, $A_x(\tau)$ and $A_y(\tau)$ are complex-valued amplitudes which depend only on the slow timescale $\tau$ and for which we wish to obtain an equation. Here again ${\rm c.c.}$ denotes the complex conjugate of the preceding terms. \\

At $O(\epsilon^2)$ we obtain the equation
\begin{equation}\label{hopf_Oeps2}
\mc{L}[\Psi_2] = -\bu_1\bcdot\bnabla\Psi_1 - \divp\big(({\bf I}-\bp\bp^{\rm T}(\bnabla\bu_1\bp)\Psi_1) \big) ,
\end{equation}
where the operator $\mc{L}$ is as defined in \eqref{hopf_Oeps}. Using \eqref{hopf_psi1}, we can explicitly calculate the expression on the right hand side of \eqref{hopf_Oeps2} up to terms of $O(D_R)$ (see Appendix \ref{app:hopf}, equation \eqref{Oeps2_hopf_RHS} for the full expression). As in the immotile case, using the form of the right hand side as a guide, we look for $(\Psi_2,\bu_2)$ of the form 
\begin{equation}\label{hopf_psi2}
\begin{aligned}
\Psi_2 &= \psi_{2,1}{\rm e}^{{\rm i}(x+y)+2{\rm i}b_Tt}A_xA_y +\psi_{2,2}{\rm e}^{{\rm i}(x-y)}A_x\overline A_y + \psi_{2,3}A_x^2{\rm e}^{2{\rm i}x+2{\rm i}b_Tt} +\psi_{2,4}A_y^2{\rm e}^{2{\rm i}y+2{\rm i}b_Tt} \\
&\hspace{3cm}+ \psi_{2,5}\abs{A_x}^2+\psi_{2,6}\abs{A_y}^2 + {\rm c.c.}, \\
\bu_2 &= - \frac{\rm i}{8\upi}\int_0^{2\upi}\bigg(\psi_{2,1}{\rm e}^{{\rm i}(x+y)+2{\rm i}b_Tt}A_xA_y(\be_x-\be_y) +\psi_{2,2}{\rm e}^{{\rm i}(x-y)}A_x\overline A_y(\be_x+\be_y)\bigg)(\cos^2\theta -\sin^2\theta) d\theta \\
&\qquad -\frac{\rm i}{4\upi}\int_0^{2\upi}\bigg( \psi_{2,3}{\rm e}^{2{\rm i}x+2{\rm i}b_Tt} A_x^2\be_y + \psi_{2,4}{\rm e}^{2{\rm i}y+2{\rm i}b_Tt} A_y^2\be_x \bigg) \sin\theta\cos\theta\,d\theta +{\rm c.c.},
\end{aligned}
\end{equation}
where $\psi_{2,j}=\psi_{2,j}(\theta)$. Inserting the ansatz \eqref{hopf_psi2} into the left hand side of \eqref{hopf_Oeps2}, we then match exponents with the right hand side and solve for each of the coefficients $\psi_{2,j}$. Further details are contained in Appendix \ref{app:hopf}, equations \eqref{hopf_psi1234} and \eqref{hopf_psi56}, but we obtain that both $\psi_{2,5}$ and $\psi_{2,6}$ are $O(D_R^{-1})$ for small $D_R$, while each of the other coefficients are $O(1)$ in $D_R$ as $D_R\to 0$. Thus for sufficiently small $D_R$, the coefficients $\psi_{2,5}$ and $\psi_{2,6}$ dominate the behavior of $\Psi_2$. We may solve for $\psi_{2,5}$ and $\psi_{2,6}$ explicitly up to terms of $O(1)$ in $D_R$:
\begin{equation}\label{hopf_psij}
\begin{aligned}
\psi_{2,5} &= \frac{c_x^2}{D_R}\bigg[a_1\cos\theta +a_2(2\cos^2\theta-1) + a_3\cos^3\theta \\
&\quad + a_4\bigg( \log(D_T+{\rm i}b_T+{\rm i}\beta_T\cos\theta) - \frac{1}{2\upi}\int_0^{2\upi}\log(D_T+{\rm i}b_T+{\rm i}\beta_T\cos\theta)\, d\theta \bigg) + O(D_R)\bigg]  \\
\psi_{2,6} &= \frac{c_y^2}{D_R}\bigg[a_1\sin\theta +a_2(2\sin^2\theta-1)+a_3\sin^3\theta \\
 &\quad + a_4\bigg(\log(D_T+{\rm i}b_T+{\rm i}\beta_T\sin\theta) - \frac{1}{2\upi}\int_0^{2\upi}\log(D_T+{\rm i}b_T+{\rm i}\beta_T\sin\theta)\, d\theta \bigg) +O(D_R)\bigg] 
\end{aligned}
\end{equation}
where
\begin{equation}\label{a_vals}
a_1 = -\frac{{\rm i}(D_T+{\rm i}b_T)^2}{2\beta_T^3}, \quad 
a_2 = -\frac{D_T+{\rm i}b_T}{8\beta_T^2}, \quad 
a_3 = \frac{\rm i}{6\beta_T}, \quad
a_4 = \frac{(D_T+{\rm i}b_T)^3}{2\beta_T^4}.
\end{equation}
Note that we must enforce $\int_0^{2\upi}\psi_{2,5}\,d\theta =\int_0^{2\upi}\psi_{2,6}\,d\theta =0$ in order to have $\int_{\T^2}\int_0^{2\upi}\Psi_2\,d\theta\,d\bx=0$, i.e. to enforce that the total concentration of particles in the system is preserved at $1/(2\upi)$. \\

We may thus rewrite \eqref{hopf_psi2} as 
\begin{align*}
\Psi_2 &= \psi_{2,5}(\theta)\abs{A_x}^2+\psi_{2,6}(\theta)\abs{A_y}^2 + O(D_R^0), \quad \bu_2= O(D_R^0). 
\end{align*}

At $O(\epsilon^3)$ we obtain the equation 
\begin{equation}\label{hopf_Oeps3}
\begin{aligned}
\mc{L}[\Psi_3] &= -\p_\tau \Psi_1 +\bp\bcdot\bnabla\Psi_1 -\bu_1\bcdot\bnabla\Psi_2 -\bu_2\bcdot\bnabla\Psi_1\\
&\quad - \divp\big(({\bf I}-\bp\bp^{\rm T})(\bnabla\bu_2\bp)\Psi_1)\big)  - \divp\big(({\bf I}-\bp\bp^{\rm T})(\bnabla\bu_1\bp)\Psi_2 \big).
\end{aligned}
\end{equation}

Using \eqref{hopf_psi1} and \eqref{hopf_psi2}, we may calculate the form of the right hand side. First, we note that 
\begin{equation}\label{hopf_psiterms} 
\begin{aligned}
 \p_\tau\Psi_1 &= c_x\psi_{x,1}(\theta)(\p_\tau A_x){\rm e}^{{\rm i}x+{\rm i}b_Tt} +c_y\psi_{y,1}(\theta)(\p_\tau A_y){\rm e}^{{\rm i}y+{\rm i}b_Tt}  +{\rm c.c.} \\
 \bp\bcdot\bnabla\Psi_1 &= {\rm i} c_x\cos\theta\psi_{x,1}(\theta)A_x{\rm e}^{{\rm i}x+{\rm i}b_Tt} + {\rm i} c_y\sin\theta\psi_{y,1}(\theta)A_y{\rm e}^{{\rm i}y+{\rm i}b_Tt} +{\rm c.c.},
 \end{aligned} 
 \end{equation}
 where $\psi_{x,1}$ and $\psi_{y,1}$ are as in \eqref{hopf_psi1}. The remaining terms on the right hand side of \eqref{hopf_Oeps3} may be written 
 \begin{equation}\label{hopf_cubic_in_A}
 \begin{aligned}
 -\bu_1\bcdot\bnabla\Psi_2 &-\bu_2\bcdot\bnabla\Psi_1 - \divp\big(({\bf I}-\bp\bp^{\rm T})(\bnabla\bu_2\bp)\Psi_1)\big)  - \divp\big(({\bf I}-\bp\bp^{\rm T})(\bnabla\bu_1\bp)\Psi_2 \big)  \\
  &= \mc{R}_1(\bx,\theta,t,\tau)+\mc{R}_3(\bx,\theta,t,\tau)+{\rm c.c.}, \\
\mc{R}_1(\bx,\theta,t,\tau)&= \frac{1}{D_R}\bigg(R_{xx}(\theta)c_x^3\abs{A_x}^2 A_x{\rm e}^{{\rm i}x+{\rm i}b_Tt} +R_{xy}(\theta)c_xc_y^2\abs{A_y}^2 A_x{\rm e}^{{\rm i}x+{\rm i}b_Tt}\\ 
&\quad +R_{yx}(\theta)c_yc_x^2\abs{A_x}^2 A_y{\rm e}^{{\rm i}y+{\rm i}b_Tt} + R_{yy}(\theta)c_y^3\abs{A_y}^2 A_y{\rm e}^{{\rm i}y+{\rm i}b_Tt}\bigg) \\
\mc{R}_3(\bx,\theta,t,\tau) &= R_{2x^+}(\theta,\tau){\rm e}^{{\rm i}(2x+y)+{\rm i}3b_Tt} + R_{2y^+}(\theta,\tau){\rm e}^{{\rm i}(x+2y)+{\rm i}3b_Tt} + R_{2x^-}(\theta,\tau){\rm e}^{{\rm i}(2x-y)+{\rm i}b_Tt} \\
&\quad + R_{2y^-}(\theta,\tau) {\rm e}^{{\rm i}(-x+2y)+{\rm i}b_Tt} + R_{3x}(\theta,\tau){\rm e}^{{\rm i}3x+{\rm i}3b_Tt} + R_{3y}(\theta,\tau){\rm e}^{{\rm i}3y+{\rm i}3b_Tt}. 
\end{aligned}
\end{equation}
The expressions for $R_{xx}$, $R_{xy}(\theta)$, $R_{yx}(\theta)$, and $R_{yy}(\theta)$ are written up to $O(D_R)$ in Appendix \ref{app:hopf}, equation \eqref{hopf_Rthetas}. As in the immotile setting, the exact form of the coefficients in $\mc{R}_3(\bx,\theta,t,\tau)$ will not be important due to the solvability condition for equation \eqref{hopf_Oeps3}. \\

In particular, in order for the $O(\epsilon^3)$ equation \eqref{hopf_Oeps3} to have a solution $\Psi_3$, the right hand side of \eqref{hopf_Oeps3} must satisfy the same Fredholm condition \eqref{noswim_fredholm} as in the immotile case, except now the operator $\mc{L}$ defined in \eqref{hopf_Oeps} is no longer self-adjoint. The adjoint $\mc{L}^*$ is given by 
\begin{align*}
\mc{L}^*[\Phi] = -\p_t\Phi - \beta_T\bp\bcdot\bnabla\Phi - 2\bp^{\rm T}\bnabla\bu\bp -D_T\Delta \Phi - D_R\Delta_p \Phi,
\end{align*}

and, by the same calculation as in Sections \ref{subsec:linearstab} and \ref{subsec:DR}, any $\Phi$ satisfying $\mc{L}^*[\Phi]=0$ has the form
\begin{equation}\label{hopf_LstarPhi}
\begin{aligned}
\Phi = \alpha_x&\overline\psi_{x,1}(\theta){\rm e}^{{\rm i}x+{\rm i}b_Tt} + \alpha_y\overline\psi_{y,1}(\theta){\rm e}^{ iy+{\rm i}b_Tt} +{\rm c.c.} , \qquad \alpha_x^2+\alpha_y^2=1 \\
\overline\psi_{x,1} &= \frac{\cos\theta\sin\theta}{D_T- {\rm i}b_T - {\rm i}\beta_T\cos\theta}+ O(D_R), \quad 
\overline\psi_{y,1}=\frac{\cos\theta\sin\theta}{D_T- {\rm i}b_T - {\rm i}\beta_T\sin\theta}+ O(D_R).
\end{aligned}
\end{equation}

Due to the form of $\mc{R}_3$ \eqref{hopf_cubic_in_A}, we have that 
\begin{align*}
\int_{\T^2}\mc{R}_3(\bx,\theta,t,\tau)\Phi(\bx,\theta,t)\, d\bx =0
\end{align*}
for $\Phi$ as in \eqref{hopf_LstarPhi}, so the Fredholm condition \eqref{noswim_fredholm} is automatically satisfied. Thus it remains to ensure that 
\begin{align*}
\int_0^{2\upi}\int_{\T^2}\bigg(-\p_\tau\Psi_1 +\bp\bcdot\bnabla\Psi_1 +\mc{R}_1 \bigg)\Phi\, d\bx d\theta &=0 
\end{align*}
for any $\Phi$ as in \eqref{hopf_LstarPhi}. Using the forms of \eqref{hopf_psiterms}, \eqref{hopf_cubic_in_A}, and \eqref{hopf_LstarPhi}, this leads to the following coupled system of equations for the amplitudes $A_x$ and $A_y$: 
\begin{equation}\label{hopf_amplitudes}
\begin{aligned}
c_x\big( \quad M_0(\p_\tau A_x) &= M_3A_x + \frac{1}{D_R}(c_x^2M_1\abs{A_x}^2 +c_y^2M_2\abs{A_y}^2)A_x \quad \big) \\
c_y\big( \quad M_0(\p_\tau A_y) &= M_3A_y + \frac{1}{D_R}(c_y^2M_1\abs{A_y}^2 + c_x^2M_2\abs{A_x}^2)A_y \quad)
\end{aligned}
\end{equation}
where the complex constants $M_0$, $M_1$, $M_2$, and $M_3$ are given by
\begin{equation}\label{hopf_Mjs}
\begin{aligned}
M_0 &= \int_0^{2\upi}\psi_{x,1}^2(\theta) \, d\theta = \int_0^{2\upi}\psi_{y,1}^2(\theta) \, d\theta \\
M_1 &= \int_0^{2\upi}R_{xx}(\theta)\psi_{x,1}(\theta) \, d\theta=\int_0^{2\upi}R_{yy}(\theta)\psi_{y,1}(\theta) \, d\theta\\
M_2 &= \int_0^{2\upi}R_{xy}(\theta)\psi_{x,1}(\theta) \, d\theta =
\int_0^{2\upi}R_{yx}(\theta)\psi_{y,1}(\theta) \, d\theta \\
M_3 &= {\rm i}\int_0^{2\upi}\psi_{x,1}^2(\theta) \cos\theta\, d\theta = {\rm i}\int_0^{2\upi}\psi_{y,1}^2(\theta) \sin\theta\, d\theta .
\end{aligned}
\end{equation}
Explicit expressions for each $M_j$ depending on $D_T$, $b_T$, and $\beta_T$ are given in Appendix \eqref{app:hopf}, equation \eqref{hopf_Mjs_expr}. Note that although $\psi_{x,1}(\theta)\neq\psi_{y,1}(\theta)$, etc., the coefficients $M_j$ for the $x$ and $y$ directions are equal.  \\

In the Hopf setting, each coefficient $M_j$ is now complex-valued. Writing $A_x(\tau)=\rho_x(\tau){\rm e}^{{\rm i}\varphi_x(\tau)}$ and $A_y(\tau)=\rho_y(\tau){\rm e}^{{\rm i}\varphi_y(\tau)}$, we may separate \eqref{hopf_amplitudes} into equations for the magnitudes $\rho_x$, $\rho_y$ and the phases $\varphi_x$, $\varphi_y$. We then look for conditions on the coefficients $M_j$ such that \eqref{hopf_amplitudes} admits a nontrivial limit cycle satisfying $\p_\tau\rho_x=\p_\tau\rho_y=0$. \\

We find that if $M_0\neq 0$, Re$\big( (M_1+M_2)/M_0\big)\neq 0$, and
\begin{equation}\label{hopf_condition}
\frac{{\rm Re}\big(\frac{M_3}{M_0}\big)}{{\rm Re}\big(\frac{M_1+M_2}{M_0}\big)} < 0,
\end{equation}
then \eqref{hopf_amplitudes} gives rise to a nontrivial limit cycle with magnitudes $\rho_x$ and $\rho_y$ given by 
\begin{equation}\label{hopf_limcyc_real}
\rho_x = \pm\frac{\sqrt{D_R}}{c_x}\sqrt{-\frac{{\rm Re}\big(\frac{M_3}{M_0}\big)}{{\rm Re}\big(\frac{M_1+M_2}{M_0}\big)}}, \quad \rho_y = \pm\frac{\sqrt{D_R}}{c_y}\sqrt{-\frac{{\rm Re}\big(\frac{M_3}{M_0}\big)}{{\rm Re}\big(\frac{M_1+M_2}{M_0}\big)}},
\end{equation}
while the phases satisfy 
\begin{equation}\label{hopf_limcyc_im}
\p_\tau\varphi_x = \p_\tau\varphi_y = {\rm Im}\bigg(\frac{M_3}{M_0}\bigg) - {\rm Im}\bigg(\frac{M_1+M_2}{M_0}\bigg)\frac{{\rm Re}\big(\frac{M_3}{M_0}\big)}{{\rm Re}\big(\frac{M_1+M_2}{M_0}\big)} =:\varphi_T. 
\end{equation}

In particular, if \eqref{hopf_condition} holds, the stable limit cycle arising just after the bifurcation, to leading order in $\epsilon=\sqrt{\beta_T-
\beta}$, is given by 
\begin{equation}\label{hopf_solution}
\Psi = \frac{1}{2\upi} \bigg(1 \pm \epsilon \sqrt{D_R}\sqrt{-\frac{{\rm Re}\big(\frac{M_3}{M_0}\big)}{{\rm Re}\big(\frac{M_1+M_2}{M_0}\big)}}\bigg(\psi_{x,1}(\theta){\rm e}^{{\rm i}x+{\rm i}(b_T+\epsilon^2\varphi_T)t} \pm \psi_{y,1}(\theta){\rm e}^{{\rm i}y+{\rm i}(b_T+\epsilon^2\varphi_T)t} \bigg) \bigg).
\end{equation}

If the condition \eqref{hopf_condition} does not hold, the bifurcation at $\beta=\beta_T$ is subcritical and, while the system behavior for $\beta<\beta_T$ is less predictable, we may expect to see hysteresis if $\beta$ is then increased beyond $\beta_T$. In particular, the system may remain in a stable, nontrivial state well after the uniform isotropic state has also become stable. \\

As in the immotile setting, we also consider the effects of an initial perturbation in the $x$-direction only. When $c_y=0$, we obtain the single equation 
\begin{equation}\label{hopf_amp_xonly}
 M_0(\p_\tau A_x) = M_3A_x + M_1\abs{A_x}^2A_x 
\end{equation}
for the $x$-direction amplitude $A_x$. In this case a stable limit cycle emerges beyond the bifurcation if $M_0\neq0$, Re$\big(M_1/M_0\big)\neq 0$, and 
\begin{equation}\label{hopf_cond_xonly}
\frac{{\rm Re}\big(\frac{M_3}{M_0}\big)}{{\rm Re}\big(\frac{M_1}{M_0}\big)} <0.
\end{equation}
If these conditions are satisfied, the magnitude and phase of the emerging limit cycle are given by 
\begin{equation}\label{hopf_limcyc_xonly}
\rho_x = \pm\sqrt{D_R}\sqrt{-\frac{{\rm Re}\big(\frac{M_3}{M_0}\big)}{{\rm Re}\big(\frac{M_1}{M_0}\big)}}, \quad \p_\tau\varphi_x = {\rm Im}\bigg(\frac{M_3}{M_0}\bigg) - {\rm Im}\bigg(\frac{M_1}{M_0}\bigg)\frac{{\rm Re}\big(\frac{M_3}{M_0}\big)}{{\rm Re}\big(\frac{M_1}{M_0}\big)}=:\varphi_T,
\end{equation}
and, to leading order in $\epsilon$, the emerging solution after the bifurcation has the form
\begin{equation}\label{hopf_solution_xonly}
\Psi = \frac{1}{2\upi} \bigg(1 \pm \epsilon \sqrt{D_R}\sqrt{-\frac{{\rm Re}\big(\frac{M_3}{M_0}\big)}{{\rm Re}\big(\frac{M_1}{M_0}\big)}}\psi_{x,1}(\theta){\rm e}^{{\rm i}x+{\rm i}(b_T+\epsilon^2\varphi_T)t} \bigg)
\end{equation}

\begin{figure}
\centering
 \begin{subfigure}{.45\textwidth}
   \includegraphics[scale=0.13]{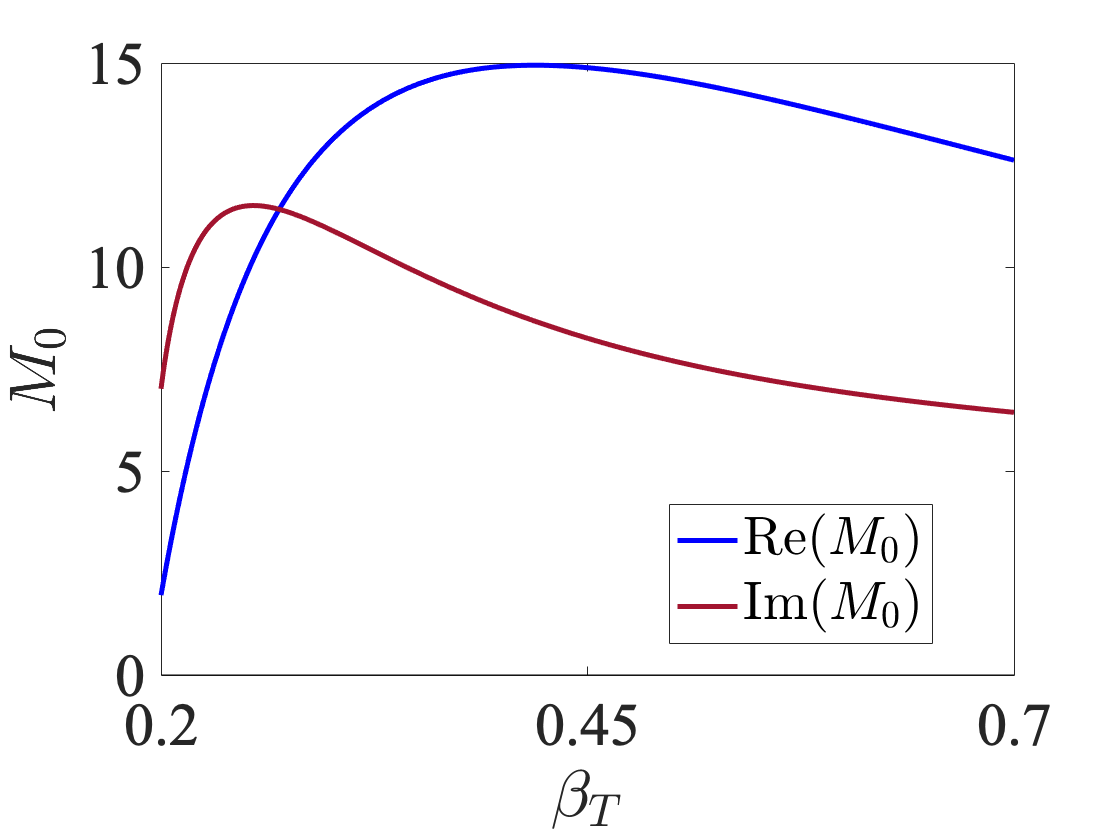}
    \caption{}
   \label{subfig:M0}
   \end{subfigure}
    \begin{subfigure}{.45\textwidth}
   \includegraphics[scale=0.13]{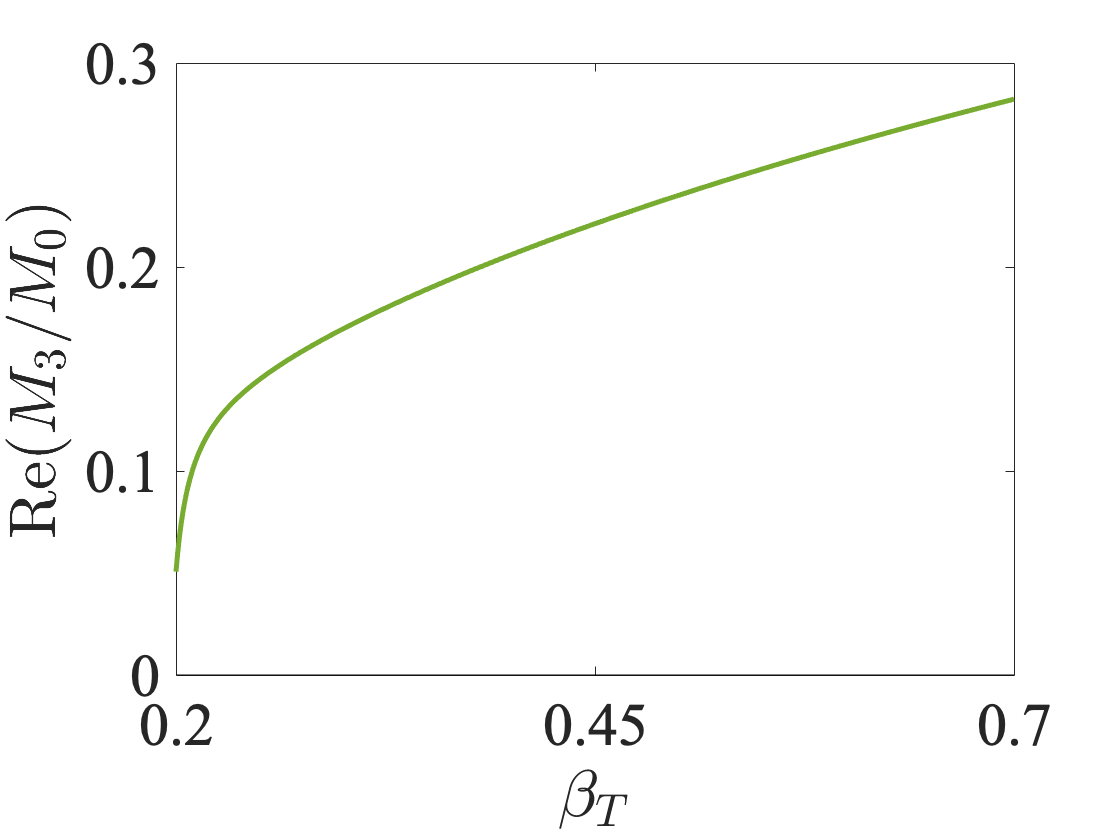}
   \caption{}
   \label{subfig:M3overM0}
   \end{subfigure} \\
   \begin{subfigure}{.45\textwidth}
   \includegraphics[scale=0.045]{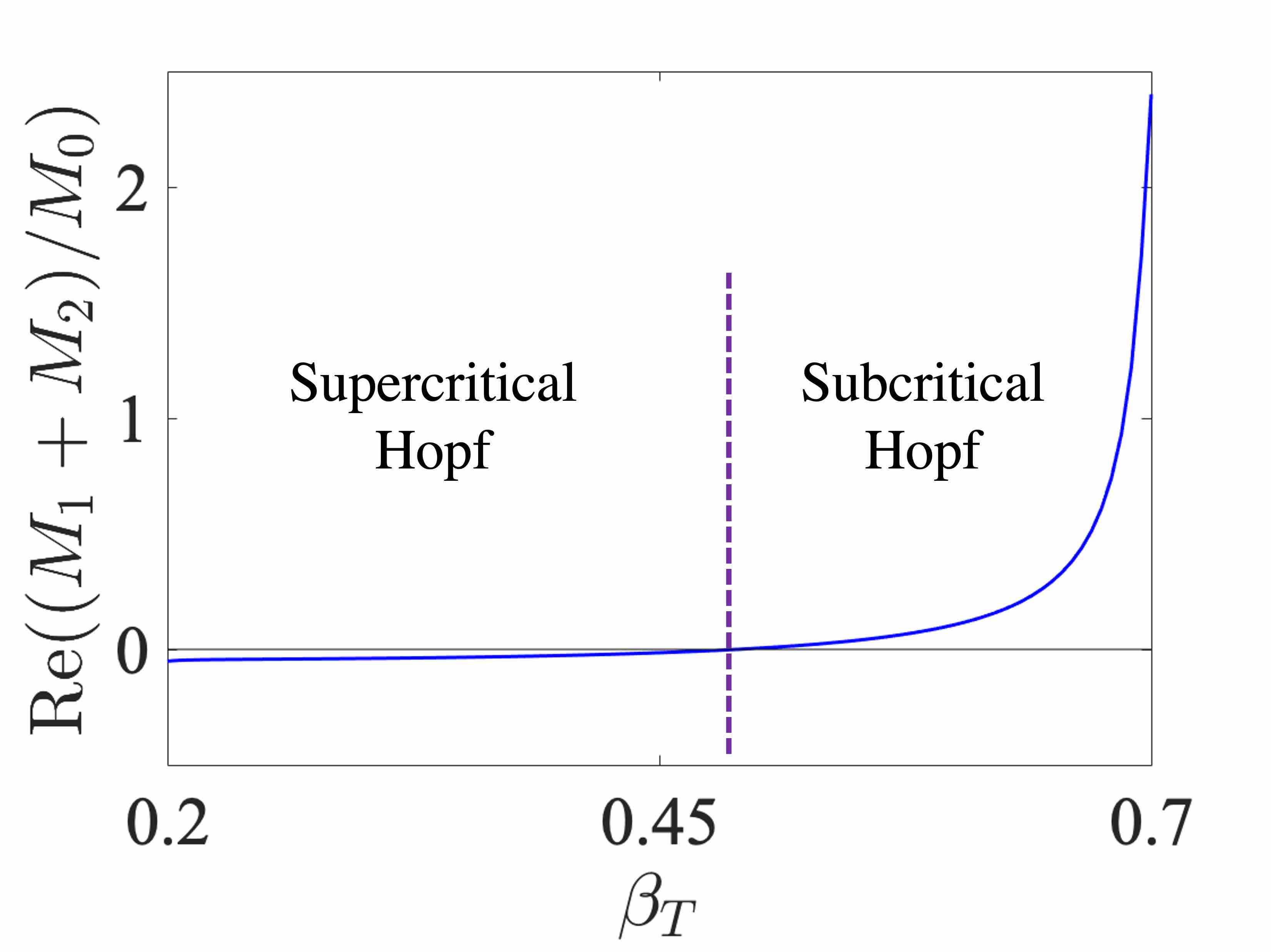}
   \caption{}
   \label{subfig:M1M2overM0}
   \end{subfigure}
     \begin{subfigure}{.45\textwidth}
   \includegraphics[scale=0.045]{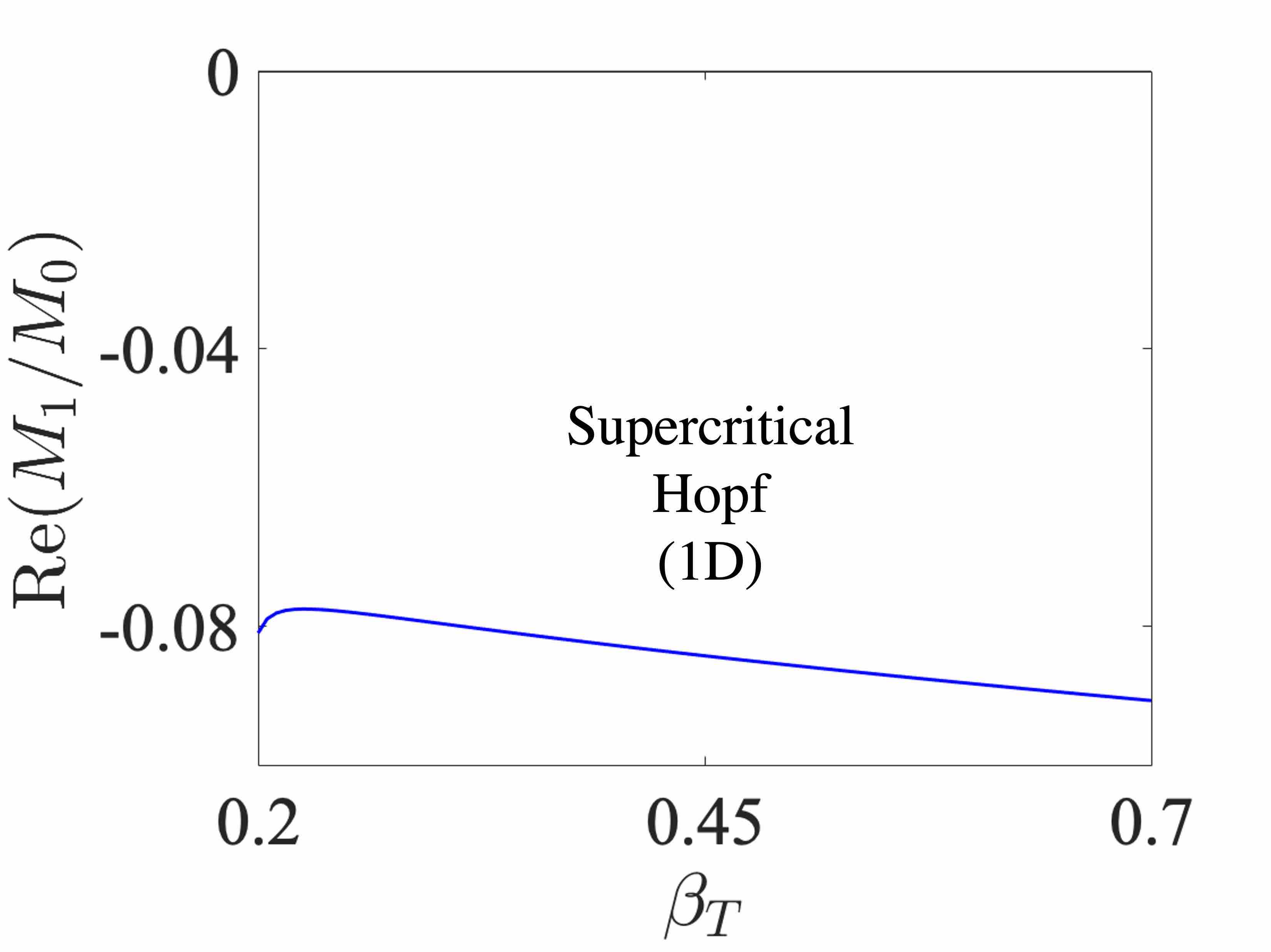}
   \caption{}
   \label{subfig:M1overM0}
   \end{subfigure}
\caption{ The relevant relationships among the coefficients $M_0$, $M_1$, $M_2$, and $M_3$ of the amplitude equations \eqref{hopf_amplitudes} are plotted over the Hopf bifurcation range $\beta_T\in[0.2,0.7]$. In particular, since the curves for (a) $M_0$ and (b) Re$(M_3/M_0)$ are both strictly positive for all such $\beta_T$, the type of Hopf bifurcation is determined by (c) Re$\big((M_1+M_2)/M_0\big)$ (for 2D initial perturbations) or (d) Re$\big(M_1/M_0\big)$ (for 1D initial perturbations). 
In the 2D case, there is a transition from supercritical to subcritical at some value of $\beta_T\in[0.2,0.7]$, indicated by the vertical dotted line in (c).  
} 
\label{fig:hopf_coeffs}
\end{figure}

In figure \ref{fig:hopf_coeffs}, we plot each of $M_0$, Re$\big(M_3/M_0\big)$, Re$\big((M_1+M_2)/M_0\big)$, and Re$\big(M_1/M_0\big)$ using the perturbed dispersion relation \eqref{DR_expansion} (see figure \ref{fig:disp_pert}) with $D_R=0.001$ for $\beta_T\in[0.2,0.7]$. This range of $\beta_T$ essentially covers all $\beta_T$ for which a Hopf bifurcation exists and for which the perturbative expression in $D_R$ is valid. \\

From figure \ref{subfig:M0}, we see that $M_0\neq0$ for all values of $\beta_T$ in the region of interest, so division by $M_0$ always makes sense. Furthermore, from figure \ref{subfig:M3overM0}, we see that Re$\big(M_3/M_0\big)$ is always positive. Therefore Re$\big((M_1+M_2)/M_0\big)$ and Re$\big(M_1/M_0\big)$ will determine the sign of the quantities of interest in conditions \eqref{hopf_condition} and \eqref{hopf_cond_xonly}, respectively. 
Interestingly, figure \ref{subfig:M1M2overM0} indicates that Re$\big((M_1+M_2)/M_0\big)$ changes sign for some $\beta_T\in[0.2,0.7]$. Note that since $M_1$ and $M_2$ are calculated only up to $O(D_R)$, the precise location where Re$\big((M_1+M_2)/M_0\big)=0$ cannot be determined since the location may depend on lower order terms in $D_R$. 
 However, for a sufficiently small bifurcation value $\beta_T$ (determined by choosing $D_T$ sufficiently large), we should see a stable limit cycle emerge beyond the bifurcation, while for $\beta_T$ sufficiently large ($D_T$ sufficiently small), the bifurcation should be subcritical. \\

In contrast, Re$\big(M_1/M_0\big)<0$ for all $\beta_T\in[0.2,0.7]$ (see figure \ref{subfig:M1overM0}), indicating that in the Hopf region, for initial perturbations in the $x$-direction only, a stable limit cycle should always arise immediately beyond the bifurcation. \\

In the following sections we explore these predictions numerically. 


\subsection{Subcritical region and bistability}\label{subsec:hopf_sub}
We first consider the subcritical region for initial perturbations in both $x$ and $y$, where we find strong numerical evidence of hysteresis. 
For the following simulations, we fix $D_R=0.001$ and $D_T=0.02$, so the bifurcation occurs at roughly $\beta_T\approx 0.63$. According to figure \ref{subfig:M1M2overM0}, this value of $\beta_T$ lies well within the subcritical region for generic initial perturbations with both $x$ and $y$ components. \\

Similar to the immotile bifurcation study in figure \ref{fig:noswim_supercrit}, we take our initial condition to be a small random perturbation in both $x$ and $y$ to the uniform isotropic state, and begin by running the simulation with $\beta=0.48$ for $500t$, allowing the system to move away from the isotropic state. We then increase $\beta$ by $0.03$ every $100t$ until $\beta=0.93$, well beyond the bifurcation value of $\beta_T\approx 0.63$. We then decrease $\beta$ by $0.05$ every $100t$ until we reach $\beta=0.53$, again passing through the bifurcation point. The bifurcation value $\beta_T\approx0.63$ is reached from below at $t=1000$ and again from above at $t=2600$. \\

\begin{figure}
\centering
  \begin{subfigure}{.32\textwidth}
\includegraphics[scale=0.04]{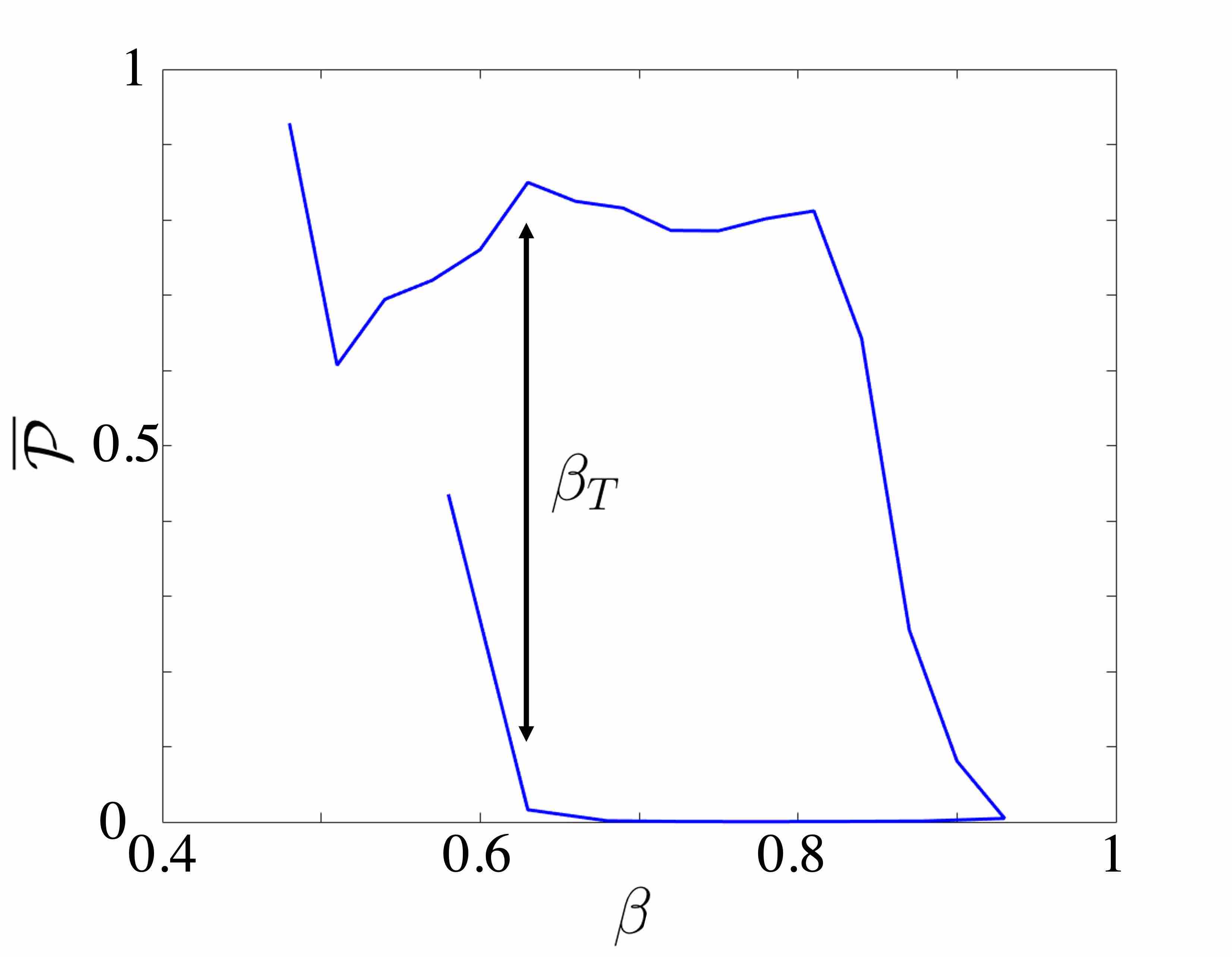} 
    \caption{}
   \label{subfig:hopf_063_hyst1}
   \end{subfigure}
   \begin{subfigure}{0.33\textwidth}
    \includegraphics[scale=0.04]{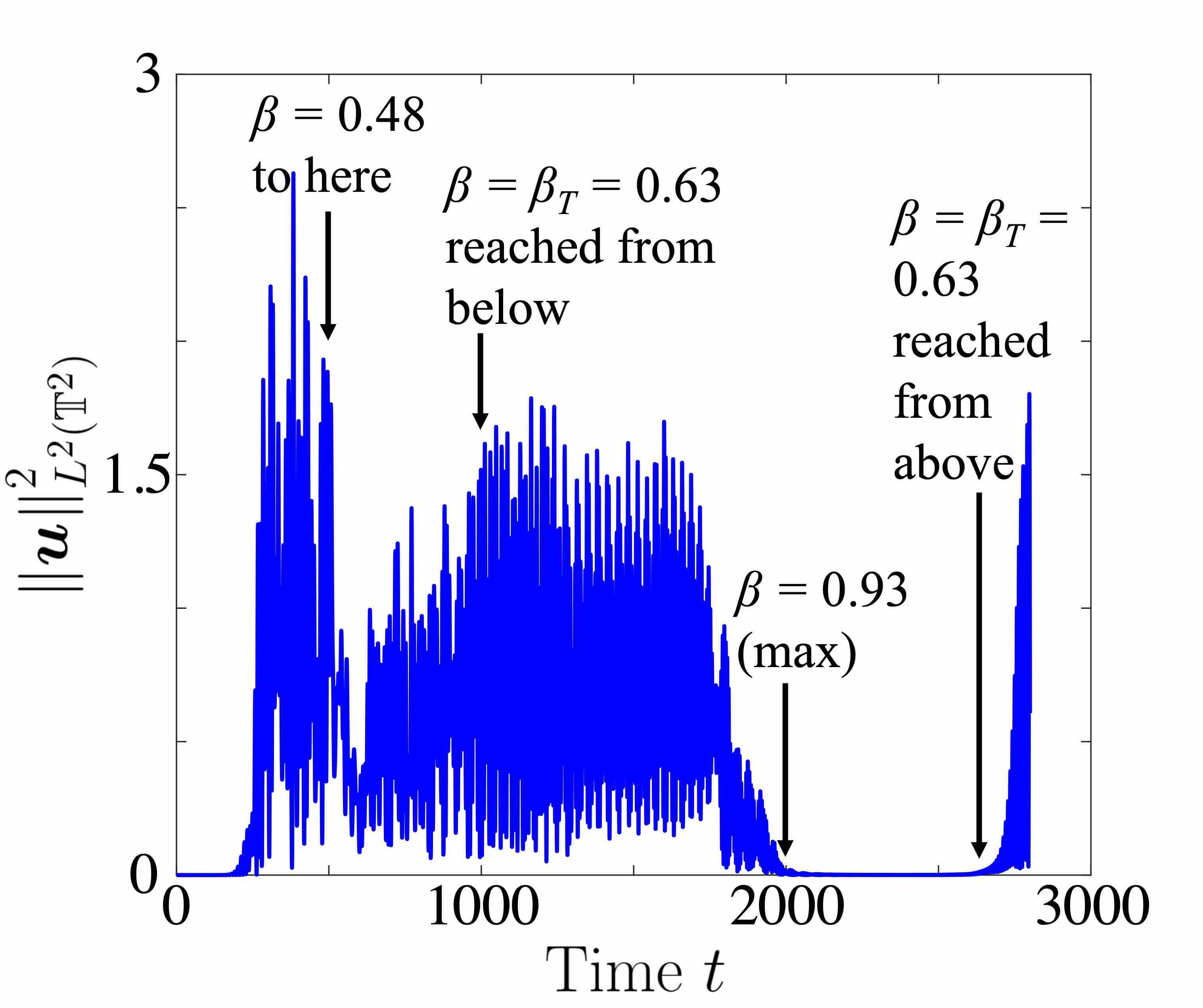}
    \caption{}
   \label{subfig:hopf_063_L2_1}
   \end{subfigure} 
     \begin{subfigure}{0.32\textwidth}
   \includegraphics[scale=0.04]{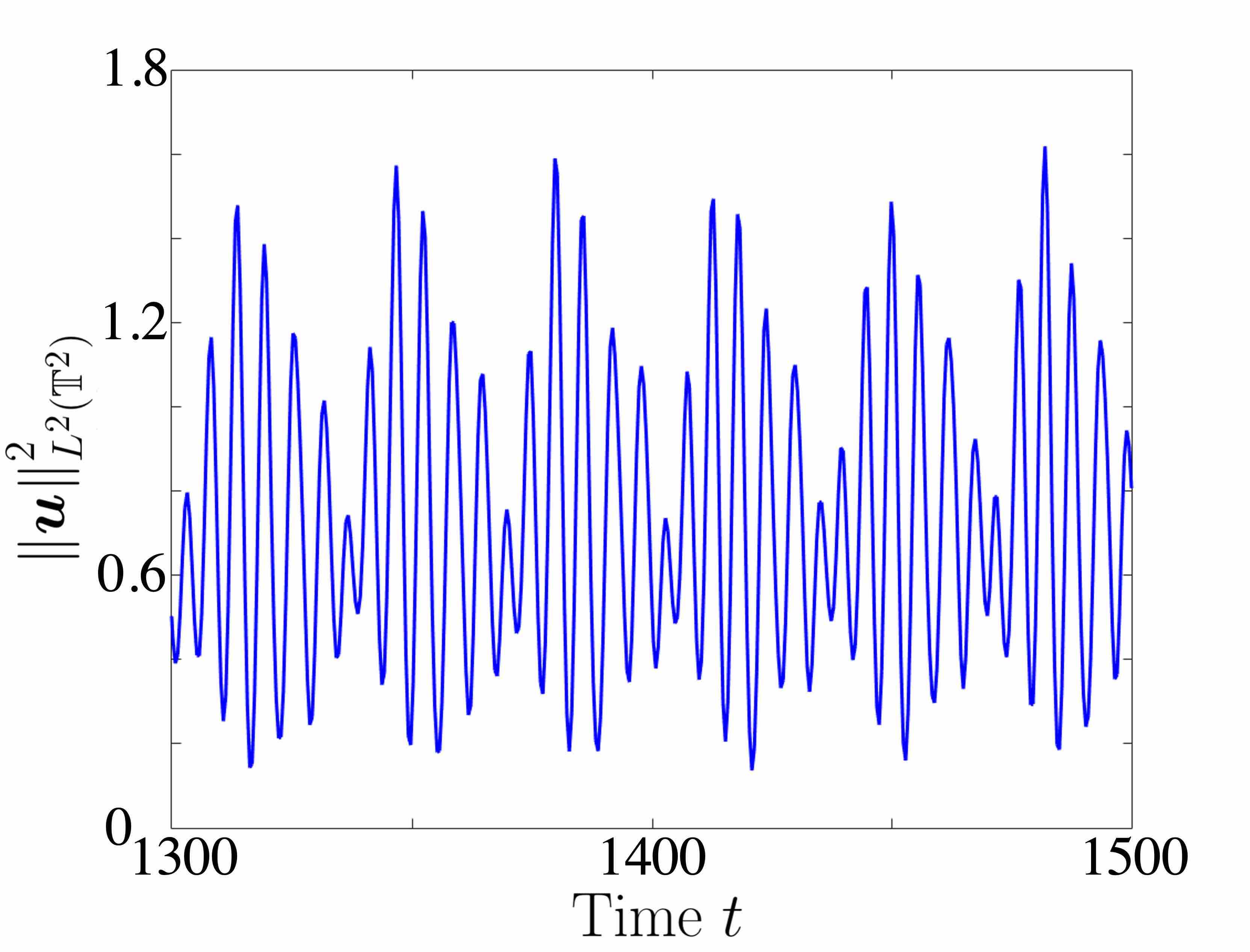} 
    \caption{}
   \label{subfig:hopf_063_L2_3}
   \end{subfigure}
\caption{(a) The plot of the time-averaged viscous dissipation $\overline{\mc{P}}$ versus $\beta$ displays bistability in the system above the subcritical Hopf bifurcation at $\beta_T\approx0.63$. (b) Hysteresis is also evident in the plot of $\norm{\bu}_{L^2(\T^2)}^2$ over the entire simulation described above. The bifurcation value of $\beta_T\approx0.63$ is reached from below at $t=1000$ and again from above at $t=2600$. After the bifurcation is passed from below, the system remains in a nontrivial state well beyond the bifurcation value. (c) An almost periodic structure appears after the bifurcation value is passed, which persists until $\beta=0.84$ at $t=1700$. Here we plot $\norm{\bu}_{L^2(\T^2)}^2$ from $t=1300$ to 1500, where $\beta=0.72$ ($t=1300$ to 1400) and $\beta=0.75$ ($t=1400$ to 1500).
} 
\label{fig:hopf_063_hyst}
\end{figure}

Again we keep track of the time-averaged rate of viscous dissipation $\overline{\mc{P}}$ \eqref{timeavg_VD}, except now the average is taken over each constant value of $\beta$ throughout the simulation. We plot $\overline{\mc{P}}$ versus $\beta$ in figure \ref{subfig:hopf_063_hyst1}. In contrast to the supercritical bifurcation seen in the immotile setting (figure \ref{fig:noswim_supercrit}), here we can see clear hysteresis: as $\beta$ increases past the bifurcation at $\beta_T\approx0.63$, the system remains in a nontrivial state -- away from the uniform, isotropic state -- well beyond the bifurcation point (up to about $\beta=0.81$) before finally dropping down to the uniform, isotropic state ($\overline{\mc{P}}=0$). Then as $\beta$ is decreased, the system remains in the uniform isotropic steady state until the uniform state loses stability at $\beta=0.63$, after which the system transitions to a nontrivial state again. This apparent region of bistability between $0.63<\beta<0.81$ is characteristic of a subcritical bifurcation. \\

We also plot the $L^2$ norm of the velocity field $\norm{\bu}_{L^2(\T^2)}^2$ over the entire course of the simulation in figure \ref{subfig:hopf_063_L2_1}. Note that fluctuations in the velocity field persist well past the bifurcation point, which is reached at $t=1000$ ($\beta=\beta_T=0.63$). The fluctuations remain until $t=1700$ (where $\beta$ is increased to 0.84), when they quickly begin to decay. After $\beta$ reaches a maximum of $\beta=0.93$ from $t=2000$ to 2100, the velocity field stays motionless as $\beta$ is decreased again. After $\beta=0.63$ is reached, $\norm{\bu}_{L^2(\T^2)}^2$ begins to increase again. 
We note that this hysteretic behavior is replicated even if we start the above procedure much closer to the bifurcation point -- e.g. at $\beta=0.6$. This indicates that the bistability we are seeing directly relates to the subcritical nature of the bifurcation. \\

Looking more closely at the region immediately following the bifurcation at $t=1000$ (see figure \ref{subfig:hopf_063_L2_3}), the system develops a very regular, nearly periodic structure, especially from $t=1300$ to 1500 ($\beta=0.72$ and $0.75$). This structure persists until $\beta$ is increased to 0.84 at $t=1700$. Snapshots of the nematic order parameter $\mc{N}(\bx,t)$ and particle concentration field $c(\bx,t)$ along this upper solution branch at $\beta=0.75$ are displayed in figure \ref{fig:hopf_040_sop}. In addition, the supplementary videos \texttt{Movie2} and \texttt{Movie3} show the quasiperiodic nature of the dynamics along this upper branch.  \\

\begin{figure}
\centering
 \begin{subfigure}{0.6\textwidth}
  \includegraphics[scale=0.1]{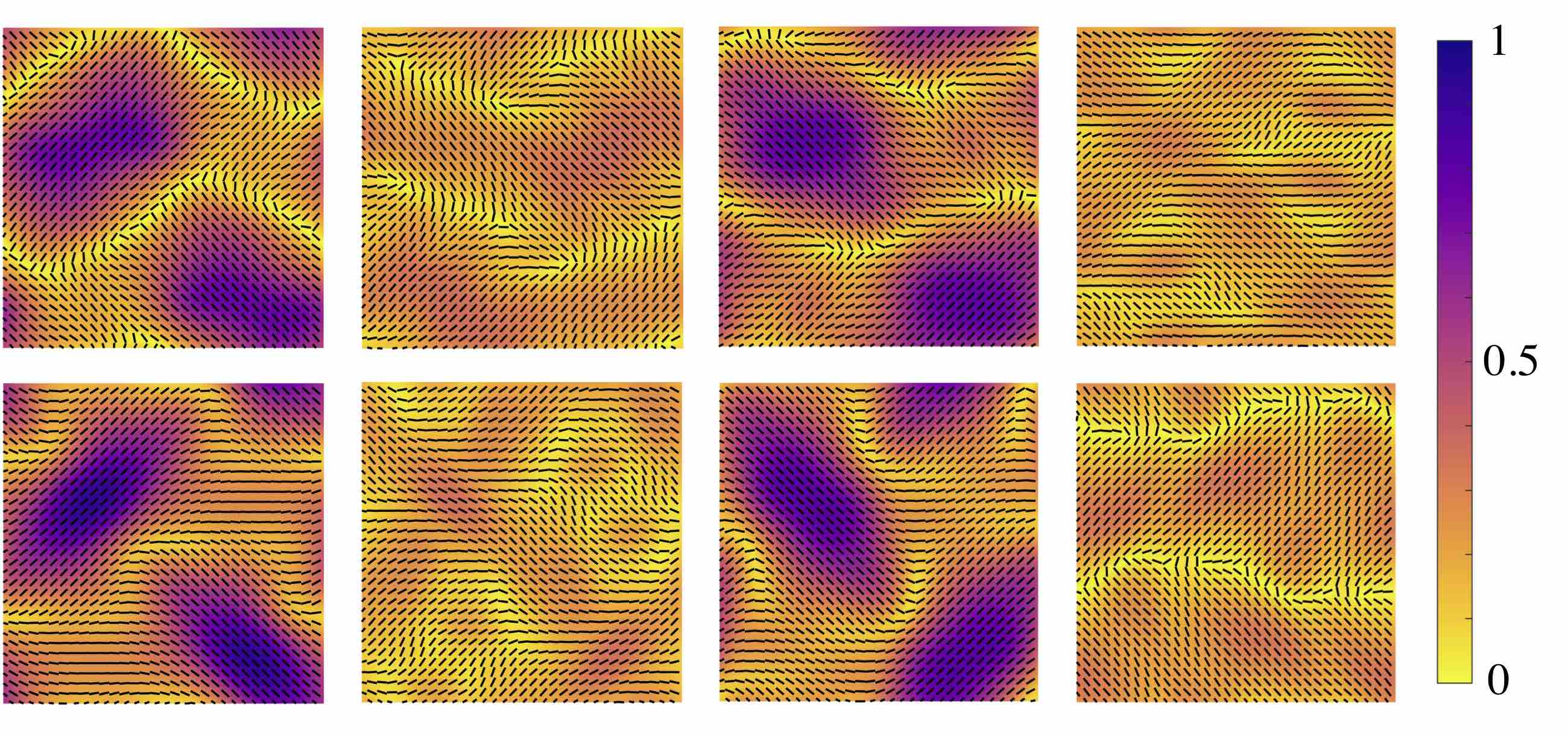} 
        \caption{}
   \label{subfig:hopf_063_sop_a}
      \end{subfigure} 
      \begin{subfigure}{0.6\textwidth}
  \includegraphics[scale=0.1]{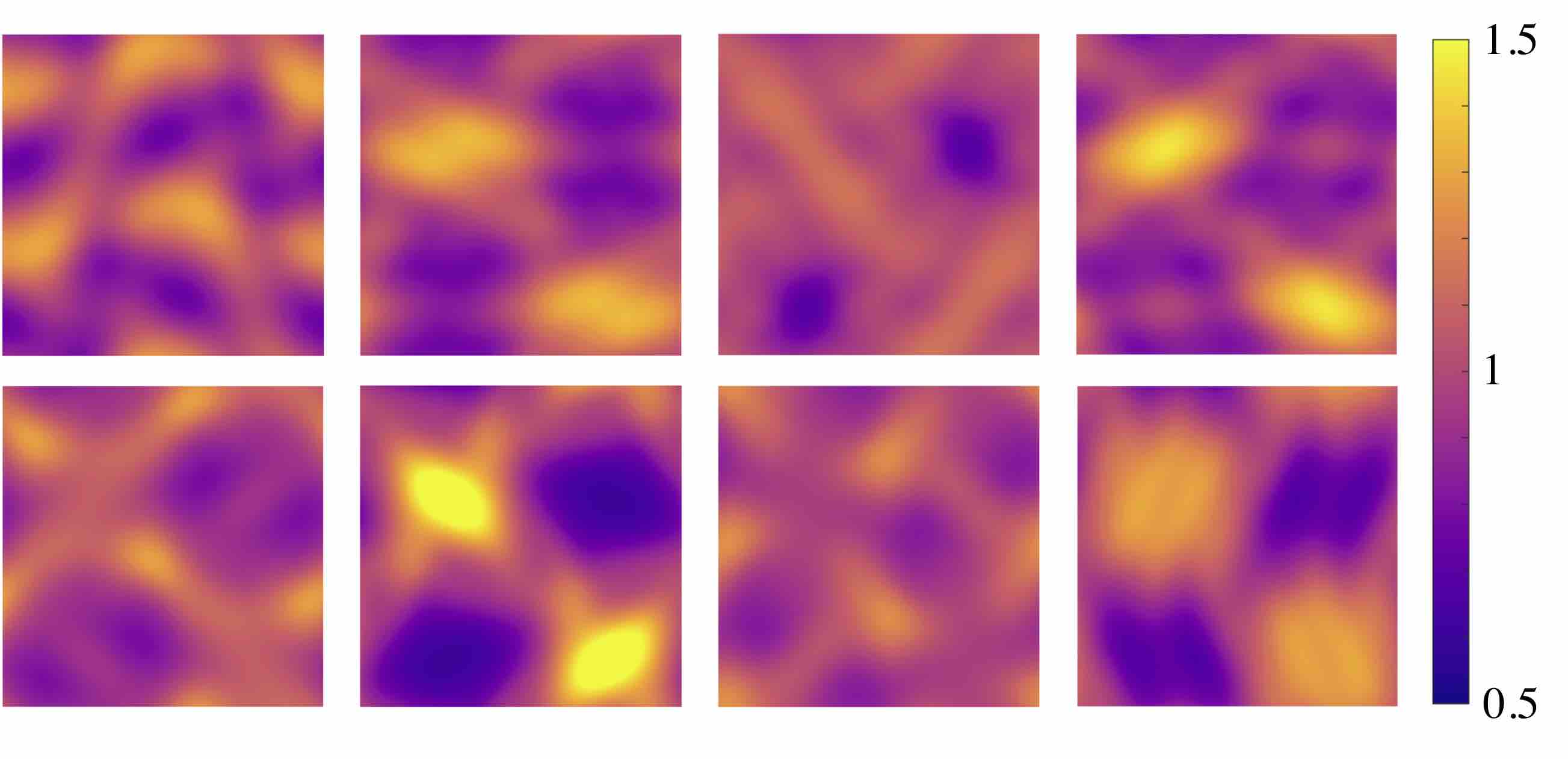}
        \caption{}
   \label{subfig:hopf_063_sop_b}
      \end{subfigure} 
\caption{Snapshots of (a) the nematic order parameter $\mc{N}(\bx,t)$ and the direction of local nematic alignment, and (b) the concentration field $c(\bx,t)$ along the nontrivial upper solution branch which emerges above the subcritical Hopf bifurcation at $\beta_T=0.63$. Here $\beta=0.75$, and figures are taken at successive local peaks and valleys in the velocity $L^2$ norm (every 5 to $5.5t$), starting with a peak.
} 
\label{fig:hopf_063_sop}
\end{figure}

\begin{figure}
\centering
    \begin{subfigure}{.32\textwidth}
   \includegraphics[scale=0.11]{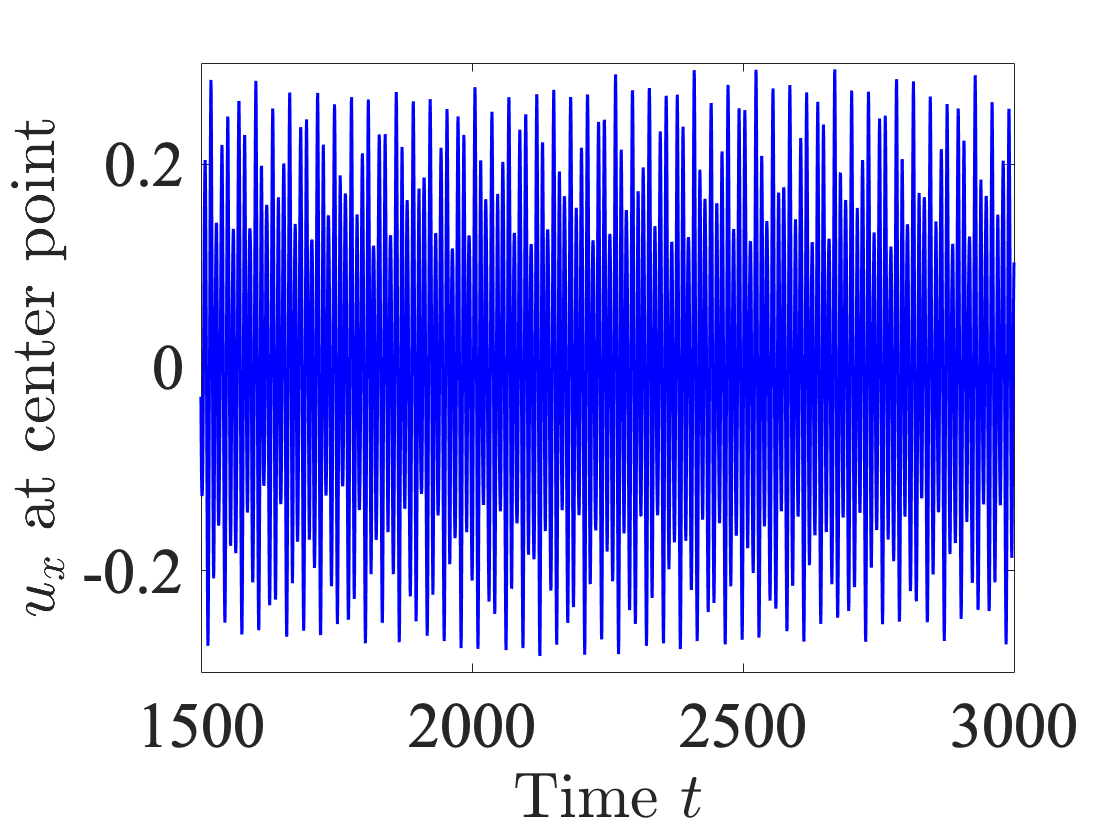}
          \caption{}
   \label{subfig:hopf_063_signal1}
   \end{subfigure}
    \begin{subfigure}{.32\textwidth}
   \includegraphics[scale=0.11]{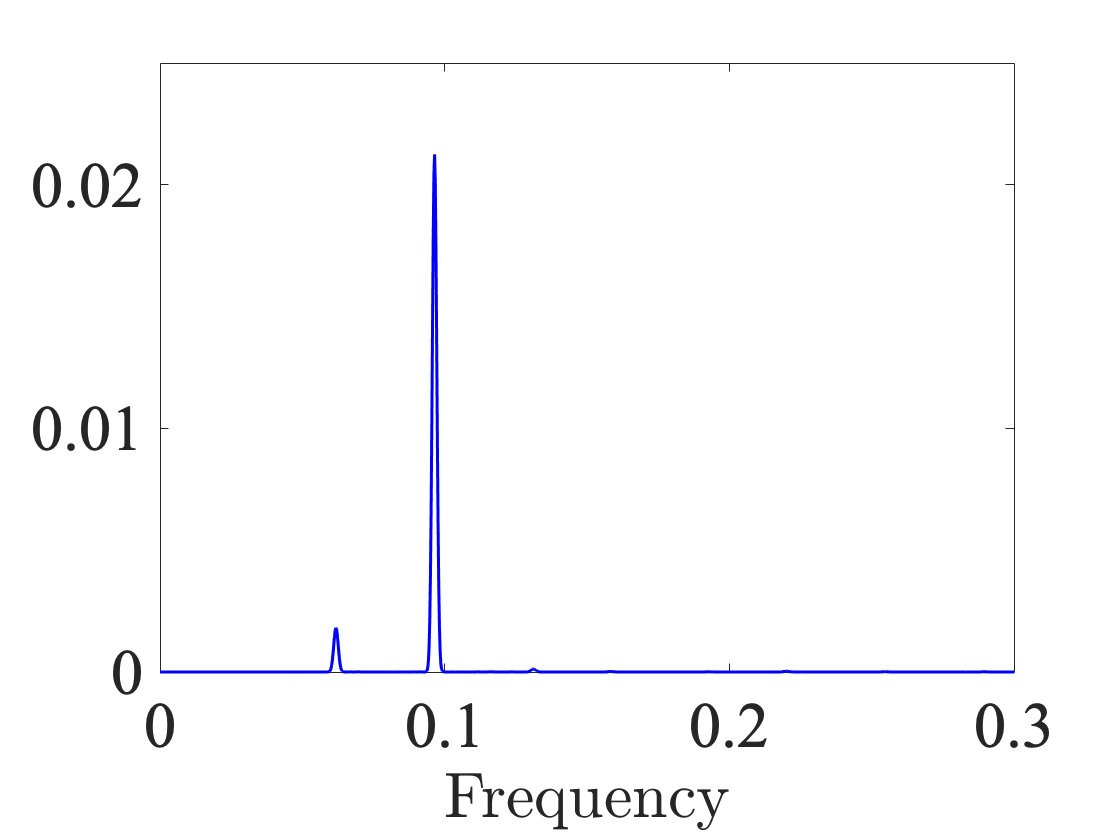}
              \caption{}
   \label{subfig:hopf_063_signal2}
 \end{subfigure}
 \begin{subfigure}{.32\textwidth}
   \includegraphics[scale=0.11]{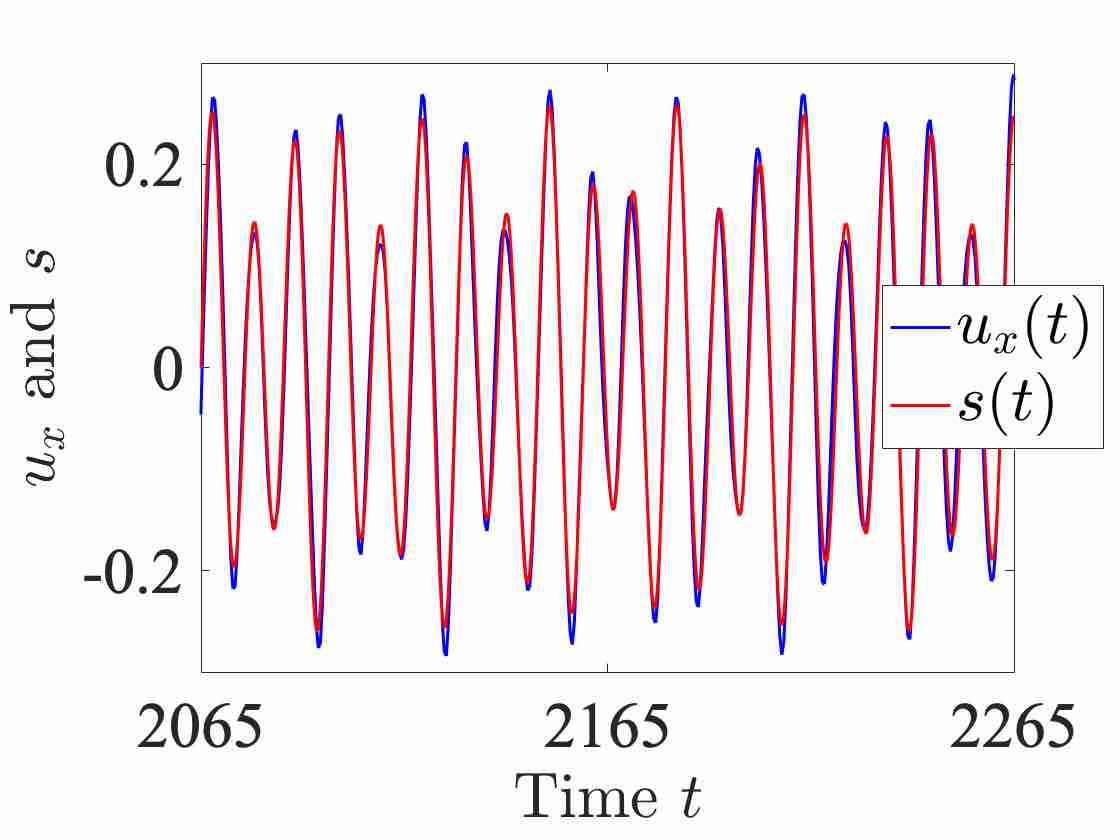} 
              \caption{}
   \label{subfig:hopf_063_signal3}
   \end{subfigure}
\caption{The surprisingly regular temporal dynamics in the nontrivial hysteretic state at $\beta=0.75$. Figure (a) displays the near-periodic dynamics of the $x$-coordinate $u_x$ of the velocity field $\bu(\bx,t)$ evaluated at the center point of the computational domain. Figure (b) is the power spectrum of $u_x$ over the time interval plotted in (a). The signal decomposes into just two temporal modes. In figure (c), we plot $u_x(t)$ along with the simple signal $s(t)$ \eqref{st_eqn} composed of the two modes in (b). The agreement is nearly perfect. 
} 
\label{fig:hopf_power_spec}
\end{figure}

The unexpectedly regular structure of the temporal dynamics in figures \ref{subfig:hopf_063_L2_3} and \ref{fig:hopf_063_sop} prompts a closer look at the nontrivial hysteretic state at $\beta=0.75$. We simulate this state over a long time and, among other values, record the velocity $\bu(\bx,t)$ evaluated at the center point of the computational domain. The value of the $x$-coordinate $u_x$ over 1500 time units is plotted in figure \ref{subfig:hopf_063_signal1}; the $y$-coordinate $u_y$ behaves similarly. The near-perfect periodicity here is striking. We plot the power spectrum of $u_x(t)$ in figure \ref{subfig:hopf_063_signal2} and note that, remarkably, $u_x(t)$ decomposes into essentially just two temporal modes: a large mode at frequency $0.096=1/10.4$ and a small mode at $0.62=1/16.2$. In figure \ref{subfig:hopf_063_signal3} we plot the signal 
\begin{equation}\label{st_eqn}
s(t)=  0.2\bigg(\sin\bigg(\frac{2\upi(t-2065)}{10.4}\bigg)+0.3\sin\bigg(\frac{2\upi(t-2065)}{16.2}\bigg) \bigg)
\end{equation}
on top of $u_x(t)$ for $200t$, where the value $2065$ was chosen to qualitatively match with $u_x$. The overlap is nearly exact. The simplicity of the signal in Figure \ref{fig:hopf_power_spec} is surprising given that these dynamics occur in a region beyond the predictive scope of the preceding weakly nonlinear analysis, where we do not necessarily expect such a regular structure.  \\

\subsection{Supercriticality for 1D initial perturbations }\label{subsec:super1D}
\begin{figure}
\centering
  \begin{subfigure}{.32\textwidth}
  \includegraphics[scale=0.034]{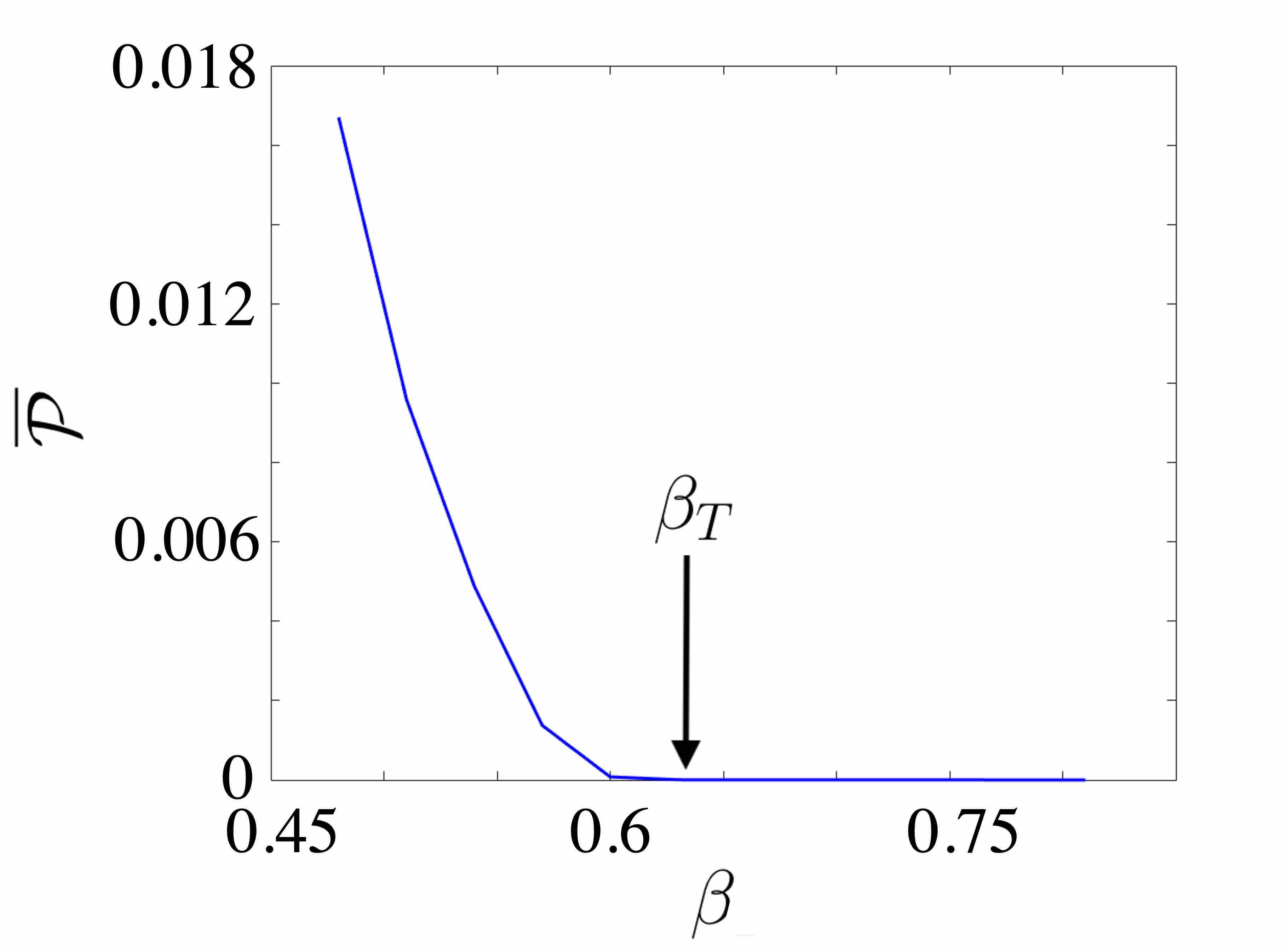} 
   \caption{}
   \label{subfig:hopf_063_xonlyVD}
   \end{subfigure}
\begin{subfigure}{.32\textwidth}
 \includegraphics[scale=0.22]{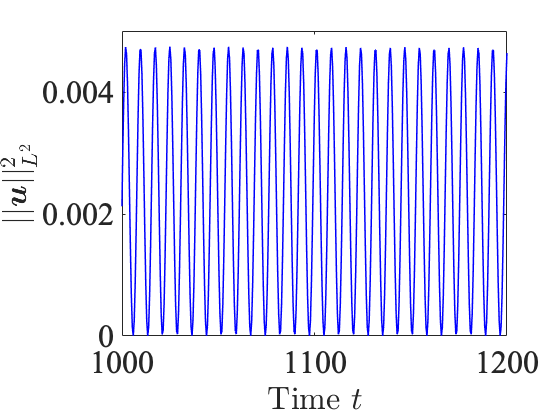}
          \caption{}
   \label{subfig:hopf_063_xonly_limcyc_L2}
   \end{subfigure}
    \begin{subfigure}{.32\textwidth}
 \includegraphics[scale=0.22]{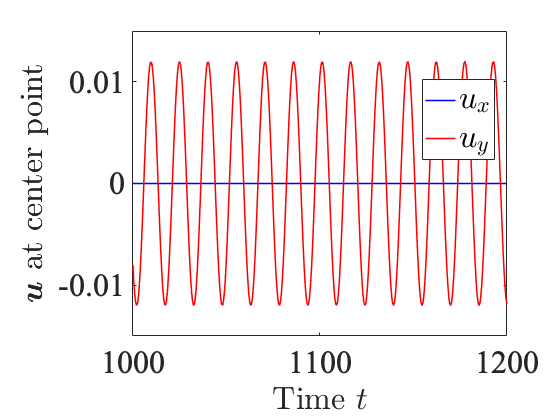}
          \caption{}
   \label{subfig:hopf_063_xonly_limcyc_uxuy}
   \end{subfigure} \\
   \begin{subfigure}{.65\textwidth}
 \includegraphics[scale=0.1]{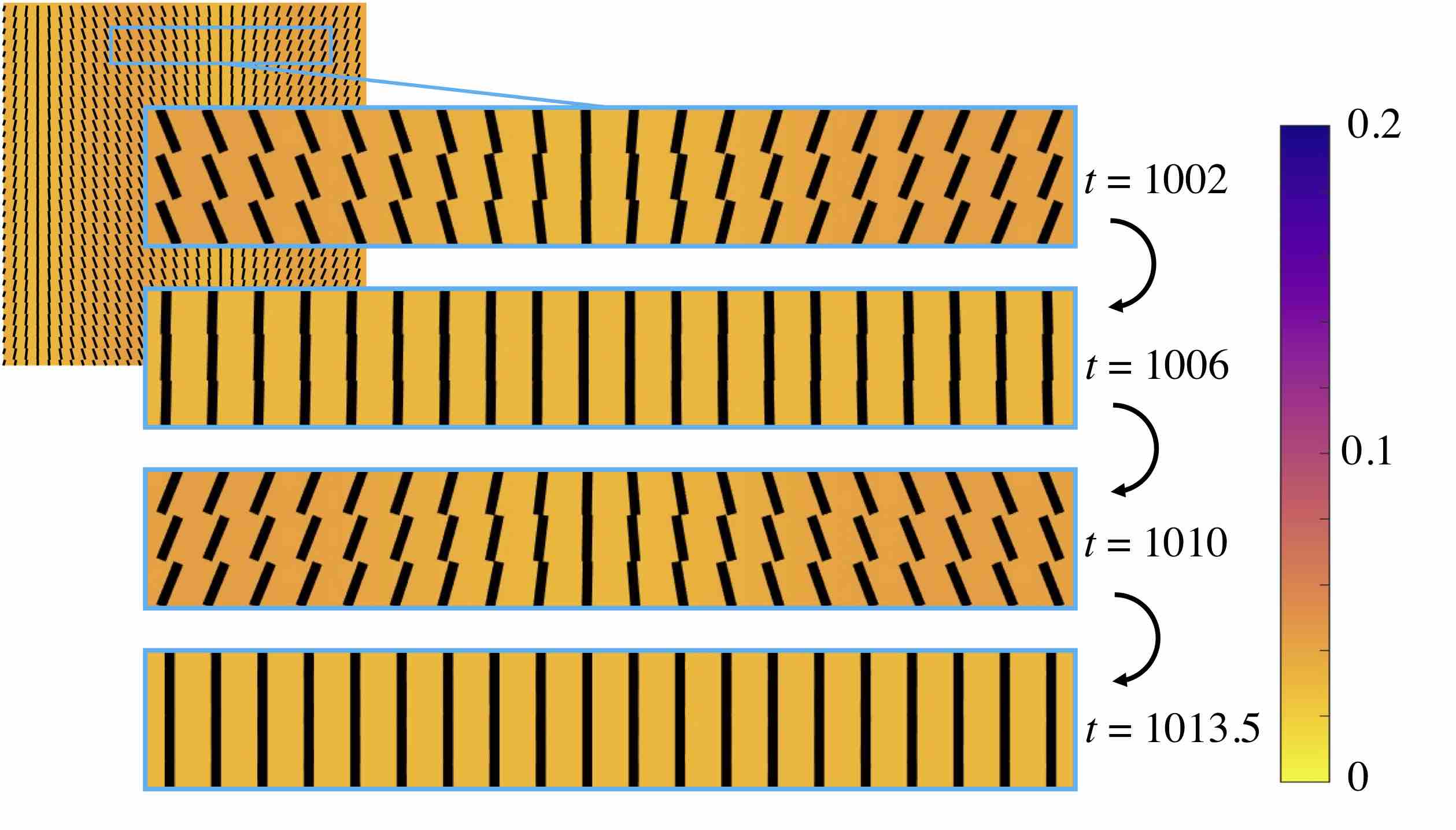}
           \caption{}
   \label{subfig:hopf_063_xonly_SOP1}
   \end{subfigure}
\caption{
 (a) Plot of $\overline{\mc{P}}$ versus $\beta$ when $D_R=0.001$ and $D_T=0.02$ (so $\beta_T\approx0.63$) for initial perturbations in the $x$-direction only. The behavior here can be contrasted with figure \ref{fig:hopf_063_hyst} for 2D ($x$ and $y$) initial perturbations.
 (b) Plot of $\norm{\bu}_{L^2(\T^2)}^2$ over time and (c) plot of the $x$-component (blue) and $y$-component (red) of the velocity field $\bu$ evaluated at the center point of the computational domain. Here $\beta=0.6$ is fixed. Both plots show that a stable limit cycle develops following an $x$-only initial perturbation. 
 (d) Snapshots of $\mc{N}(\bx,t)$ over the period of one limit cycle. The particles are very weakly aligned here, but their preferred direction oscillates. 
} 
\label{fig:hopf_063_xonly_sop}
\end{figure}

While an initial perturbation in both $x$ and $y$ gives rise to a subcritical bifurcation at $\beta_T=0.63$, as predicted by figure \ref{fig:hopf_coeffs}, an initial perturbation in only the $x$-direction should result in a supercritical bifurcation. Indeed, if we perform the same numerical test as in figure \ref{fig:hopf_063_hyst} but with an initial perturbation in only the $x$-direction instead, the resulting relationship between the average viscous dissipation $\overline{\mc{P}}$ and $\beta$ (figure \ref{subfig:hopf_063_xonlyVD}) is characteristic of a supercritical bifurcation. In particular, $\overline{\mc{P}}$ smoothly decreases to zero as the bifurcation is approached from below, similar to the behavior seen in the immotile bifurcation (figure \ref{fig:noswim_supercrit}). Furthermore, we can numerically locate the limit cycle which emerges just below the bifurcation value of $\beta_T=0.63$. Snapshots from a single period of this limit cycle are shown in figure \ref{fig:hopf_063_xonly_sop}, and a few periods of the cycle are documented in the supplementary video \texttt{Movie4}. The alignment among particles is very weak, but they display a clear preferred direction which oscillates over time. Note that the period of the limit cycle corresponds to every other peak of the velocity $L^2$ norm (figure \ref{subfig:hopf_063_xonly_limcyc_L2}); roughly $15$--$15.5t$. This may be compared with the predicted period of $2\upi/b_T=2\upi/0.43\approx 14.6t$.  \\

\subsection{Supercritical region (2D)}\label{subsec:hopf_super}
\begin{figure}
\centering
  \begin{subfigure}{.45\textwidth}
   \includegraphics[scale=0.13]{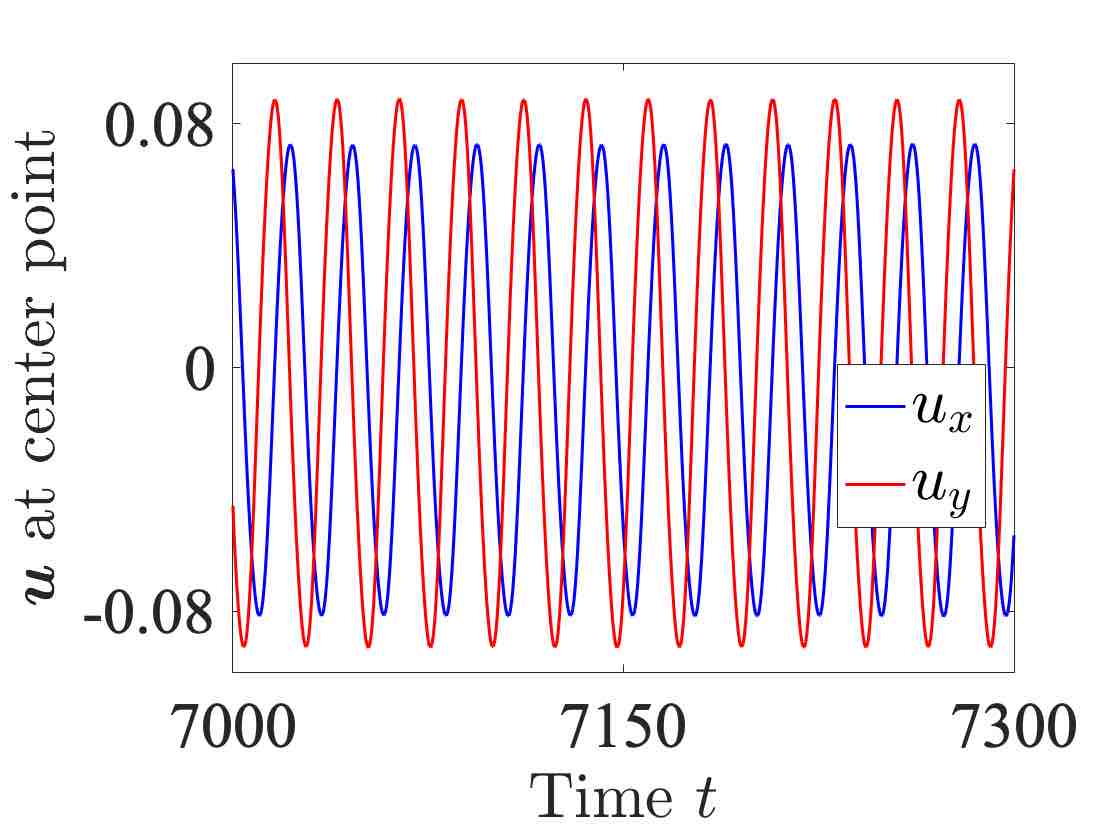}
         \caption{}
   \label{subfig:hopf_040_vels_a}
   \end{subfigure}
    \begin{subfigure}{.45\textwidth}
   \includegraphics[scale=0.13]{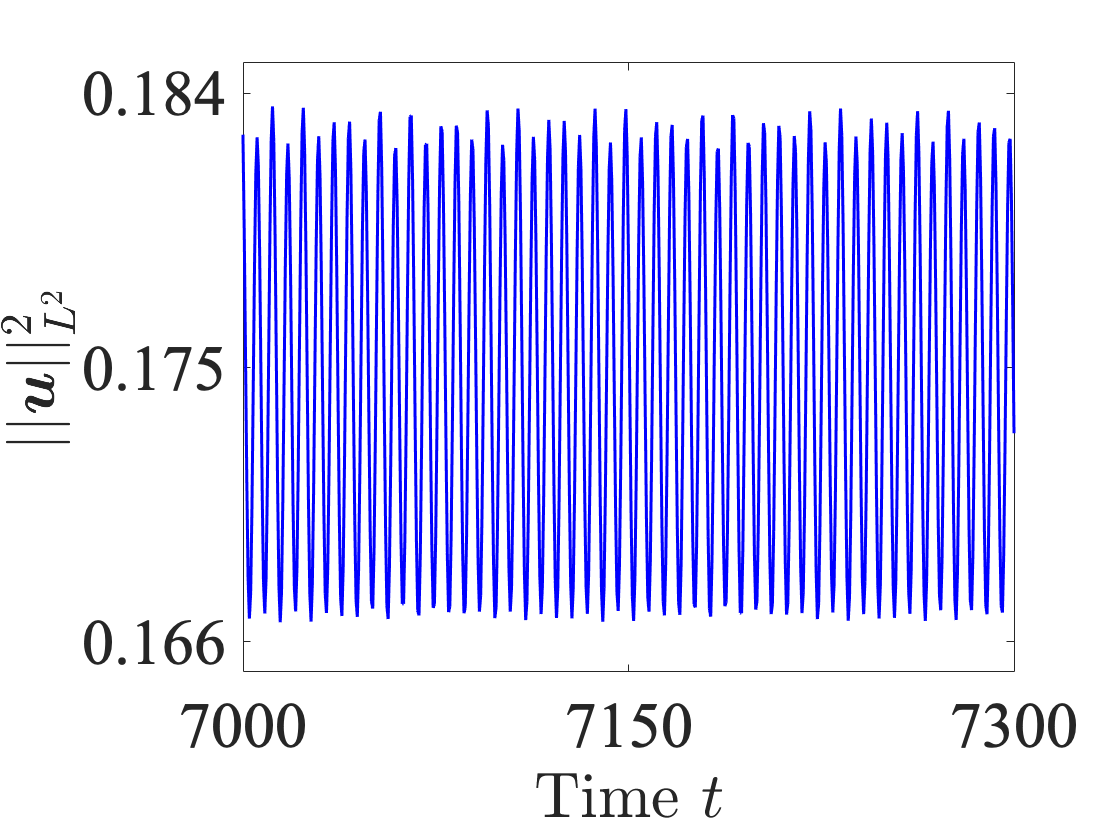}
            \caption{}
   \label{subfig:hopf_040_vels_b}
   \end{subfigure}
\caption{The stable limit cycle which develops just below the Hopf bifurcation at $\beta_T=0.40$. Figure (a) shows the $x$-component (blue) and $y$-component (red) of the velocity $\bu$ evaluated at the center point of the computational domain. Figure (b) shows the $L^2$ norm of the velocity field $\norm{\bu}_{L^2(\T^2)}^2$ over time.  
} 
\label{fig:hopf_040_vels}
\end{figure}

While initial perturbations in only the $x$-direction give rise to a supercritical bifurcation for all $\beta_T\in [0.2,0.7]$, generic 2D perturbations in both $x$ and $y$ should transition from subcritical to supercritical near $\beta_T=0.5$. We fix $D_R=0.001$ and $D_T=0.075$ so that $\beta_T\approx0.40$, which according to figure \ref{subfig:M1M2overM0} lies within the supercritical region for 2D perturbations. Setting $\beta=0.38$, we simulate the longtime system dynamics starting with a 2D initial perturbation. After an initial period of slow growth, a stable limit cycle develops, as shown in figure \ref{fig:hopf_040_vels}). \\

\begin{figure}
\centering
  \begin{subfigure}{.6\textwidth}
   \includegraphics[scale=0.1]{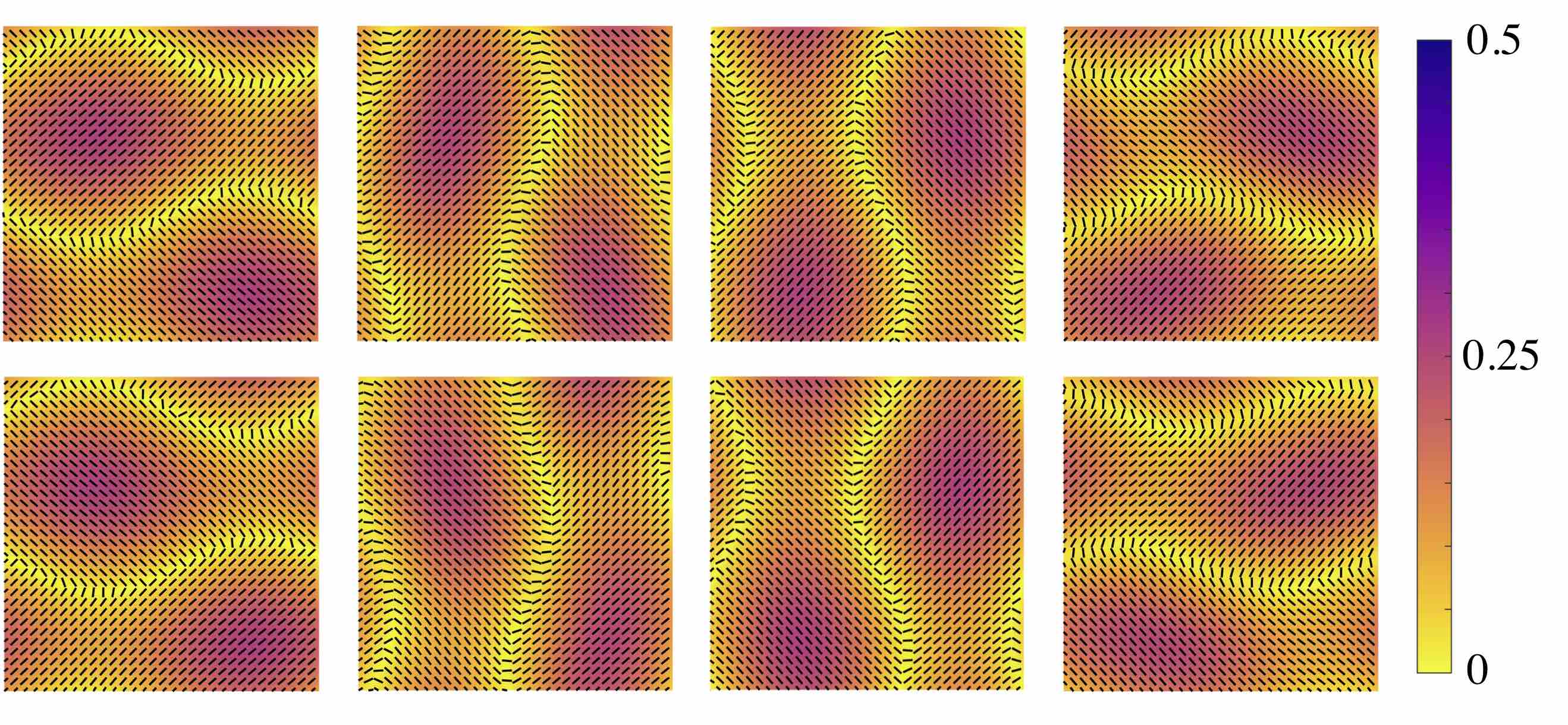}
            \caption{}
   \label{subfig:hopf_040_sop_a}
   \end{subfigure}
  \begin{subfigure}{.6\textwidth}
   \includegraphics[scale=0.1]{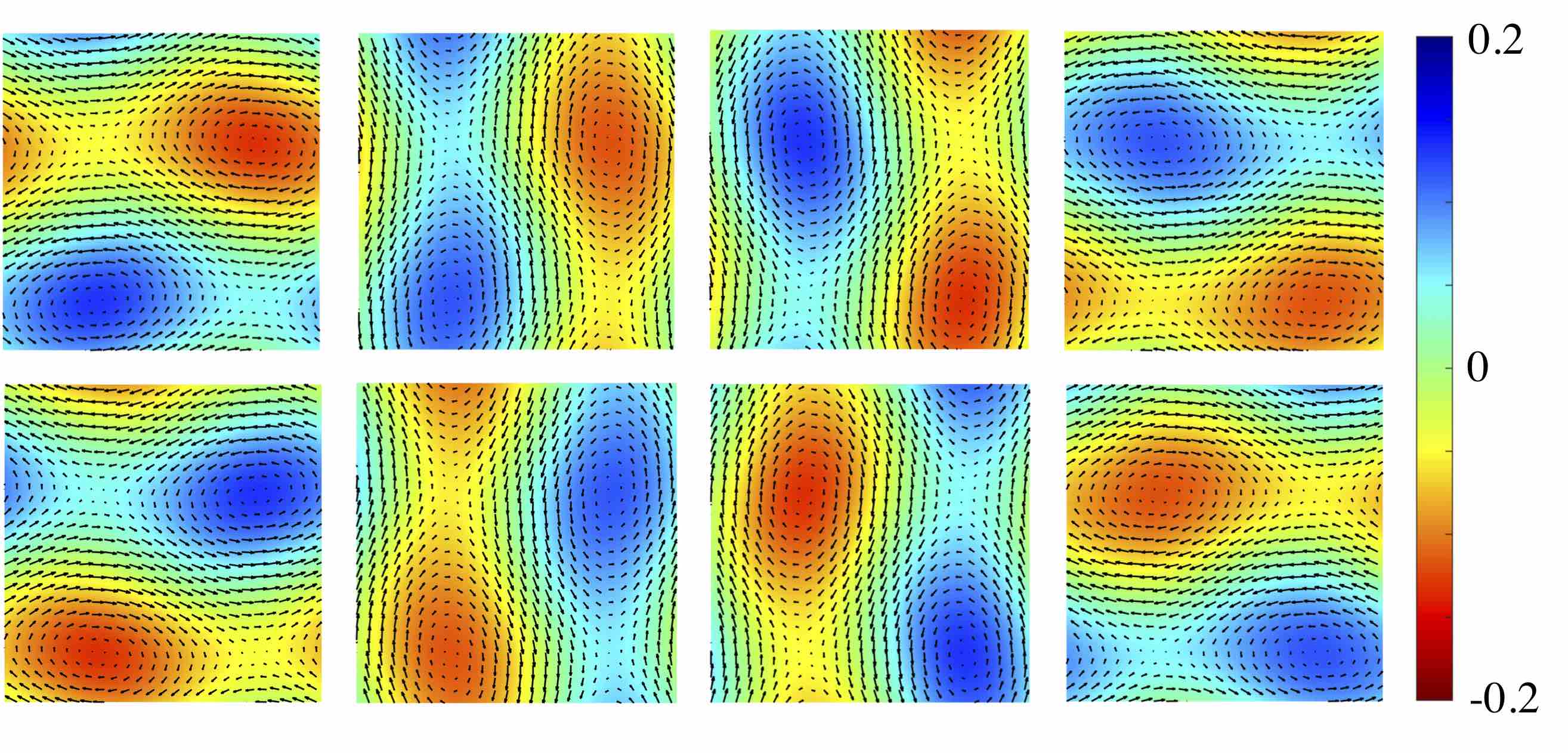}
            \caption{}
   \label{subfig:hopf_040_sop_b}
   \end{subfigure}
\caption{Snapshots of (a) the nematic order parameter $\mc{N}(\bx,t)$ and the direction of local nematic alignment and (b) the fluid vorticity and velocity fields over one period of the limit cycle. The snapshots alternate between peaks and valleys in $\norm{\bu}_{L^2(\T^2)}^2$ in figure \ref{fig:hopf_040_vels}, starting with a peak. As in figure \ref{fig:xy_pert_differentDR}, the extensile flow produced by the aligned dipoles is clear. 
} 
\label{fig:hopf_040_sop}
\end{figure}

Although the peaks in $\norm{\bu}_{L^2(\T^2)}^2$ over time are not perfectly equal in height, they occur at regular intervals ($\sim 6t$), and give rise to a regular, repeated pattern in the particle alignment and generated velocity field, which can be viewed in the supplementary videos \texttt{Movie5} (nematic order parameter) and \texttt{Movie6} (vorticity field and velocity direction).  Snapshots of a single period of the cycle are plotted in figure \ref{fig:hopf_040_sop}.
\\

From figure \ref{fig:hopf_040_sop} we can see that the period of the limit cycle corresponds to every 4 peaks in $\norm{\bu}_{L^2(\T^2)}^2$, so every $24t$. This can be compared with the predicted imaginary part of the growth rate $b_T\approx0.24$, which yields a period of $2\upi/b_T\approx 26t$. \\

The periodic behavior displayed in figure \ref{fig:hopf_040_sop} may be compared to the noisy quasiperiodic behavior along the upper solution branch in the bistable region for the subcritical Hopf bifurcation in figure \ref{fig:hopf_063_sop}. The dynamics in the supercritical case are much more regular over time. \\

\section{Motile particles: pitchfork bifurcation}\label{sec:real_eval}
Keeping $D_R$ fixed and small, we now fix translational diffusion within the range $1/9<D_T<1/4$ such that one of the (now purely real) eigenvalues corresponding to the $\abs{k}=1$ modes crosses zero for some $\beta_T\in(0,\sqrt{3}/9)$. Again, the subscript $T$ is used to denote that the bifurcation value of $\beta$ depends on the choice of translational diffusion $D_T$. \\

In this parameter regime, the same weakly nonlinear calculations as in the Hopf setting may be performed (see Section \ref{subsec:hopf_wnl}), except now $b_T=0$. We thus arrive at the same form of amplitude equations as \eqref{hopf_amplitudes}:
\begin{equation}\label{realeig_amplitude}
\begin{aligned}
c_x\big( \quad M_0(\p_\tau A_x) &= M_3A_x + \frac{1}{D_R}(c_x^2M_1\abs{A_x}^2 +c_y^2M_2\abs{A_y}^2)A_x \quad \big) \\
c_y\big( \quad M_0(\p_\tau A_y) &= M_3A_y + \frac{1}{D_R}(c_y^2M_1\abs{A_y}^2 + c_x^2M_2\abs{A_x}^2)A_y \quad),
\end{aligned}
\end{equation}
but now $M_0$, $M_1$, $M_2$, and $M_3$ given in Appendix \ref{app:hopf}, equation \eqref{hopf_Mjs_expr} are all real-valued for each $(D_T,\beta_T)$ in the region of interest. Since the coefficients $M_j$ are real-valued, similar to the immotile setting, we may look for conditions under which \eqref{realeig_amplitude} admits a nontrivial steady state solution $\p_\tau A_x=\p_\tau A_y=0$. If the coefficients $M_j$ satisfy $M_0\neq0$, $M_1+M_2\neq0$, and
\begin{equation}\label{realeig_cond_xy}
\frac{M_3}{M_1+M_2} < 0
\end{equation}
for initial perturbations in both $x$ and $y$, or $M_0\neq0$, $M_1\neq0$, and
\begin{equation}\label{realeig_cond_xonly}
\frac{M_3}{M_1} < 0
\end{equation}
for initial perturbations in $x$ only, then \eqref{realeig_amplitude} admits nontrivial steady states of the form
\begin{equation}\label{realeig_steady_xy}
A_x = \pm\frac{\sqrt{D_R}}{c_x} \sqrt{-\frac{M_3}{M_1+M_2}},\quad 
A_y = \pm\frac{\sqrt{D_R}}{c_y} \sqrt{-\frac{M_3}{M_1+M_2}}
\end{equation}
for 2D ($x$ and $y$) initial perturbations or
\begin{equation}\label{realeig_steady_xonly}
A_x = \pm\sqrt{D_R}\sqrt{-\frac{M_3}{M_1}}
\end{equation}
for $x$-only perturbations. Then, to leading order in $\epsilon=\sqrt{\beta_T-\beta}$, a stable steady state emerges after the real eigenvalue crossing of the form
\begin{equation}\label{realeig_steady_psi}
\Psi = \frac{1}{2\upi}\bigg(1\pm \epsilon\sqrt{D_R}\sqrt{-\frac{M_3}{M_1+M_2}}\big( \psi_{x,1}(\theta){\rm e}^{{\rm i}x}\pm \psi_{y,1}(\theta){\rm e}^{{\rm i}y}\big) \bigg) + {\rm c.c.}
\end{equation}
for initial perturbations in both $x$ and $y$ and 
\begin{equation}\label{realeig_steady_psi_xonly}
\Psi = \frac{1}{2\upi}\bigg(1\pm \epsilon\sqrt{D_R}\sqrt{-\frac{M_3}{M_1}}\psi_{x,1}(\theta){\rm e}^{{\rm i}x}\bigg) + {\rm c.c.}
\end{equation}
for initial perturbations in the $x$-direction only. If the conditions \eqref{realeig_cond_xy} and \eqref{realeig_cond_xonly} do not hold for 2D and 1D perturbations, respectively, then the bifurcation is subcritical and the system behavior beyond the bifurcation value is less predictable. \\

\begin{figure}
\centering
    \begin{subfigure}{.32\textwidth}
   \includegraphics[scale=0.118]{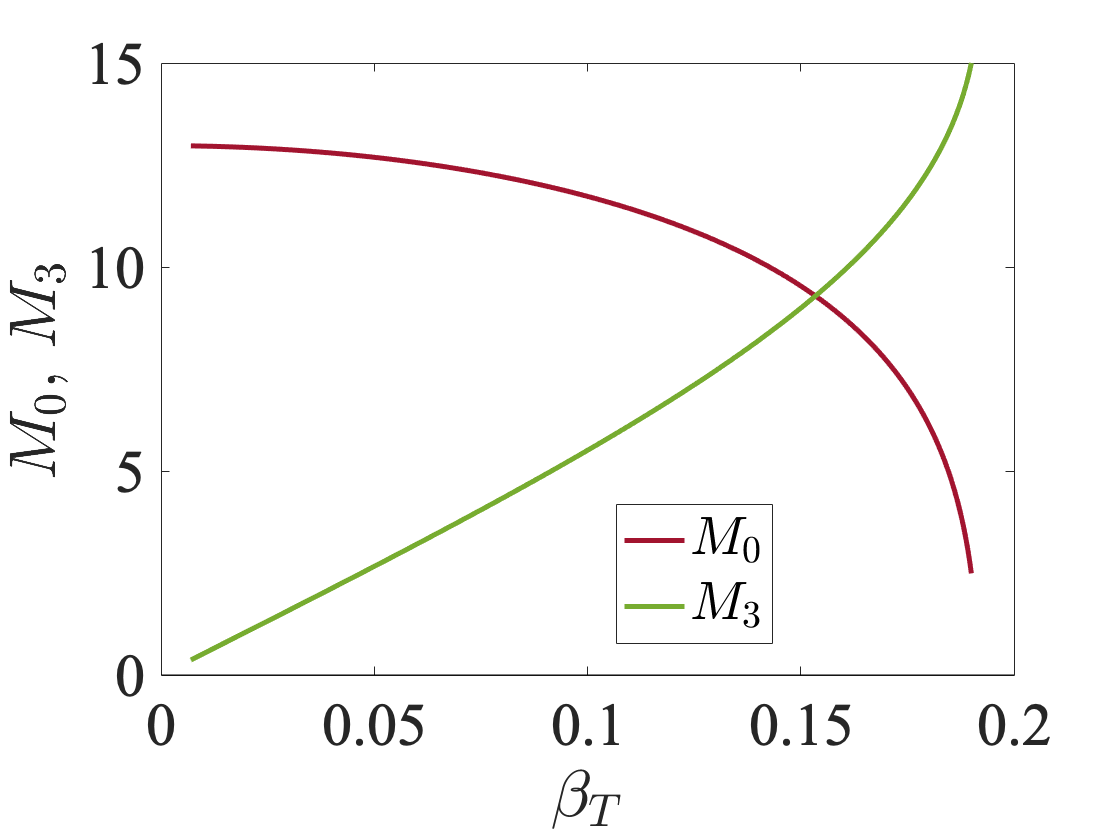}
      \caption{}
   \label{subfig:M0_realeig}
   \end{subfigure}
     \begin{subfigure}{.32\textwidth}
   \includegraphics[scale=0.04]{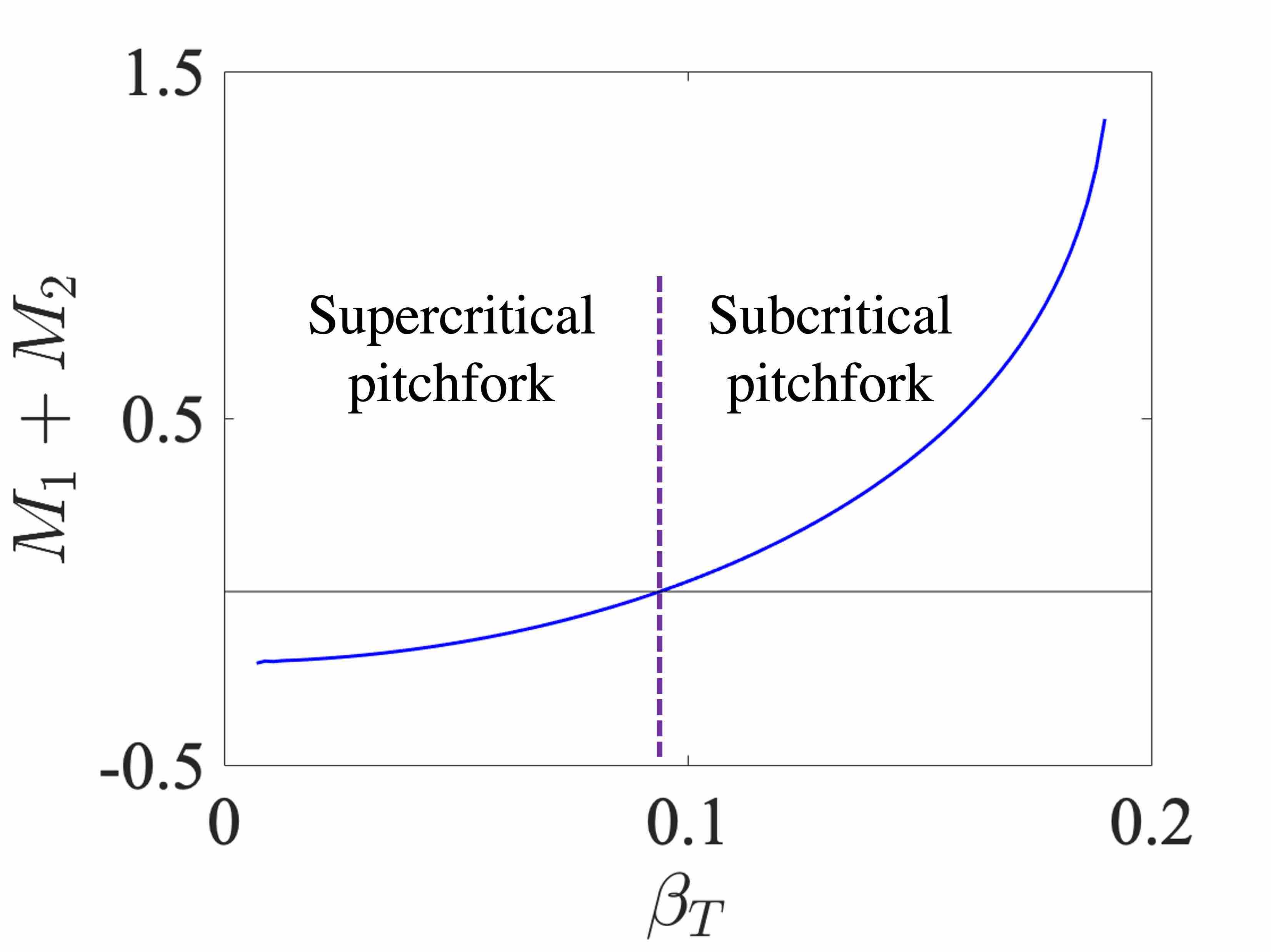}
   \caption{}
   \label{subfig:M1M2_realeig}
   \end{subfigure}
     \begin{subfigure}{.32\textwidth}
   \includegraphics[scale=0.04]{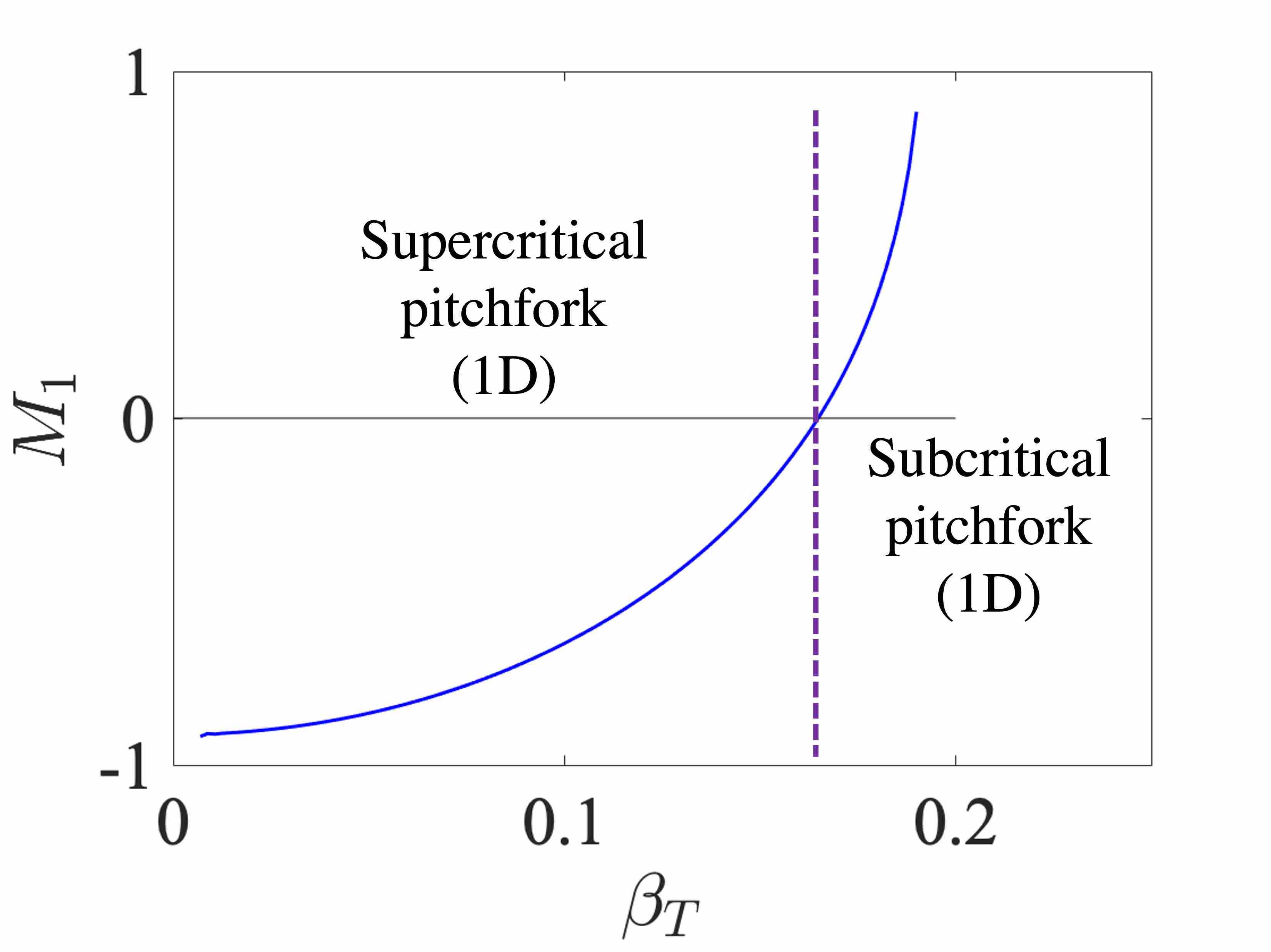}
  \caption{}
   \label{subfig:M1_realeig}
   \end{subfigure}
\caption{Plots of the coefficients (a) $M_0$ and $M_3$, (b) $M_1+M_2$, and (c) $M_1$ given by the expressions \eqref{hopf_Mjs} in the region $0<\beta_T<\sqrt{3}/9$ (or $1/9<D_T<1/4$), where each $M_j$ is real-valued. The vertical dotted lines in (b) and (c) indicate the value of $\beta_T$ where the pitchfork bifurcation transitions from supercritical to subcritical for 2D and 1D initial perturbations, respectively.
} 
\label{fig:realeig_coeffs}
\end{figure}

As in the Hopf setting, using the expressions from Appendix \ref{app:hopf}, equation \eqref{hopf_Mjs_expr}, we plot $M_0$, $M_3$, $M_1+M_2$, and $M_1$ over the desired range of $(D_T,\beta_T)$ in figure \ref{fig:realeig_coeffs}. Here we again use the perturbed dispersion relation \eqref{DR_expansion} of Section \ref{subsec:DR} (see figure \ref{fig:disp_pert}) with $D_R=0.001$. We find that both $M_0$ and $M_3$ are always positive within our region of interest, although $M_0$ appears to approach 0 as $\beta_T\to \sqrt{3}/9\approx0.192$ while $M_3\to 0$ as $\beta_T\to 0$. Thus, as in the Hopf case, the existence of a new stable steady state following the real eigenvalue crossing is determined by the sign of $M_1+M_2$ (for 2D initial perturbations in $x$ and $y$) or the sign of $M_1$ (for $x$-only perturbations). From figures \ref{subfig:M1M2_realeig} and \ref{subfig:M1_realeig}, we see that both $M_1+M_2$ and $M_1$ are negative for small $\beta_T$, indicating the emergence of a nontrivial stable steady state after the bifurcation, but both $M_1+M_2$ and $M_1$ are positive for larger $\beta_T$, indicating that the bifurcation type switches to subcritical somewhere in the interval $0<\beta_T<\sqrt{3}/9$. \\

We again explore the supercritical and subcritical regions numerically in the following sections. 

\subsection{Supercritical region}\label{subsec:realeig_super}
When $\beta_T$ is small, we expect dynamics near the bifurcation to look very similar to the immotile case, where the isotropic steady state loses stability to a supercritical pitchfork bifurcation. We fix $D_R=0.001$ and $D_T=0.23$, so the bifurcation occurs at roughly $\beta_T\approx0.09$. According to figure \ref{fig:realeig_coeffs}, this value of $\beta_T$ lies within the supercritical region for both 2D ($x$ and $y$) and 1D ($x$-only) initial perturbations.  \\

\begin{figure}
\centering
\begin{subfigure}{.32\textwidth}
   \includegraphics[scale=0.035]{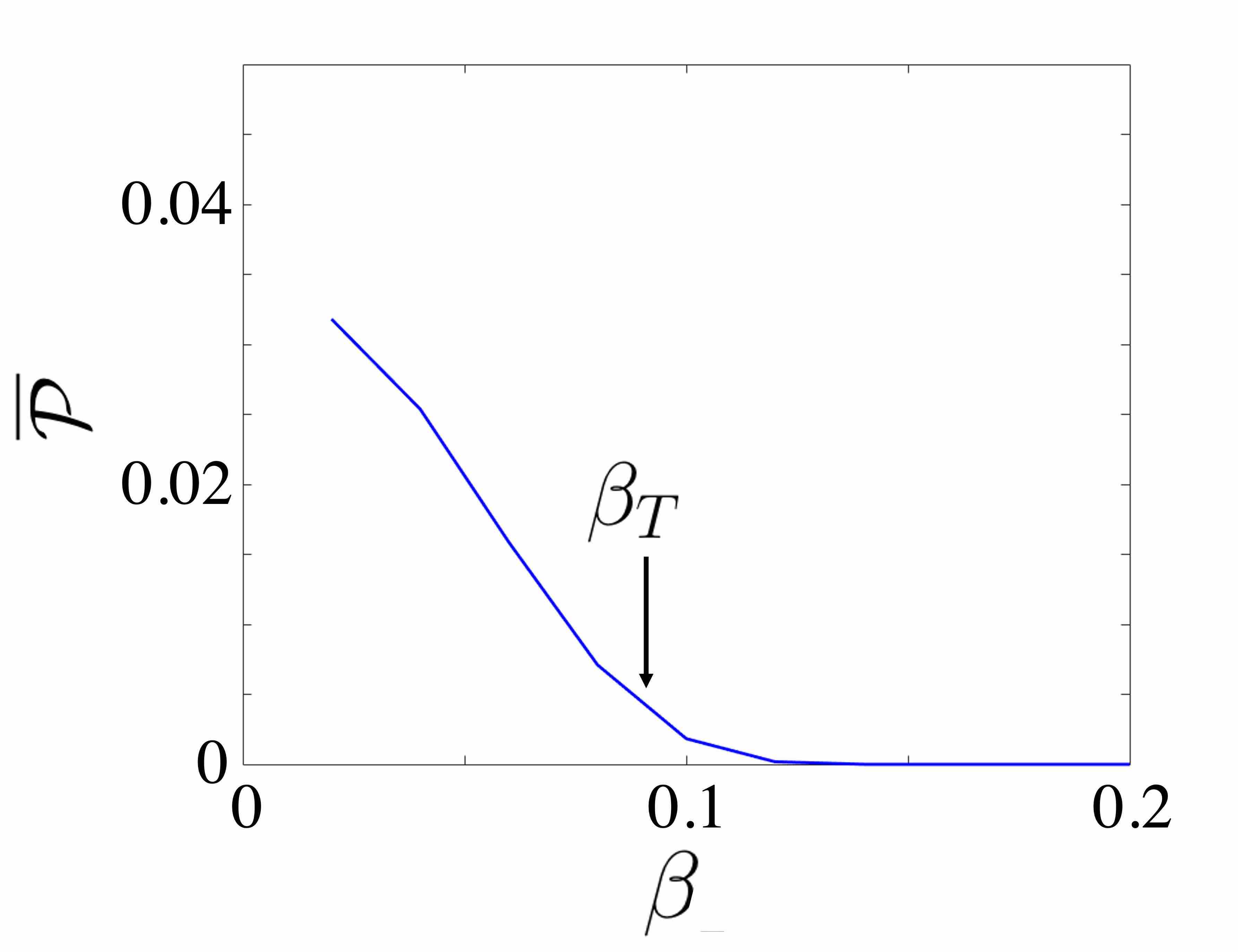}
      \caption{}
   \label{subfig:realeig_superxy_VD}
   \end{subfigure}
   \begin{subfigure}{.32\textwidth}
   \includegraphics[scale=0.275]{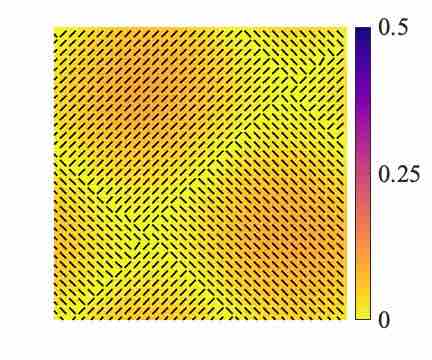}
     \caption{}
   \label{subfig:realeig_superxy_ss}
      \end{subfigure} 
   \begin{subfigure}{.32\textwidth}
   \includegraphics[scale=0.135]{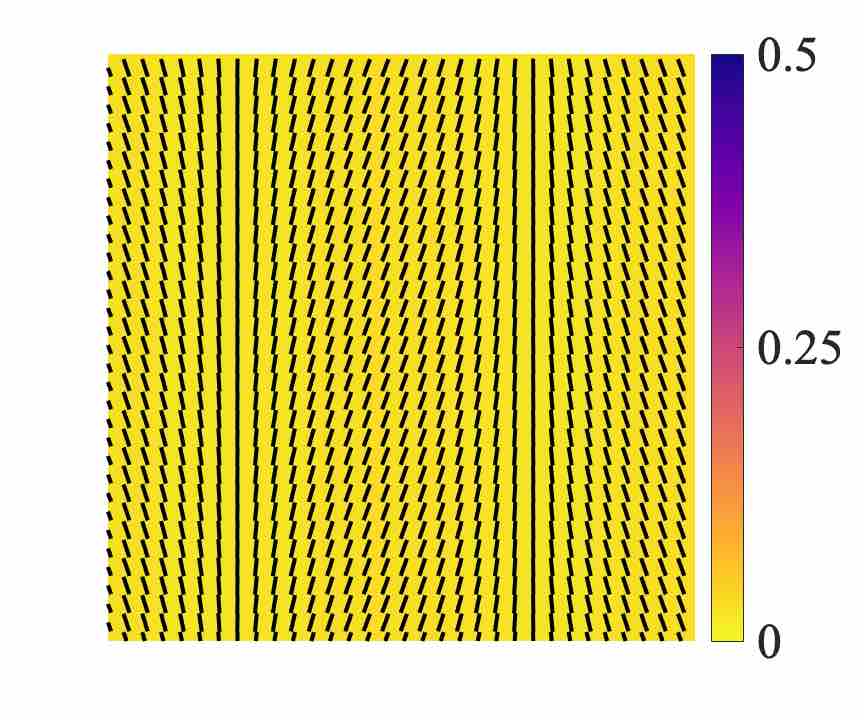}
      \caption{}
   \label{subfig:realeig_superxonly_ss}
   \end{subfigure}
\caption{(a) Plot of $\overline{\mc{P}}$ versus $\beta$ using $D_R=0.001$, $D_T=0.23$ (so $\beta_T\approx0.09$), and a 2D ($x$ and $y$) initial perturbation. The bifurcation is supercritical, and we plot the nematic order parameter $\mc{N}(\bx,t)$ and preferred local alignment direction for the stable state which emerges just beyond the bifurcation in the case of (b) a 2D initial perturbation and (c) an $x$-only initial perturbation. 
} 
\label{fig:realeig_supercrit_xy}
\end{figure}

We begin with a small random perturbation to the uniform, isotropic state in both $x$ and $y$. We initialize the simulation with $\beta=0$ until $t=1500$ and then increase $\beta$ by 0.02 every $100t$ until $\beta=0.2$. We plot the average viscous dissipation $\overline{\mc{P}}$ \eqref{timeavg_VD} versus each different value of $\beta$ in figure \ref{subfig:realeig_superxy_VD}. Again, the relationship between $\overline{\mc{P}}$ and $\beta$ supports the expectation of supercriticality. We note that the viscous dissipation $\overline{\mc{P}}$ behaves qualitatively the same as figure \ref{subfig:realeig_superxy_VD} using an $x$-only initial perturbation. \\

 The emergent stable steady state for $\beta<\beta_T$ is plotted in figure \ref{subfig:realeig_superxy_ss} for a 2D initial perturbation and in figure \ref{subfig:realeig_superxonly_ss} for an initial $x$-only perturbation. Not surprisingly, in both cases the steady state essentially looks like the stable states which arise following the immotile bifurcation (figure \ref{fig:xy_pert_differentDR}). The calculation in the immotile case helps to explain the very weak alignment seen in the emerging steady state, since the rotational diffusion $D_R$ is very small in this setting. \\

\subsection{Subcritical region}\label{subsec:realeig_sub}
We next fix $D_R=0.001$ and $D_T=0.13$, so $\beta_T\approx0.19$ is the bifurcation value. According to figure \ref{fig:realeig_coeffs}, both 2D ($x$ and $y$) and 1D ($x$-only) initial perturbations to the isotropic state give rise to a subcritical bifurcation at this value of $\beta_T$.  \\

\begin{figure}
\centering
  \begin{subfigure}{.45\textwidth}
   \includegraphics[scale=0.26]{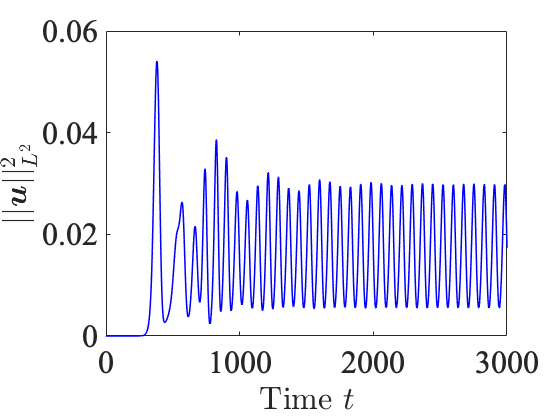}
        \caption{}
   \label{subfig:realeig_sub_L2}
   \end{subfigure}
   \begin{subfigure}{.45\textwidth}
   \includegraphics[scale=0.26]{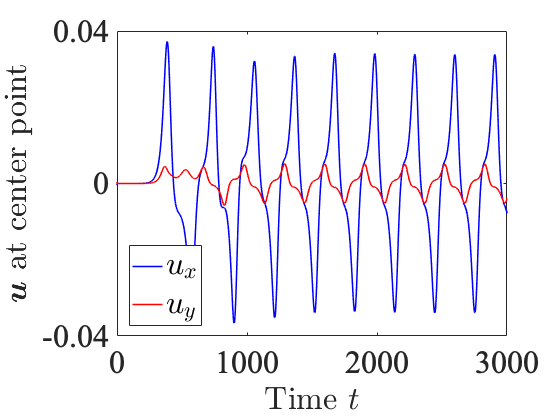} 
          \caption{}
   \label{subfig:realeig_sub_uxuy}
   \end{subfigure}
   \begin{subfigure}{.6\textwidth}
   \includegraphics[scale=0.1]{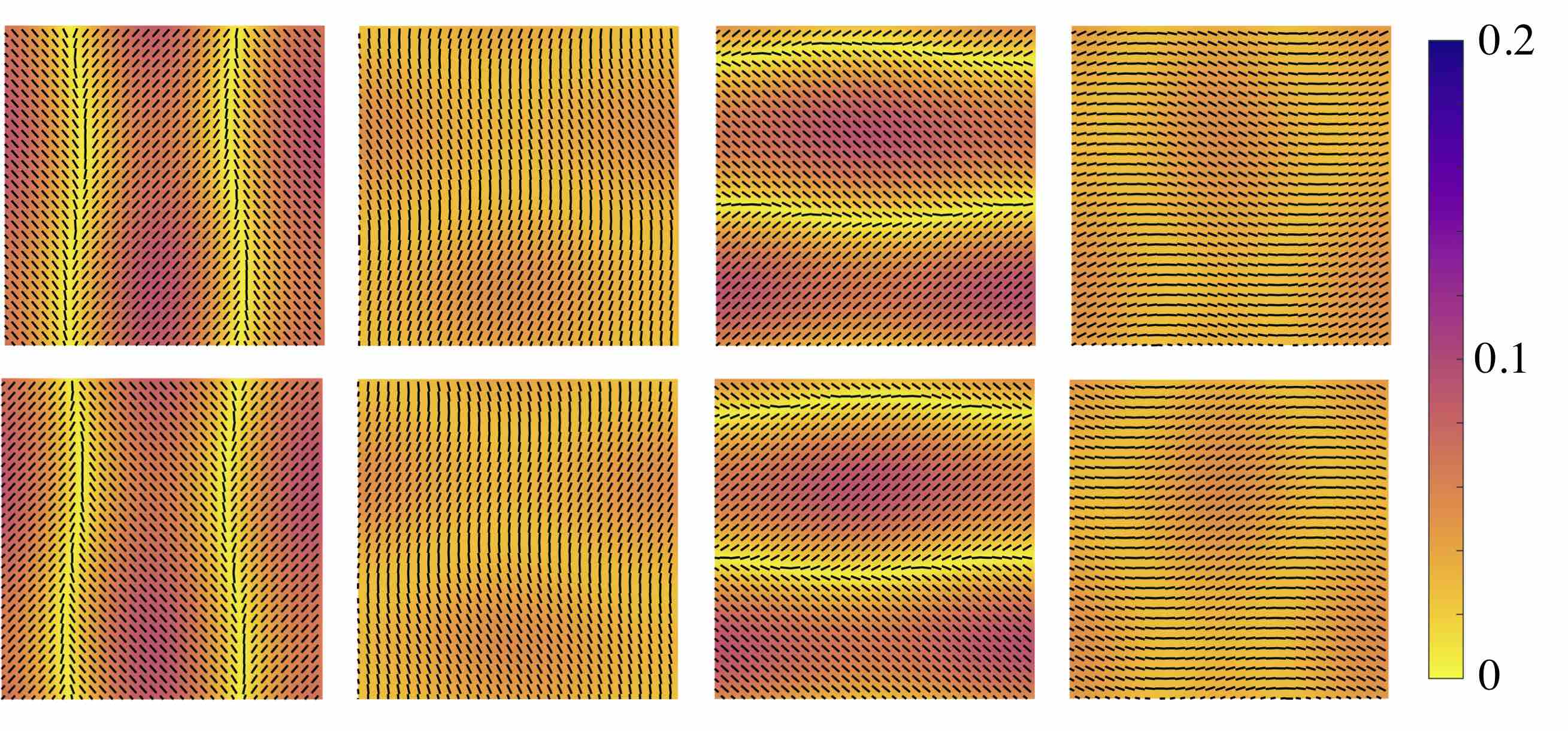}
    \caption{}
   \label{subfig:realeig_sub_SOP}
   \end{subfigure}
\caption{For an initial perturbation in $x$ and $y$, after a quick spike in activity, the system settles into a limit cycle following the subcritical pitchfork bifurcation. Figure (a) shows $\norm{\bu}_{L^2(\T^2)}^2$ over time. Figure (b) shows the $x$-component (blue) and $y$-component (red) of $\bu$ evaluated at the center of the domain over time. Figure (c) displays snapshots of the nematic order parameter $\mc{N}(\bx,t)$ and direction of local nematic alignment at successive peaks and valleys in $\norm{\bu}_{L^2(\T^2)}^2$ over one period of the cycle, starting with a peak.
} 
\label{fig:realeig_018_sub_xy}
\end{figure}

Using a small, random 2D perturbation to the uniform, isotropic state as our initial condition, we simulate the model dynamics until $t=3000$ using $\beta=0.17$, just on the unstable side of $\beta_T$. As predicted by the amplitude equation coefficients in figure \ref{fig:realeig_coeffs}, and in contrast to the supercritical setting, the system does not settle into a stable nontrivial steady state. Rather, the system undergoes a relatively quick spike in activity before settling into what appears to be a stable limit cycle, as shown in figure \ref{fig:realeig_018_sub_xy} as well as the supplementary video \texttt{Movie7}). \\

The fast initial spike in activity may be contrasted with the supercritical Hopf bifurcation, where the oscillations grow slowly in amplitude over a much longer time before reaching the near-periodic dynamics shown in figure \ref{fig:hopf_040_vels}. Since the isotropic state is predicted to lose stability through a subcritical pitchfork bifurcation for this parameter combination, we may be seeing the results of a second bifurcation. Unlike the subcritical Hopf setting, this behavior does not persist beyond the bifurcation value -- the system appears to quickly return to the isotropic state when $\beta$ is increased above $\beta_T$. \\

\begin{figure}
\centering
  \begin{subfigure}{.45\textwidth}
   \includegraphics[scale=0.12]{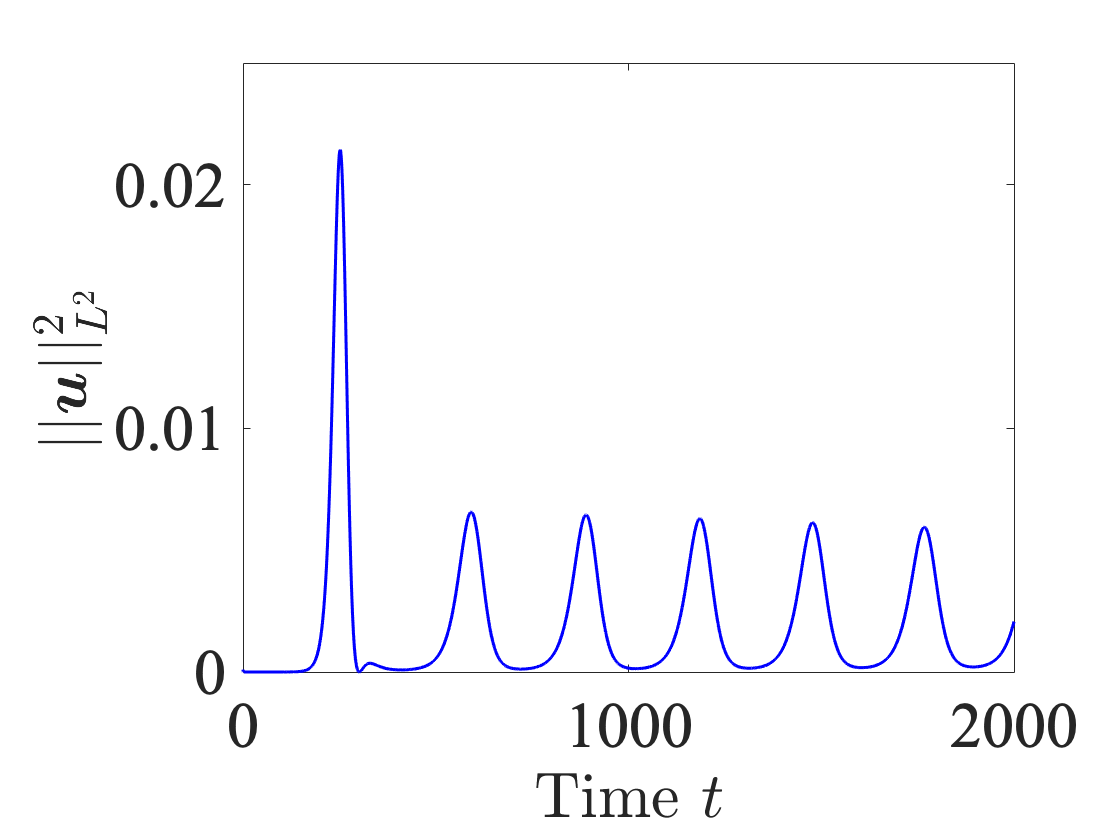}
         \caption{}
   \label{subfig:realeig_sub_L2_xonly}
  \end{subfigure}
    \begin{subfigure}{.45\textwidth}
   \includegraphics[scale=0.12]{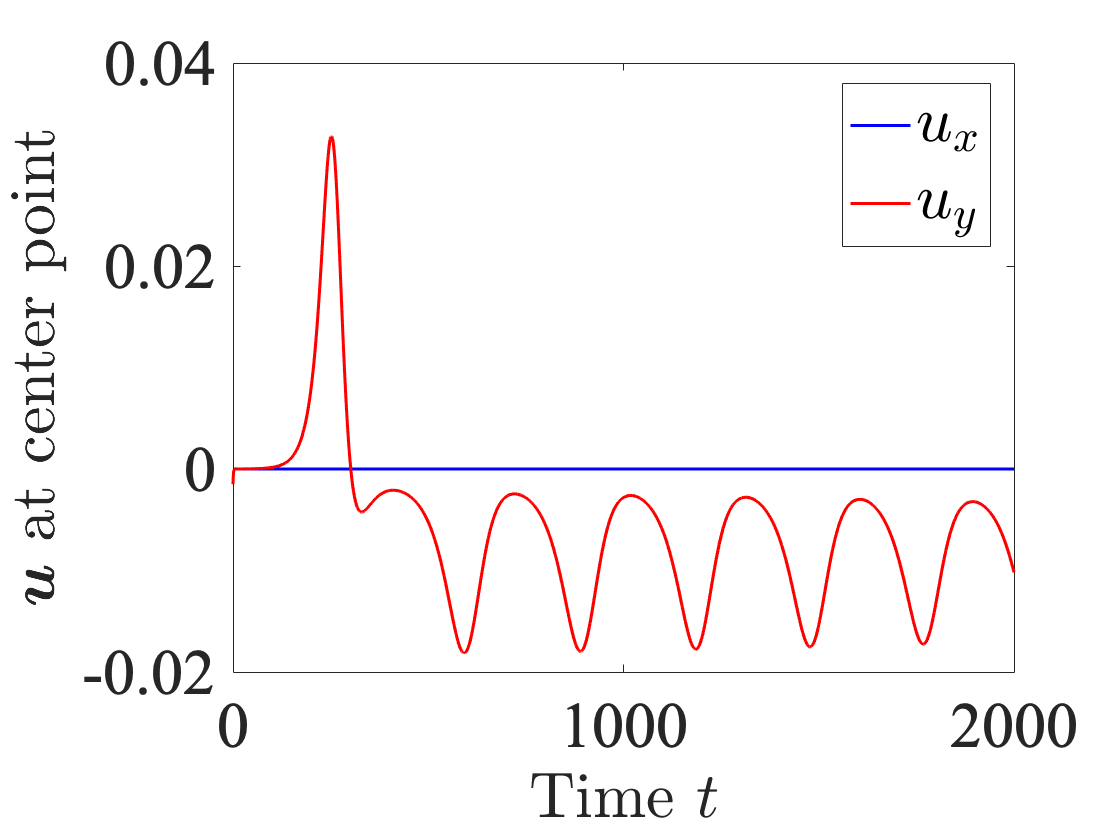} 
            \caption{}
   \label{subfig:realeig_sub_uxuy_xonly}
   \end{subfigure}
   \begin{subfigure}{.55\textwidth}
    \includegraphics[scale=0.1]{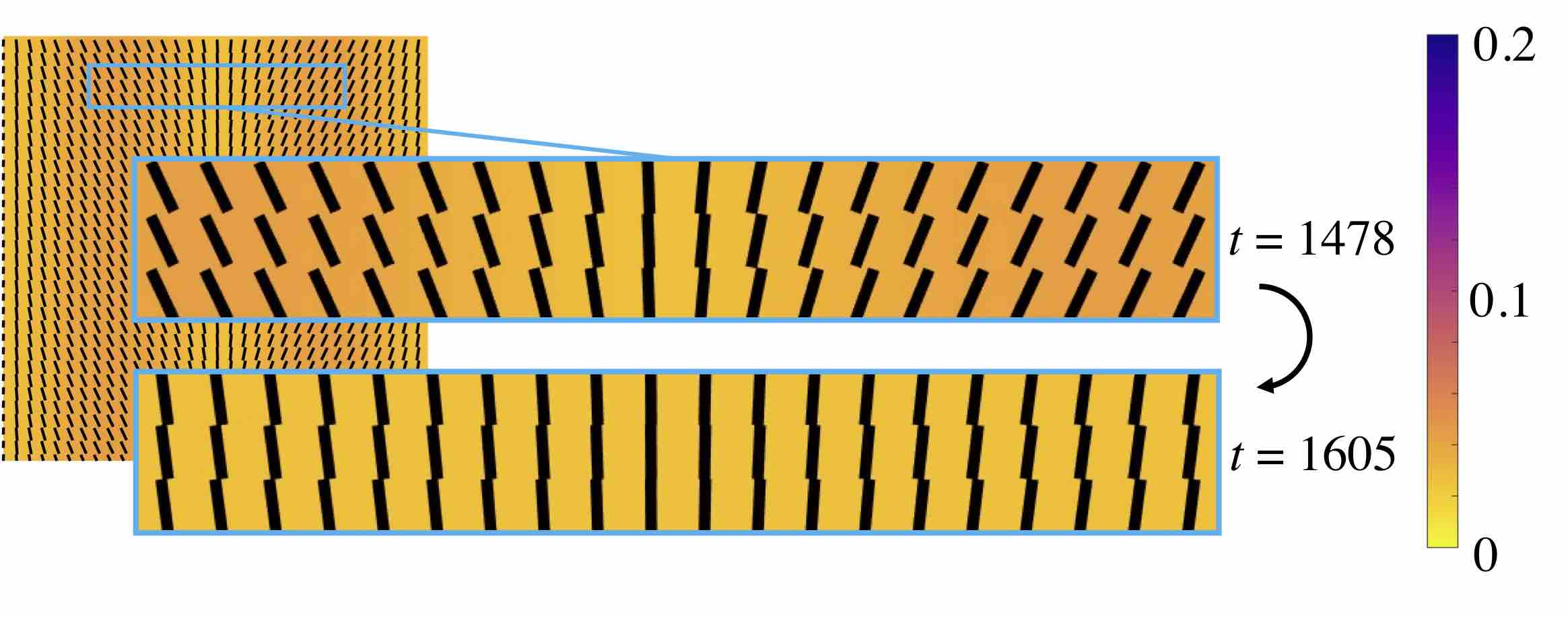}
                \caption{}
   \label{subfig:realeig_sub_SOP_xonly}
   \end{subfigure}
\caption{An initial $x$-only perturbation also results in a limit cycle following the subcritical pitchfork bifurcation. Figure (a) shows $\norm{\bu}_{L^2(\T^2)}^2$ over time. Figure (b) shows the $x$-component (blue) and $y$-component (red) of $\bu$ evaluated at the center of the domain over time. Figure (c) shows the nematic order parameter $\mc{N}(\bx,t)$ and direction of local nematic alignment at the $L^2$ norm peak and valley, respectively. The system oscillates slowly between the two states pictured.
} 
\label{fig:realeig_018_sub_xonly}
\end{figure}

Similar behavior is observed for $x$-only initial perturbations, as shown in figure \ref{fig:realeig_018_sub_xonly}. In particular, using the same parameter values as in the 2D setting, we see a quick initial spike in activity which then settles into what appears to nearly be a limit cycle (the amplitude here is very slightly decreasing over time). Again, this near-periodic behavior may be the result of a second bifurcation beyond the subcritical pitchfork. 
In both the 2D ($x$ and $y$) and 1D ($x$-only) cases, however, we find no evidence of any type of hysteresis like what is seen in the subcritical Hopf setting.  \\

\section{Discussion}\label{sec:disc}
 We have determined exactly how the uniform isotropic steady state loses stability in the Saintillan--Shelley kinetic model of a dilute suspension of active rodlike particles. Our weakly nonlinear analysis reveals a surprisingly complex array of possible bifurcations that the system may undergo as the nondimensional swimming speed (ratio of swimming speed to concentration and active stress magnitude) is decreased below the instability threshold. The type of bifurcation depends on the particle diffusivity, and passes from supercritical pitchfork (relatively high diffusion or immotile particles) to subcritical pitchfork to supercritical Hopf to subcritical Hopf as the diffusivity is decreased.
 The analysis is supported by numerical examples of each different bifurcation in the model. The numerical examples include uncovering surprising structures in the hysteretic solution that is bistable with the uniform isotropic state in the subcritical Hopf region, as well as locating different 1D and 2D limit cycles and steady states which emerge following the supercritical Hopf and pitchfork bifurcations.  \\

 The bifurcation analysis presented here provides mathematical insight into the onset of collective particle motion and active turbulence. It would be interesting to see if the full range of different transitions to collective motion can be realized experimentally.
The patterns observed here may depend strongly on the aspect ratio of the periodic domain, as evidenced especially by the difference in behavior for 1D versus 2D initial perturbations in the Hopf region (section \ref{subsec:hopf_sub}). In particular, a long, thin 2D domain may give rise to patterns more similar to the 1D case. Furthermore, while a similar linear stability picture to section \ref{subsec:linearstab} holds in 3D \citep{hohenegger2010stability}, the effects of an additional spatial and orientational dimension on sub- versus supercriticality remain unclear. These questions both warrant further study. 
\\

  From a modeling perspective, it would be useful to perform a similar bifurcation analysis for different closure models derived from the kinetic theory to see to what extent different closures can capture the complexity of the transition to collective behavior seen in the full kinetic model. It also may be interesting to analyze how the inclusion of steric interactions among particles \citep{gao2017analytical} affects the nature of the possible types of bifurcations to collective motion. 
From a mathematical perspective, it would be interesting to prove that the uniform isotropic steady state is indeed stable above the bifurcation at $\beta=\beta_T$ and that this stability boundary is sharp. Our numerical tests indicate that the eigenvalues calculated in Section \ref{subsec:linearstab} provide a fairly full picture of the stability boundaries, but it would be reassuring to verify this rigorously, particularly in the absence of translational or rotational diffusion. See forthcoming work \citep{albritton2022stabilizing} addressing the stabilizing effects of the swimming term in the Saintillan--Shelley model.


\backsection[Supplementary data]{\label{SupMat}Supplementary movies are available online. } 

\backsection[Acknowledgements]{We thank Scott Weady for supplying a base code for numerics.}

\backsection[Funding]{LO acknowledges support from NSF postdoctoral fellowship DMS-2001959. MJS acknowledges support from NSF grants DMR-2004469 and DMR-1420073 (NYU-MRSEC). }

\backsection[Declaration of interests]{ The authors report no conflict of interest.}





\appendix
\section{Immotile bifurcation}\label{app:noswim}
Using the expressions for $\Psi_1$ and $\bu_1$ \eqref{noswim_psi1}, we can calculate the right hand side of equation \eqref{Oeps2_noswim}, given by  
\begin{equation}\label{Oeps2_noswim_RHS}
\begin{aligned}
-\bu_1\bcdot\bnabla\Psi_1 &- \divp\big(({\bf I}-\bp\bp^{\rm T})(\bnabla\bu_1\bp)\Psi_1) \big) \\
&= G_1(\theta){\rm e}^{{\rm i}(x+y)}A_xA_y+ G_2(\theta){\rm e}^{{\rm i}(x-y)}A_x\overline A_y \\
&\qquad + G_3(\theta)({\rm e}^{2{\rm i}x}A_x^2+\abs{A_x}^2) + G_4(\theta)({\rm e}^{2{\rm i}y} A_y^2+\abs{A_y}^2) + {\rm c.c.},
\end{aligned}
\end{equation}
where 
\begin{align*}
G_1 &= -\frac{c_xc_y}{8}\big( \sin^4\theta+\cos^4\theta-6\cos^2\theta\sin^2\theta+ 2\cos\theta\sin\theta \big),\\
G_2 &= -\frac{c_xc_y}{8}\big( \sin^4\theta+\cos^4\theta-6\cos^2\theta\sin^2\theta- 2\cos\theta\sin\theta \big), \\
G_3 &= -\frac{c_x^2}{8}\big( \cos^4\theta- 3\cos^2\theta\sin^2\theta \big), 
\quad G_4 = -\frac{c_y^2}{8}\big(\sin^4\theta - 3\sin^2\theta\cos^2\theta\big).
\end{align*}

The form of \eqref{Oeps2_noswim_RHS} leads us to consider the ansatz \eqref{noswim_psi2} for $\Psi_2$ and $\bu_2$. Plugging this ansatz into the operator $\mc{L}$, the left hand side of the $O(\epsilon^2)$ equation \eqref{Oeps2_noswim} takes the form 
\begin{equation}\label{Oeps2_noswim_LHS}
\begin{aligned}
\mc{L}[\Psi_2] 
&= -\frac{1}{4\upi}(\cos^2\theta - \sin^2\theta)\int_0^{2\upi}\bigg(\psi_{2,1}{\rm e}^{{\rm i}(x+y)}A_xA_y+\psi_{2,2}{\rm e}^{{\rm i}(x-y)}A_x\overline A_y\bigg)(\cos^2\theta -\sin^2\theta) d\theta \\
&\quad - \frac{1}{\upi}\cos\theta\sin\theta\int_0^{2\upi}\bigg( \psi_{2,3}{\rm e}^{2{\rm i}x} A_x^2 +\psi_{2,4}{\rm e}^{2{\rm i}y} A_y^2\bigg)\sin\theta\cos\theta d\theta \\
&\quad +\bigg(\frac{1}{4}-4D_R\bigg)\big(2\psi_{2,1}{\rm e}^{{\rm i}(x+y)}A_xA_y+ 2\psi_{2,2}{\rm e}^{{\rm i}(x-y)}A_x\overline A_y + 4\psi_{2,3}{\rm e}^{2{\rm i}x}A_x^2 +4\psi_{2,4} {\rm e}^{2{\rm i}y}A_y^2\big) \\
&\quad - D_R\bigg((\p_{\theta\theta}\psi_{2,1}){\rm e}^{{\rm i}(x+y)}A_xA_y+ (\p_{\theta\theta}\psi_{2,2}){\rm e}^{{\rm i}(x-y)}A_x\overline A_y+ (\p_{\theta\theta}\psi_{2,3}){\rm e}^{2{\rm i}x}A_x^2 \\
&\hspace{2cm} + (\p_{\theta\theta}\psi_{2,4}) {\rm e}^{2{\rm i}y}A_y^2 + (\p_{\theta\theta}\psi_{2,5})\abs{A_x}^2 + (\p_{\theta\theta}\psi_{2,6})\abs{A_y}^2\bigg) + {\rm c.c.}
\end{aligned}
\end{equation}

Matching the exponents in \eqref{Oeps2_noswim_RHS} to \eqref{Oeps2_noswim_LHS}, we obtain a series of six independent integrodifferential ODEs which may each be solved to yield the coefficients $\psi_{2,j}$ listed in \eqref{noswim_O2coeffs}. \\

We also need to keep track of the components of $R_x(\theta,\tau){\rm e}^{{\rm i}x}$ and $R_y(\theta,\tau){\rm e}^{{\rm i}y}$ appearing in the expression $\mc{R}$ \eqref{noswim_RHSform} for the right hand side of the $O(\epsilon^3)$ equation \eqref{Oeps3_noswim}. \\

Each term of $\mc{R}$ contributes the following to $R_x(\theta,\tau){\rm e}^{{\rm i}x}$ and $R_y(\theta,\tau){\rm e}^{{\rm i}y}$:
\begin{equation}\label{noswim_RxRy}
\begin{aligned}
\p_\tau \Psi_1 :& \quad \cos\theta\sin\theta\big(c_x (\p_\tau A_x){\rm e}^{{\rm i}x} +c_y (\p_\tau A_y){\rm e}^{{\rm i}y} \big) + {\rm c.c.} \\
\Delta \Psi_1 :& \quad -\cos\theta\sin\theta\big(c_x A_x{\rm e}^{{\rm i}x} +c_yA_y{\rm e}^{{\rm i}y} \big) +{\rm c.c.}\\
\bu_1\bcdot\bnabla\Psi_2 
:& \quad  -\frac{\sin\theta\cos\theta}{64D_R}\bigg( c_xc_y^2 {\rm e}^{{\rm i}x}A_x\abs{A_y}^2 + c_x^2c_y {\rm e}^{{\rm i}y}\abs{A_x}^2A_y  \bigg)+{\rm c.c.} \\
\divp\big( ({\bf I}-\bp\bp^{\rm T})(\bnabla\bu_1\bp)\Psi_2 \big) :& \\
&\hspace{-3cm} 
\bigg( r_1\cos\theta\sin\theta(4\sin^2\theta -6\sin^4\theta -1) +r_2\cos^3\theta\sin\theta (2 -3\cos^2\theta) \\
&\hspace{-3cm}   - (r_3+r_4)\sin\theta\cos\theta \bigg) \bigg(c_xc_y^2{\rm e}^{{\rm i}x}A_x\abs{A_y}^2+c_x^2c_y{\rm e}^{{\rm i}y}\abs{A_x}^2A_y\bigg) \\
&\hspace{-3cm}  - \cos\theta\sin\theta\bigg(3r_5\cos^4\theta + 2r_6\cos^2\theta + r_7 \bigg)\bigg(c_x^3{\rm e}^{{\rm i}x}\abs{A_x}^2A_x  +c_y^3{\rm e}^{{\rm i}y}\abs{A_y}^2A_y\bigg) +{\rm c.c.}
\end{aligned}
\end{equation}
where
\begin{align*}
r_1 &= -\frac{1}{2(1+16D_R)}, 
\quad r_2 = -\frac{1}{64D_R}, 
\quad r_3 = \frac{3}{8(1+16D_R)}, 
\quad r_4 = \frac{3}{512D_R},  \\
 r_5 &= -\frac{1+8D_R}{64D_R}, 
\quad r_6 = \frac{3(1-16D_R)}{32(1-12D_R)}, 
\quad r_7 = 3\frac{32D_R^2-12D_R+1}{512D_R(1-12D_R)} .
\end{align*}

The Fredholm condition \eqref{noswim_fredholm} then requires that
\begin{align*}
\alpha_xF[R_x(\theta,\tau)]{\rm e}^{{\rm i}x} + \alpha_yF[R_y(\theta,\tau)]{\rm e}^{{\rm i}y} +{\rm c.c.}=0, \qquad \alpha_x^2+\alpha_y^2=1,
\end{align*}
where again $F[\bcdot]=\int_0^{2\upi}\bcdot \, \cos\theta\sin\theta \,d\theta$. In particular, we need $F[R_x]=F[R_y]=0$. The contribution of each term in \eqref{noswim_RxRy} to $F[R_x]$ is given by
\begin{align*}
&\p_\tau \Psi_1 : \quad \frac{\upi}{4}c_x(\p_\tau A_x), \qquad \Delta \Psi_1 : \quad -\frac{\upi}{4}c_x A_x,  \qquad \bu_1\bcdot\bnabla\Psi_2 : \quad \frac{\upi}{4}\frac{c_xc_y^2}{8(1-8D_R)}\abs{A_y}^2A_x,\\
 &\divp\big(({\bf I}-\bp\bp^{\rm T})(\bnabla\bu_1\bp)\Psi_2 \big)
: \quad  \frac{\upi}{4}\bigg(-\frac{c_xc_y^2(7+48D_R)}{1024 D_R (1 +16D_R)} \abs{A_y}^2  + c_x^3\frac{3(3 - 28D_R - 32D_R^2) }{1024 D_R (1 - 12 D_R)}\abs{A_x}^2\bigg)A_x . 
\end{align*}

Similarly, the contribution of each term in \eqref{noswim_RxRy} to $F[R_y]$ is given by
\begin{align*}
&\p_\tau \Psi_1 : \quad \frac{\upi}{4}c_y (\p_\tau A_y), \qquad
\Delta \Psi_1 :\quad -\frac{\upi}{4}c_yA_y, \qquad 
\bu_1\bcdot\bnabla\Psi_2 :\quad \frac{\upi}{4}\frac{c_x^2c_y}{8(1-8D_R)}\abs{A_x}^2A_y, \\
&\divp\big(({\bf I}-\bp\bp^{\rm T})(\bnabla\bu_1\bp)\Psi_2 \big)
: \quad  \frac{\upi}{4}\bigg(-\frac{c_x^2c_y(7+48D_R)}{1024 D_R (1 +16D_R)}\abs{A_x}^2 + c_y^3\frac{3(3 - 28D_R - 32D_R^2) }{1024 D_R (1 - 12 D_R)} \abs{A_y}^2\bigg)A_y.
\end{align*}


\section{Hopf bifurcation}\label{app:hopf}
As in the immotile case, we may use the expressions \eqref{hopf_psi1} for $\Psi_1$ and $\bu_1$ to compute the right hand side of equation \eqref{hopf_Oeps2} up to $O(D_R)$: 
\begin{equation}\label{Oeps2_hopf_RHS}
\begin{aligned}
-\bu_1\bcdot\bnabla\Psi_1 &- \divp\big(({\bf I}-\bp\bp^{\rm T})(\bnabla\bu_1\bp)\Psi_1) \big) \\
&= H_1(\theta){\rm e}^{{\rm i}(x+y)+2{\rm i}b_Tt}A_xA_y+ H_2(\theta){\rm e}^{{\rm i}(x-y)}A_x\overline A_y \\
&\qquad + H_3(\theta)({\rm e}^{2{\rm i}x+2{\rm i}b_Tt}A_x^2+\abs{A_x}^2) + H_4(\theta)({\rm e}^{2{\rm i}y+2{\rm i}b_Tt} A_y^2+\abs{A_y}^2) + {\rm c.c.},
\end{aligned}
\end{equation}
where 
\begin{align*}
H_1 &= -\frac{c_xc_y}{2}\bigg[ \frac{ -3\cos^2\theta\sin^2\theta + \sin^4\theta +\cos\theta\sin\theta}{D_T+ {\rm i}b_T+ {\rm i}\beta_T\cos\theta} - {\rm i}\beta_T\frac{\sin^4\theta\cos\theta }{(D_T+ {\rm i}b_T+ {\rm i}\beta_T\cos\theta)^2} \\
&\qquad + \frac{-3\cos^2\theta\sin^2\theta + \cos^4\theta+\cos\theta\sin\theta}{D_T+ {\rm i}b_T+ {\rm i}\beta_T\sin\theta} - {\rm i}\beta_T\frac{ \cos^4\theta\sin\theta}{(D_T+ {\rm i}b_T+ {\rm i}\beta_T\sin\theta)^2} \bigg] + O(D_R) \\
H_2 &= -\frac{c_xc_y}{2}\bigg[ \frac{-3\cos^2\theta\sin^2\theta + \sin^4\theta -\cos\theta\sin\theta}{D_T+ {\rm i}b_T+ {\rm i}\beta_T\cos\theta}
 - {\rm i}\beta_T\frac{\sin^4\theta\cos\theta }{(D_T+ {\rm i}b_T+ {\rm i}\beta_T\cos\theta)^2} \\
&\qquad + \frac{-3\cos^2\theta\sin^2\theta + \cos^4\theta-\cos\theta\sin\theta}{D_T- {\rm i}b_T- {\rm i}\beta_T\sin\theta}
 + {\rm i}\beta_T\frac{\cos^4\theta\sin\theta}{(D_T- {\rm i}b_T- {\rm i}\beta_T\sin\theta)^2} \bigg] + O(D_R) \\
H_3 &= -\frac{c_x^2}{2}\bigg[ \frac{-3\cos^2\theta\sin^2\theta + \cos^4\theta }{D_T+ {\rm i}b_T+ {\rm i}\beta_T\cos\theta} - {\rm i}\beta_T\frac{-\cos^3\theta\sin^2\theta}{(D_T+ {\rm i}b_T+ {\rm i}\beta_T\cos\theta)^2}\bigg]+ O(D_R) \\
H_4 &= -\frac{c_y^2}{2}\bigg[\frac{-3\cos^2\theta\sin^2\theta + \sin^4\theta }{D_T+ {\rm i}b_T+ {\rm i}\beta_T\sin\theta} - {\rm i}\beta_T\frac{-\sin^3\theta\cos^2\theta }{(D_T+ {\rm i}b_T+ {\rm i}\beta_T\sin\theta)^2} \bigg] +O(D_R).
\end{align*}

Again, as in the immotile case, the form of the right hand side expression \eqref{Oeps2_hopf_RHS} leads us to consider the ansatz \eqref{hopf_psi2} for $(\Psi_2,\bu_2)$. Using this ansatz, the left hand side of the $O(\epsilon^2)$ equation \eqref{hopf_Oeps2} takes the following form: 
\begin{equation}\label{hopf_Oeps2_LHS}
\begin{aligned}
\mc{L}[\Psi_2] 
&= 2{\rm i}b_T\bigg(\psi_{2,1}{\rm e}^{{\rm i}(x+y)+2{\rm i}b_Tt}A_xA_y + \psi_{2,3}{\rm e}^{2{\rm i}x+2{\rm i}b_Tt}A_x^2 +\psi_{2,4}{\rm e}^{2{\rm i}y+2{\rm i}b_Tt}A_y^2\bigg) \\
&\quad + {\rm i}\beta_T\bigg((\cos\theta+\sin\theta)\psi_{2,1}{\rm e}^{{\rm i}(x+y)+2{\rm i}b_Tt}A_xA_y + (\cos\theta-\sin\theta)\psi_{2,2}{\rm e}^{{\rm i}(x-y)}A_x\overline A_y \\
&\hspace{2cm}+2\cos\theta\psi_{2,3}{\rm e}^{2{\rm i}x+2{\rm i}b_Tt}A_x^2 +2\sin\theta\psi_{2,4} {\rm e}^{2{\rm i}y+2{\rm i}b_Tt}A_y^2\bigg)  \\
&\quad - \frac{1}{4\upi}(\cos^2\theta - \sin^2\theta)\int_0^{2\upi}\bigg(\psi_{2,1}{\rm e}^{{\rm i}(x+y)+2{\rm i}b_Tt}A_xA_y+\psi_{2,2}{\rm e}^{{\rm i}(x-y)}A_x\overline A_y\bigg)(\cos^2\theta -\sin^2\theta) d\theta \\
&\quad - \frac{1}{\upi}\cos\theta\sin\theta\int_0^{2\upi}\bigg( \psi_{2,3}{\rm e}^{2{\rm i}x+2{\rm i}b_Tt} A_x^2 + \psi_{2,4}{\rm e}^{2{\rm i}y+2{\rm i}b_Tt} A_y^2\bigg)\sin\theta\cos\theta d\theta \\
&\quad  + D_T\bigg(2\psi_{2,1}{\rm e}^{{\rm i}(x+y)+2{\rm i}b_Tt}A_xA_y+ 2\psi_{2,2}{\rm e}^{{\rm i}(x-y)}A_x\overline A_y+ 4\psi_{2,3}{\rm e}^{2{\rm i}x+2{\rm i}b_Tt}A_x^2 +4\psi_{2,4}{\rm e}^{2{\rm i}y+2{\rm i}b_Tt}A_y^2\bigg) \\
&\quad - D_R\bigg( (\p_{\theta\theta}\psi_{2,1}){\rm e}^{{\rm i}(x+y)+2{\rm i}b_Tt}A_xA_y+ (\p_{\theta\theta}\psi_{2,2}){\rm e}^{{\rm i}(x-y)}A_x\overline A_y +(\p_{\theta\theta}\psi_{2,3}){\rm e}^{2{\rm i}x+2{\rm i}b_Tt}A_x^2 \\
&\hspace{2cm} +(\p_{\theta\theta}\psi_{2,4}) {\rm e}^{2{\rm i}y+2{\rm i}b_Tt}A_y^2 +(\p_{\theta\theta}\psi_{2,5})\abs{A_x}^2  +(\p_{\theta\theta}\psi_{2,6})\abs{A_y}^2\bigg) +{\rm c.c.}
\end{aligned}
\end{equation}

Equating exponents from the right hand side \eqref{Oeps2_hopf_RHS} with the left hand side \eqref{hopf_Oeps2_LHS}, we obtain six independent integrodifferential ODEs for the coefficients $\psi_{2,j}(\theta)$. \\

For each of $\psi_{2,j}$, $j=1,2,3,4$, it can be shown that (small) $D_R>0$ results in a small perturbation of the $D_R=0$ solution to the following set of equations: 
\begin{equation}\label{hopf_psi1234}
\begin{aligned}
&\big(2D_T +2{\rm i}b_T +{\rm i}\beta_T(\cos\theta+\sin\theta)\big)\psi_{2,1} - \frac{\cos^2\theta - \sin^2\theta}{4\upi} \int_0^{2\upi}\psi_{2,1}(\cos^2\theta - \sin^2\theta) d\theta - D_R\p_{\theta\theta}\psi_{2,1} \\
&\qquad = -\frac{c_xc_y}{2}\bigg[ \frac{ -3\cos^2\theta\sin^2\theta + \sin^4\theta +\cos\theta\sin\theta}{D_T+ {\rm i}b_T+ {\rm i}\beta_T\cos\theta} -{\rm i}\beta_T\frac{\sin^4\theta\cos\theta }{(D_T+{\rm i}b_T+{\rm i}\beta_T\cos\theta)^2} \\
&\hspace{2cm} + \frac{-3\cos^2\theta\sin^2\theta + \cos^4\theta+\cos\theta\sin\theta}{D_T+ {\rm i}b_T+ {\rm i}\beta_T\sin\theta} -{\rm i}\beta_T\frac{ \cos^4\theta\sin\theta}{(D_T+ {\rm i}b_T+ {\rm i}\beta_T\sin\theta)^2} \bigg]  +O(D_R), \\
& \big(2D_T + {\rm i}\beta_T(\cos\theta-\sin\theta)\big)\psi_{2,2}
- \frac{\cos^2\theta - \sin^2\theta}{4\upi}\int_0^{2\upi}\psi_{2,2}(\cos^2\theta -\sin^2\theta) d\theta -D_R\p_{\theta\theta}\psi_{2,2}\\
&\qquad = -\frac{c_xc_y}{2}\bigg[ \frac{ -3\cos^2\theta\sin^2\theta + \sin^4\theta +\cos\theta\sin\theta}{D_T+ {\rm i}b_T+ {\rm i}\beta_T\cos\theta}- {\rm i}\beta_T\frac{\sin^4\theta\cos\theta }{(D_T+ {\rm i}b_T+ {\rm i}\beta_T\cos\theta)^2} \\
&\hspace{2cm} + \frac{-3\cos^2\theta\sin^2\theta + \cos^4\theta+\cos\theta\sin\theta}{D_T+ {\rm i}b_T+ {\rm i}\beta_T\sin\theta}- {\rm i}\beta_T\frac{ \cos^4\theta\sin\theta}{(D_T+ {\rm i}b_T+ {\rm i}\beta_T\sin\theta)^2} \bigg] +O(D_R),\\
 &\big(4D_T +2{\rm i}b_T +2{\rm i}\beta_T\cos\theta\big)\psi_{2,3} -\frac{1}{\upi}\cos\theta\sin\theta \int_0^{2\upi}\psi_{2,3} \sin\theta\cos\theta \, d\theta - D_R\p_{\theta\theta}\psi_{2,3}\\
 &\qquad = -\frac{c_x^2}{2}\bigg[ \frac{-3\cos^2\theta\sin^2\theta + \cos^4\theta }{D_T+{\rm i}b_T+ {\rm i}\beta_T\cos\theta} -{\rm i}\beta_T\frac{-\cos^3\theta\sin^2\theta}{(D_T+{\rm i}b_T+ {\rm i}\beta_T\cos\theta)^2}\bigg] +O(D_R), \\
&\big(4D_T + 2{\rm i}b_T+ 2{\rm i}\beta_T\sin\theta\big)\psi_{2,4} -\frac{1}{\upi}\cos\theta\sin\theta \int_0^{2\upi}\psi_{2,4} \sin\theta\cos\theta \, d\theta  - D_R\p_{\theta\theta}\psi_{2,4} \\
&\qquad= -\frac{c_y^2}{2}\bigg[\frac{-3\cos^2\theta\sin^2\theta + \sin^4\theta }{D_T+ {\rm i}b_T+ {\rm i}\beta_T\sin\theta} - {\rm i}\beta_T\frac{-\sin^3\theta\cos^2\theta }{(D_T+ {\rm i}b_T+ {\rm i}\beta_T\sin\theta)^2} \bigg] +O(D_R).
\end{aligned}
\end{equation}
In particular, following a similar approach to Section \ref{subsec:DR}, it may be shown that the solutions to each of the above four equations is bounded independent of $D_R$ as $D_R\to 0$ for all values of the triple $(D_T,b_T,\beta_T)$ of interest.  \\

In contrast, the remaining two coefficients $\psi_{2,5}$ and $\psi_{2,6}$ blow up like $1/D_R$ as $D_R\to 0$:
\begin{equation}\label{hopf_psi56}
\begin{aligned}
\p_{\theta\theta}\psi_{2,5}  
&=\frac{c_x^2}{2D_R}\bigg[ \frac{-3\cos^2\theta\sin^2\theta + \cos^4\theta }{D_T+ {\rm i}b_T+ {\rm i}\beta_T\cos\theta} + {\rm i}\beta_T\frac{\cos^3\theta\sin^2\theta}{(D_T+ {\rm i}b_T+ {\rm i}\beta_T\cos\theta)^2}+O(D_R)\bigg] \\
\p_{\theta\theta}\psi_{2,6} 
&=\frac{c_y^2}{2D_R}\bigg[\frac{-3\cos^2\theta\sin^2\theta + \sin^4\theta }{D_T+ {\rm i}b_T+ {\rm i}\beta_T\sin\theta} + {\rm i}\beta_T\frac{\sin^3\theta\cos^2\theta }{(D_T+ {\rm i}b_T+ {\rm i}\beta_T\sin\theta)^2} +O(D_R) \bigg] .
\end{aligned}
\end{equation}
In particular, for sufficiently small $D_R>0$, the behavior of $\psi_{2,5}$ and $\psi_{2,6}$ dominate over each of the other terms which are quadratic in the amplitudes $A_x$ and $A_y$. We may then integrate the equations \eqref{hopf_psi56} in $\theta$ to obtain the expressions for $\psi_{2,5}$ and $\psi_{2,6}$ given in equation \eqref{hopf_psij}. \\


Moving on to the $O(\epsilon^3)$ equation \eqref{hopf_Oeps3}, we note that out of all the cubic-in-$A$ terms in the right hand side expression \eqref{hopf_cubic_in_A}, each term is $O(1)$ in $D_R$ except for $-\divp\big(({\bf I}-\bp\bp^{\rm T})(\bnabla\bu_1\bp)\Psi_2 \big)$, which is $O(D_R^{-1})$. For sufficiently small $D_R$, this term determines the behavior of the cubic-in-$A$ terms and gives rise to the following expressions for $R_{xx}(\theta)$, $R_{xy}(\theta)$, $R_{yx}(\theta)$, and $R_{yy}(\theta)$:
\begin{equation}\label{hopf_Rthetas}
\begin{aligned}
R_{xx} &= \frac{3}{2}a_1\cos^2\theta\sin\theta +4a_2\cos^3\theta\sin\theta-a_2\cos\theta\sin\theta +\frac{5}{2} a_3\cos^4\theta\sin\theta  \\
&\quad + \frac{a_4}{2}\frac{{\rm i}\beta_T\sin\theta\cos^2\theta}{D_T+{\rm i}b_T+{\rm i}\beta_T\cos\theta} + a_4\cos\theta\sin\theta \, L(\cos\theta) + O(D_R) \\
R_{xy} &= \frac{3}{2}a_1\sin^2\theta\cos\theta-\frac{a_1}{2}\cos\theta -4a_2\sin\theta\cos^3\theta+a_2\sin\theta\cos\theta +\frac{5}{2}a_3\cos\theta\sin^4\theta \\
 &\quad -\frac{3}{2}a_3\cos\theta\sin^2\theta - \frac{a_4}{2}\frac{{\rm i}\beta_T\cos^3\theta}{D_T+{\rm i}b_T+{\rm i}\beta_T\sin\theta} + a_4\sin\theta\cos\theta \, L(\sin\theta)  + O(D_R)\\ 
R_{yx} &= \frac{3}{2}a_1\cos^2\theta\sin\theta-\frac{a_1}{2}\sin\theta -4a_2\cos\theta\sin^3\theta+a_2\cos\theta\sin\theta +\frac{5}{2} a_3\cos^4\theta\sin\theta \\
&\quad  -\frac{3}{2}a_3\cos^2\theta\sin\theta - \frac{a_4}{2}\frac{{\rm i}\beta_T\sin^3\theta}{D_T+{\rm i}b_T+{\rm i}\beta_T\cos\theta} + a_4\cos\theta\sin\theta\, L(\cos\theta) +O(D_R) \\
R_{yy} &= \frac{3}{2}a_1\sin^2\theta\cos\theta +4a_2\sin^3\theta\cos\theta-a_2\sin\theta\cos\theta +\frac{5}{2}a_3\sin^4\theta\cos\theta \\
 &\quad + \frac{a_4}{2}\frac{{\rm i}\beta_T\cos\theta\sin^2\theta}{D_T+{\rm i}b_T+{\rm i}\beta_T\sin\theta} +a_4\sin\theta\cos\theta\, L(\sin\theta)  +O(D_R). 
\end{aligned}
\end{equation}
Here the constants $a_j$ are given in \eqref{a_vals}, and the notation $L(\cdot)$ is used to denote
\begin{equation}\label{Ldef}
L(z):= \log(D_T+{\rm i}b_T+{\rm i}\beta_T z) - \frac{1}{2\upi}\int_0^{2\upi}\log(D_T+{\rm i}b_T+{\rm i}\beta_T z)\, d\theta, \quad z=\sin\theta,\cos\theta.
\end{equation}

The solvability condition for equation \eqref{hopf_Oeps3} then requires integrating the expressions in \eqref{hopf_Rthetas} against the eigenmodes $\psi_{x,1}(\theta)$ or $\psi_{y,1}(\theta)$ to obtain the coefficients $M_j$ defined in \eqref{hopf_Mjs}. Most of the resulting $\theta$-integrals in \eqref{hopf_Mjs} may be evaluated analytically, leading to the following expressions for $M_0$, $M_1$, $M_2$, $M_3$: 
\begin{equation}\label{hopf_Mjs_expr}
\begin{aligned}
M_0&= -\frac{\upi}{\beta_T^4} \bigg(6 (D_T+{\rm i}b_T)^2 + \beta_T^2 
- (6(D_T+{\rm i}b_T)^2+4\beta_T^2) \frac{(D_T+{\rm i}b_T)}{\sqrt{\beta_T^2+ (D_T+{\rm i}b_T)^2}}
  \bigg) \\
M_1 &=  \frac{\upi}{96\beta_T^8} 
\bigg(5\beta_T^6 + 232\beta_T^2(D_T+{\rm i}b_T)^4 + 512(D_T+{\rm i}b_T)^6 - 28\beta_T^4(D_T+{\rm i}b_T)^2 \\
&\quad
 -\frac{8(D_T+{\rm i}b_T)^3}{\sqrt{\beta_T^2 +(D_T+{\rm i}b_T)^2}} \big(61\beta_T^2(D_T+{\rm i}b_T)^2 + 64(D_T+{\rm i}b_T)^4+ 4\beta_T^4  \big)
\bigg) \\
&\quad + \frac{(D_T+{\rm i}b_T)^3}{2\beta_T^4}L_1(D_T,b_T,\beta_T)+ O(D_R) \\
M_2 &= \frac{\upi(D_T+{\rm i}b_T)^3}{4\beta_T^8} \bigg(\sqrt{\beta_T^2+(D_T+{\rm i}b_T)^2}\bigg( 4(D_T+{\rm i}b_T)^2 +\beta_T^2 +\frac{2\beta_T^2(D_T+{\rm i}b_T)^2}{\beta_T^2+2(D_T+{\rm i}b_T)^2}  \\
&\quad  -4(D_T+{\rm i}b_T)^3 -  4\beta_T^2(D_T+{\rm i}b_T) \bigg)
\bigg) + \frac{(D_T+{\rm i}b_T)^3}{2\beta_T^4}L_2(D_T,b_T,\beta_T) +O(D_R) \\
M_3 &= \frac{2\upi(D_T+{\rm i}b_T)}{\beta_T^5} \bigg(
        4\big(D_T+{\rm i}b_T)^2+ \beta_T^2- \frac{(4(D_T+{\rm i}b_T)^2+3\beta_T^2\big)(D_T+{\rm i}b_T)}{\sqrt{\beta_T^2+(D_T+{\rm i}b_T)^2} }
        \bigg),
\end{aligned}
\end{equation}

where 
\begin{align*}
L_1(D_T,b_T,\beta_T) &= \int_0^{2\upi}\frac{\cos^2\theta\sin^2\theta}{D_T+{\rm i}b_T+{\rm i}\beta_T\cos\theta}L(\cos\theta)\, d\theta 
=\int_0^{2\upi}\frac{\cos^2\theta\sin^2\theta}{D_T+{\rm i}b_T+{\rm i}\beta_T\sin\theta}L(\sin\theta) \, d\theta\\
L_2(D_T,b_T,\beta_T) &= \int_0^{2\upi}\frac{\sin^2\theta\cos^2\theta}{D_T+{\rm i}b_T+{\rm i}\beta_T\cos\theta}L(\sin\theta) \,d\theta 
= \int_0^{2\upi}\frac{\sin^2\theta\cos^2\theta}{D_T+{\rm i}b_T+{\rm i}\beta_T\sin\theta}L(\cos\theta) \, d\theta
\end{align*}
for $L(\cdot)$ as in \eqref{Ldef}.



\bibliographystyle{jfm}
\bibliography{patternbib}

\end{document}